\documentclass[onecolumn,epjc3]{svjour3} 
 
\usepackage[T1]{fontenc}
\usepackage{epsfig,amsmath,amssymb}
\usepackage{graphicx}
\usepackage{caption}
\usepackage{subcaption}
\usepackage{xspace}
\usepackage[listings]{tcolorbox}
\usepackage{ulem}
\usepackage{slashed}
\usepackage{cite}
\usepackage{xcolor}
\usepackage{units}
\usepackage{nicefrac}
\usepackage{textcomp}
\usepackage[utf8]{inputenc}
\usepackage{feynmp}
\usepackage[colorlinks,citecolor=blue,urlcolor=blue,linkcolor=blue]{hyperref}
\usepackage[capitalise]{cleveref}

\crefname{section}{Sec.}{Secs.}
\crefname{table}{Tab.}{Tabs.}
\crefname{figure}{Fig.}{Figs.}
\crefname{equation}{Eq.}{Eqs.}
\crefname{appendix}{Appendix\ }{Appendix\ }

\usepackage[framemethod=TikZ]{mdframed}
\usepackage{tikz} 
\usetikzlibrary{chains,shapes,fit,calc,snakes}
\usetikzlibrary{positioning,arrows}
\journalname{Eur. Phys. J. C}

\definecolor{maroon}{cmyk}{0, 0.87, 0.68, 0.32}
\definecolor{halfgray}{gray}{0.55}
\definecolor{slha_frame}{RGB}{207, 207, 207}
\definecolor{slha_bg}{RGB}{247, 247, 247}
\definecolor{slha_red}{RGB}{186, 33, 33}
\definecolor{slha_green}{RGB}{0, 128, 0}
\definecolor{slha_cyan}{RGB}{64, 128, 128}
\definecolor{slha_purple}{RGB}{170, 34, 255}

\definecolor{mathematica_frame}{RGB}{207, 207, 207}
\definecolor{mathematica_bg}{RGB}{247, 247, 247}
\definecolor{mathematica_red}{RGB}{186, 33, 33}
\definecolor{mathematica_green}{RGB}{0, 128, 0}
\definecolor{mathematica_cyan}{RGB}{64, 128, 128}
\definecolor{mathematica_purple}{RGB}{170, 34, 255}

\lstnewenvironment{MIN}[1][]{%
\mdframed[roundcorner=5pt,backgroundcolor=mathematica_bg]
  \renewcommand{\thelstnumber}{In[\arabic{lstnumber}]}
  \lstset{language=MathIn,numbers=left,basicstyle=\ttfamily,#1,escapeinside=||}%
}{%
\endmdframed
}

\lstnewenvironment{MOUT}[1][]{%
\mdframed[roundcorner=5pt,backgroundcolor=mathematica_bg]
  \renewcommand{\thelstnumber}{Out[\arabic{lstnumber}]}
  \lstset{language=MathOut,numbers=left,basicstyle=\ttfamily,#1}%
}{%
\endmdframed
}

\usepackage{listings}
\lstdefinelanguage{Fortran90}{
    morekeywords={Real,Complex,Intent},%
    emph={End,Subroutine,dp,in,Function,Implicit,None},%
    emphstyle={\color{mathematica_purple}},    
    sensitive=true,%
    morecomment=[l]\#,%
    morestring=[b]',%
    morestring=[b]",%
    morestring=[s]{'''}{'''},
    morestring=[s]{"""}{"""},
    morestring=[s]{r'}{'},
    morestring=[s]{r"}{"},%
    morestring=[s]{r'''}{'''},%
    morestring=[s]{r"""}{"""},%
    morestring=[s]{u'}{'},
    morestring=[s]{u"}{"},%
    morestring=[s]{u'''}{'''},%
    morestring=[s]{u"""}{"""},%
    identifierstyle=\color{black}\ttfamily,
    commentstyle=\color{slha_cyan}\ttfamily,
    stringstyle=\color{slha_red}\ttfamily,
    keepspaces=true,
    showspaces=false,
    showstringspaces=false,
    rulecolor=\color{slha_frame},
    frame=true,
    frameround={t}{t}{t}{t},
    framexleftmargin=6mm,
    numbers=left,
    numberstyle=\tiny\color{halfgray},
    backgroundcolor=\color{slha_bg},
    basicstyle=\footnotesize,
    keywordstyle=\color{slha_green}\ttfamily,
    aboveskip=1.2em,
    belowskip=1.2em,
}

\lstdefinelanguage{SLHA}{
    morekeywords={block,Block,BLOCK,decay,Decay,DECAY},%
    sensitive=true,%
    morecomment=[l]\#,%
    morestring=[b]',%
    morestring=[b]",%
    morestring=[s]{'''}{'''},
    morestring=[s]{"""}{"""},
    morestring=[s]{r'}{'},
    morestring=[s]{r"}{"},%
    morestring=[s]{r'''}{'''},%
    morestring=[s]{r"""}{"""},%
    morestring=[s]{u'}{'},
    morestring=[s]{u"}{"},%
    morestring=[s]{u'''}{'''},%
    morestring=[s]{u"""}{"""},%
    identifierstyle=\color{black}\ttfamily,
    commentstyle=\color{slha_cyan}\ttfamily,
    stringstyle=\color{slha_red}\ttfamily,
    keepspaces=true,
    showspaces=false,
    showstringspaces=false,
    rulecolor=\color{slha_frame},
    frame=true,
    frameround={t}{t}{t}{t},
    framexleftmargin=6mm,
    numbers=left,
    numberstyle=\tiny\color{halfgray},
    backgroundcolor=\color{slha_bg},
    basicstyle=\footnotesize,
    keywordstyle=\color{slha_green}\ttfamily,
    aboveskip=1.2em,
    belowskip=1.2em,
}

\lstdefinestyle{terminal} {
  morekeywords={cp,-r,make,cd},
  numbers=left, 
  stepnumber=1, 
  numberstyle=\tiny\color{halfgray}, 
  numbersep=10pt, 
  backgroundcolor=\color{black}, 
  basicstyle=\color{white}\ttfamily,
  stringstyle=\color{white}\ttfamily,
  keywordstyle=\color{white}\ttfamily\bfseries
 }

\lstdefinelanguage{MathIn}{
    morekeywords={Simplify,Eigenvalues,epsUV,Delta,UVscaleQ},%
    emph={Start,InitUnitarity,GetScatteringDiagrams,BuildScatteringMatrix,MakeSPheno,InitMatching,EFTcoupLO,EFTcoupNLO},%
    emphstyle={\color{mathematica_purple}},
    sensitive=true,%
    morecomment=[l]\%,%
    morestring=[b]',%
    morestring=[b]",%
    morestring=[s]{'''}{'''},
    morestring=[s]{"""}{"""},
    morestring=[s]{r'}{'},
    morestring=[s]{r"}{"},%
    morestring=[s]{r'''}{'''},%
    morestring=[s]{r"""}{"""},%
    morestring=[s]{u'}{'},
    morestring=[s]{u"}{"},%
    morestring=[s]{u'''}{'''},%
    morestring=[s]{u"""}{"""},%
    identifierstyle=\color{black}\ttfamily,
    commentstyle=\color{mathematica_cyan}\ttfamily,
    stringstyle=\color{mathematica_red}\ttfamily,
    keepspaces=true,
    showspaces=false,
    showstringspaces=false,
    rulecolor=\color{mathematica_frame},
    frame=none,
    numbers=left,
    numberstyle=\tiny\color{halfgray},
    basicstyle=\footnotesize,
    keywordstyle=\color{mathematica_green}\ttfamily,
    aboveskip=0.2em,
    belowskip=0.2em
}

\lstdefinelanguage{MathOut}{
    morekeywords={Simplify,Eigenvalues},%
    sensitive=true,%
    morecomment=[l]\%,%
    morestring=[b]',%
    morestring=[b]",%
    morestring=[s]{'''}{'''},
    morestring=[s]{"""}{"""},
    morestring=[s]{r'}{'},
    morestring=[s]{r"}{"},%
    morestring=[s]{r'''}{'''},%
    morestring=[s]{r"""}{"""},%
    morestring=[s]{u'}{'},
    morestring=[s]{u"}{"},%
    morestring=[s]{u'''}{'''},%
    morestring=[s]{u"""}{"""},%
    identifierstyle=\color{black}\ttfamily,
    commentstyle=\color{mathematica_cyan}\ttfamily,
    stringstyle=\color{mathematica_red}\ttfamily,
    keepspaces=true,
    showspaces=false,
    showstringspaces=false,
    rulecolor=\color{mathematica_frame},
    frame=none,
    frameround={t}{t}{t}{t},
    framexleftmargin=10mm,
    numbers=left,
    numberstyle=\tiny\color{halfgray},
    basicstyle=\footnotesize,
    keywordstyle=\color{mathematica_green}\ttfamily,
    aboveskip=0.2em,
    belowskip=0.2em,
}

\let\origthelstnumber\thelstnumber
\makeatletter
\newcommand*\Suppressnumber{%
  \lst@AddToHook{OnNewLine}{%
    \let\thelstnumber\relax%
     \advance\c@lstnumber-\@ne\relax%
    }%
}

\newcommand*\Reactivatenumber{%
  \lst@AddToHook{OnNewLine}{%
   \let\thelstnumber\origthelstnumber%
   \advance\c@lstnumber\@ne\relax}%
}

\def\postbreak{%
  \raisebox{0ex}[0ex][0ex]{\ensuremath{\hookrightarrow\space}}}

\def\lagr{\mathcal{L}}

\def\thv[#1,#2,#3]{\left( \begin{array}{c} #1 \\ #2 \\ #3 \end{array} \right)}
\def\twv[#1,#2]{\left( \begin{array}{c} #1 \\ #2 \end{array} \right)}

\def\beq{\begin{equation}}
\def\eeq{\end{equation}}

\newcommand\SARAH{{\tt SARAH}\xspace}

\newcommand\SPheno{{\tt SPheno}\xspace}

\newcommand\Mathematica{{\tt Mathematica}\xspace} 

\newcommand{\Fortran}{\texttt{Fortran}\xspace}

\newcommand{\DR}{\ensuremath{\overline{\text{DR}}}\xspace}
\newcommand{\MS}{\ensuremath{\overline{\text{MS}}}\xspace}

\definecolor{fsblue}{rgb}{0.,.0,1.}

\begin{document}

\title{Automatised matching between two scalar sectors at the one-loop level}
\author{
   M. Gabelmann \thanksref{a2} \and
   M. M. M\"uhlleitner \thanksref{a2} \and
   F. Staub \thanksref{a2,a3}
   }

\institute{
Institute for Theoretical Physics (ITP), Karlsruhe Institute of Technology, Engesserstra{\ss}e 7, D-76128 Karlsruhe, Germany \label{a2}
\and
Institute for Nuclear Physics (IKP), Karlsruhe Institute of Technology, Hermann-von-Helmholtz-Platz 1, D-76344 Eggenstein-Leopoldshafen, Germany \label{a3}
}

\date{}

\maketitle

\begin{abstract}
Nowadays, one needs to consider seriously the possibility that a large
separation between the scale of new physics and the electroweak scale exists.
Nevertheless, there are still observables in this scenario,  in particular the
Higgs mass, which are sensitive to the properties of the UV theory. In order to
obtain reliable predictions for a model which involves very heavy degrees of
freedom, the precise matching to an effective theory is necessary. While this
has been so far only studied for a few selected examples, we present an
extension of the \Mathematica package \SARAH to perform automatically the
matching between two scalar sectors at the full one-loop level for general
models. We show that we can reproduce all important results for commonly
studied models like split- or high-scale supersymmetry. One can now easily go
beyond that and study new ideas involving very heavy states, where the
effective model can either be just the standard model or an extension of it.
Also scenarios with several matching scales can be easily considered. We
provide model files for the MSSM with seven different mass hierarchies as well
as two high-scale versions of the NMSSM. Moreover, it is explained how new
models are implemented.
\end{abstract}

\section{Introduction}
The Standard Model (SM) of particle physics is a very successful theory which
has been completed with the discovery of the Higgs boson at the Large
Hadron Collider (LHC) 
\cite{Aad:2012tfa,Chatrchyan:2012xdj}. On the other side, there are
observations like dark matter for which no viable candidate exists
within the SM. While it has been expected that solutions to the open
problems of the SM, like e.g.~supersymmetry (SUSY), exist close
to the electroweak scale, the LHC has not found any direct  
signal for new physics so far. Therefore, the possibility of a large gap
between the electroweak (EW) and the scale of new physics has been studied more
intensively in the recent years. The most prominent idea in this direction is
'split supersymmetry' (split-SUSY) in which the SUSY scalars are much heavier
than the SM particles and the SUSY fermions \cite{ArkaniHamed:2004fb,
Giudice:2004tc, ArkaniHamed:2004yi}. In this setup, most of the appealing
properties of SUSY like gauge coupling unification and a dark matter candidate
are kept, but the coloured particles are too heavy to be produced at the LHC.
Mechanisms have been proposed how split-SUSY could arise from string theory
\cite{Antoniadis:2006eb,Antoniadis:2005em}, and also the question of
naturalness has been discussed \cite{Bhattacharyya:2012ct}. Moreover, the
ansatz of high-scale SUSY, i.e.~that all SUSY particles are much heavier than
the EW scale, is taken seriously nowadays
\cite{Giudice:2011cg,Bagnaschi:2014rsa}. While it is widely believed that these
models suffer from a large fine-tuning, it has pointed out that large SUSY
scales can be combined with the relaxion mechanism to solve the big and the
small hierarchy problem simultaneously \cite{Batell:2015fma}. The idea of SUSY
with very large mass scales is not restricted to the Minimal
Supersymmetric extension of the SM (MSSM), but has also been applied
to other SUSY models like the Next-to-MSSM (NMSSM)
\cite{Zarate:2016jch} or models with Dirac gauginos 
\cite{Unwin:2012fj,Benakli:2013msa,Dudas:2013gga,Fox:2014moa,Benakli:2015ioa}.  \\

Even if states beyond the SM (BSM) are too heavy to be produced
at current colliders, they often still have an in-print in experimental results, see
e.g. Refs.~\cite{Arvanitaki:2005fa,Ibe:2013oha}. The precise measurement of the
Higgs boson mass of $m_h = 125.09$~GeV \cite{Aad:2015zhl} at the LHC has added
another very important constraint in this 
direction. Consequently, large efforts were put in a precise Higgs boson mass
calculation in split- or high-scale SUSY
\cite{Bernal:2007uv,Arvanitaki:2012ps,Giudice:2011cg,Bagnaschi:2014rsa,Vega:2015fna}. 
The reason for this endeavour is that the commonly used fixed order
calculations of the Higgs boson mass in SUSY models should only be applied in the
case of a small separation between the EW scale and the SUSY scale. Otherwise,
the presence of large logarithms introduces a large uncertainty in the prediction of the numerical value of $m_h$ 
\cite{Vega:2015fna,Athron:2016fuq,Staub:2017jnp,Allanach:2018fif}.
This can be resolved either by the standard ansatz of an effective field theory
(EFT) in which the heavy states are integrated out
\cite{Sasaki:1991qu,Carena:1995bx,Haber:1990aw,Haber:1996fp,Carena:2000dp,Carena:2000yi,Carena:2001fw,Espinosa:2001mm,Gorbahn:2009pp,Lee:2015uza},
or by a hybrid method in which the fixed-order calculation is combined with the
higher-order leading logarithms extracted from an EFT
\cite{Hahn:2013ria,Bahl:2017aev,Bahl:2016brp,Bahl:2018jom}. In both cases, one
needs to know how the couplings among the light states depend on the full
theory. In terms of the EFT ansatz this means that the full model involving
heavy and light states  must be
matched to an effective theory at the scale at which the heavy degrees of
freedom are integrated out. The matching at leading order is straight-forward
and the relations often can be read off from the tree-level Lagrangians of both
models. However, tree-level relations are usually not sufficient to obtain the necessary
precision in the Higgs boson mass prediction. Therefore, higher-order corrections are
needed. Of course, the matching procedure at the full one-loop level is already 
much more time-consuming. Depending on the details of the full and effective
model also several subtleties like infra-red divergences can occur as discussed
in Ref.~\cite{Braathen:2018htl}.
\\

In order to facilitate these studies, we have developed an automatised process 
to perform the matching between the scalar sectors of two renormalisable
theories. This feature has been implemented in the \Mathematica package \SARAH
\cite{Staub:2008uz,Staub:2009bi,Staub:2010jh,Staub:2012pb,Staub:2013tta} and
provides the functionality to obtain analytical expressions for the matching
conditions at the one-loop level. Also the interface between \SARAH and \SPheno
\cite{Porod:2003um,Porod:2011nf} has been extended to include the matching
between an EFT and  a UV-complete theory. In that way, one can obtain very quickly
numerical predictions for the Higgs boson mass but also for all kind of other
observables that concern the Higgs boson. 
It is worth to stress that this functionality is not restricted to split- or
high-scale versions of the MSSM. A large variety of SUSY, but also non-SUSY models,
with large BSM scales can be studied with the presented tool-chain. Also the
considered EFT need not be the SM, but could be a singlet extension, a
Two-Higgs-Doublet-Model (THDM), or an even more complicated model. 
Concerning the nature of the heavy states, we restrict our attention to heavy
fermions and scalars. The implementation of integrating out heavy vector bosons
at the one-loop level is reserved for future work. However, the low-energy EFT
can still contain an extended gauge sector which is also matched at the one-loop level.
Nevertheless, we will mainly concentrate in the given examples on the
established MSSM scenarios because they offer the possibility to compare our
generic approach with results available in the literature.
\\

This paper is organised as follows. In \cref{sec:matching} we explain our
generic matching procedure. In \cref{sec:implementation} the new routines in
\SARAH are explained while a comparison with the literature is done in
\cref{sec:results}. We summarise in \cref{sec:summary}

\section{Generic Matching between Two Scalar Sectors}
\label{sec:matching}
\subsection{General Ansatz}
We consider a general, renormalisable gauge theory with a set of scalars
$\{\phi_i\}$ and fermions $\{\psi_i\}$ charged under unspecified
(sub-)sets of the theories gauge group. Without loss of generality,
one can always assume that the scalars are real. The Lagrangian can be written
as 
\begin{align}
    \mathcal{L}_{UV} = & \frac12 D_\mu \phi_a D^\mu \phi_a + \psi^\dagger_a
    \slashed{D}_\mu \psi_a + \mbox{Tr}\left( F_{\mu\nu, a} F^{\mu\nu}_a \right) \nonumber \\
\label{eq:LagUV}
 & \hspace{1cm}- \frac{1}{4!}\lambda_{abcd} \phi_a \phi_b \phi_c \phi_d - \frac{1}{3!}\kappa_{abc} \phi_a  \phi_b \phi_c - \frac{1}{2} m^2_{ab} \phi_a \phi_b - (M_{ab} \psi_a \psi_b+ Y_{abc} \psi_a \psi_b \phi_c + \text{h.c.})\, ,
\end{align}
where all gauge and representation indices have been suppressed. The covariant
derivative $D^\mu$ and the gauge fields are chosen such that the field strength tensors
$\{F^{\mu\nu}_a\}$ form diagonal kinetic terms (in case of multiple gauged U(1) groups).
In the following it is always assumed that all gauge groups are broken near the scale of EW symmetry
breaking.
If particles with very different masses appear in such a theory, one can
categorise the particle content into light fields ($\{\phi_i^L\},\{\psi_i^L\}$) 
and heavy fields  ($\{\phi_i^H\}, \{\psi_i^H\})$. The Lagrangian becomes
\begin{equation}
\mathcal{L}_{UV} = \mathcal{L}_L(\phi_i^L,\psi_i^L) + \mathcal{L}_{\rm mix}(\phi_i^L,\psi_i^L,\phi_i^H,\psi_i^H) + \mathcal{L}_H(\phi_i^H,\psi_i^H) \, .
\end{equation}
Integrating out all heavy fields leads to an effective theory which contains only light degrees of freedom
\begin{equation}
    \label{eq:lageft}
\begin{aligned}
    \mathcal{L}_{EFT} = &  \frac12 D_\mu \phi^L_a D^\mu \phi^L_a +
    \psi^{L,\dagger}_a \slashed{D}_\mu \psi^L_a  + \mbox{Tr}\left(F_{\mu\nu,a}
    F^{\mu\nu}_a \right) \\
    &- \frac{1}{4!}\tilde{\lambda}_{abcd} \phi^L_a \phi^L_b \phi^L_c \phi^L_d - \frac{1}{3!}\tilde{\kappa}_{abc} \phi^L_a  \phi^L_b \phi^L_c - \frac{1}{2}\ \tilde{m}^2_{ab} \phi^L_a \phi^L_b \\ 
    & - (\tilde{M}_{ab} \psi^L_a \psi^L_b + \tilde{Y}_{abc} \psi^L_a \psi^L_b \phi^L_c + \text{h.c.}) \\
    & + \mathcal{L}_{EFT}^{d>4} \, , 
\end{aligned}
\end{equation}
where the last line contains operators with dimension greater than four.
Concerning a precise prediction of Higgs boson masses, only purely scalar
operators with ascending dimensionality may be of interest for the matching. 
However, for $d>4$, their influence on the scalar potential is of 
the order $\nicefrac{v_i}{M_j}$, where $v_i$ is the vacuum expectation value (VEV) of a light
and $M_j$ the mass of a heavy field, $v_i \ll M_j$. Supposed that the
fundamental theory is  
renormalisable, it follows from the decoupling theorem, that the higher-dimensional 
operators become unimportant if $M_j\rightarrow \infty$.
The question arises, at which scale the $\nicefrac{v_i}{M_j}$ terms are no longer
relevant for a precise Higgs boson mass calculation.
The impact of dimension-six terms, compared to ordinary threshold corrections
($d\leq4$), on the Higgs boson mass in a matching of the SM to the MSSM
was studied in Ref. \cite{Bagnaschi:2017xid}. It was found that for
$\unit[500]{GeV}< M_j < \unit[1000]{GeV}$, a two-loop matching 
of these operators yields corrections on $m_h$ in the sub-GeV range, which rapidly drop for $M_j> \unit[1]{TeV}$.
Since the focus of this work is on BSM scenarios with
$M_j \geqslant\unit[1]{TeV}$, we neglect all $\nicefrac{v_i}{M_j}$
contributions during the matching.  Thus, we assume that all VEVs responsible for 
the breaking of a low-energy gauge theory can be neglected compared to the masses of the heavy states. 
In particular, this means that all gauge bosons as well as chiral fermions are treated as massless in 
the computation of the matching conditions.
\\
\\
All information about the heavy states is encoded in the effective
couplings and masses $\tilde{\lambda}$, $\tilde{\kappa}$, $\tilde{m}^2$, $\tilde{M}$ and
$\tilde{Y}$. The purely scalar interactions $\tilde{\lambda}$, $\tilde{\kappa}$
and mass squared $\tilde{m}^2$ contain the crucial
information about the scalar sector of 
the EFT, hence, they have the biggest impact on the properties of the light
scalars. We know today, that (at least) one of these light scalars must have
couplings comparable to the predictions of an SM-like Higgs boson and
the mass must be about 125~GeV. Thus, even if the mass scale of the
heavy fields is well above the reach of the LHC, we can test if the
fundamental UV theory is consistent with the Higgs 
boson mass measurements through a precise calculation of the effective couplings at
the matching scale and the Higgs properties at the weak scale. In order to
determine the effective couplings in terms of parameters of the UV-theory, one assumes the matching condition that the $n$-loop
$m$-point amplitudes involving the same external (light) states must yield the same result in the infra-red (IR) regime of 
the UV-theory (i.e. the scale where the heavy fields are integrated out) and the EFT, 
\begin{equation}
\label{eq:matchcond}
\mathcal{M}^{(n)}(\phi^L_a,\dots,\phi^L_m)^{UV} =  \mathcal{M}^{(n)}(\phi^L_a,\dots,\phi^L_m)^{EFT}\, .
\end{equation}
Note, that the external fields in the two theories to be matched must be treated equally. 
Thus, additional wave-function renormalisations involing internal heavy fields may contribute to \cref{eq:matchcond} by also matching the first derivative of the 2-point function w.r.t the external momentum of the light fields.
\begin{figure}[tb]
\begin{eqnarray*}
&
\begin{aligned}
\includegraphics{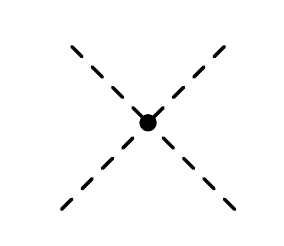}
\end{aligned}
+\left(
\begin{aligned}
\includegraphics{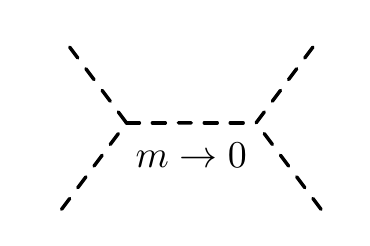}
\end{aligned}
+ t + u\right)&\\
&=
\begin{aligned}
\includegraphics{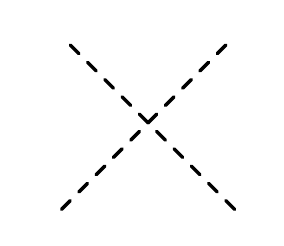}
\end{aligned}
+\left(
\begin{aligned}
\includegraphics{FeynMF/FM_MatchTree2.pdf}
\end{aligned}
+ t + u\right)
+\left(
\begin{aligned}
\includegraphics{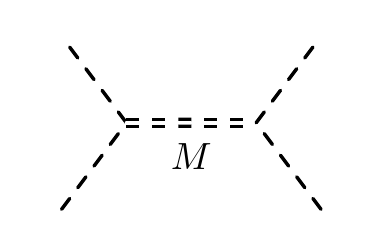}
\end{aligned}
+ t + u\right) &
\end{eqnarray*}
\caption{Matching of quartic couplings at tree-level. Light scalars are drawn
with a single dashed line, heavy scalars with a double dashed line.}
\label{fig:tree_matching}
\end{figure}
In this paper, we are going to calculate $\mathcal{M}$ using the Feynman diagrammatic approach
 neglecting all external momenta. The tree-level matching condition for a
quartic coupling, \begin{align}
\tilde{\lambda}_{abcd} - \sum_{x \in \{\phi^L\}} \left(\frac{\tilde{\kappa}_{abx} \tilde{\kappa}_{xcd}}{m_x^2\to 0} + \frac{\tilde{\kappa}_{acx} \tilde{\kappa}_{xbd}}{m_x^2\to 0} + \frac{\tilde{\kappa}_{adx} \tilde{\kappa}_{xcb}}{m_x^2\to 0}  \right) 
& = 
\lambda_{abcd} \nonumber \\
& - \sum_{x \in \{\phi^L\}} \left(\frac{{\kappa}_{abx} {\kappa}_{xcd}}{m_x^2\to 0} + \frac{{\kappa}_{acx} {\kappa}_{xbd}}{m_x^2\to 0} + \frac{{\kappa}_{adx} {\kappa}_{xcb}}{m_x^2\to 0}  \right) \nonumber \\
& - \sum_{x \in \{\phi^H\}} \left(\frac{{\kappa}_{abx} {\kappa}_{xcd}}{M_x^2} + \frac{{\kappa}_{acx} {\kappa}_{xbd}}{M_x^2} + \frac{{\kappa}_{adx} {\kappa}_{xcb}}{M_x^2}  \right)\, , \label{eq:matchingirdiv}
\end{align}
is depicted in \cref{fig:tree_matching}. Due to the assumption of vanishing external momenta and vanishing light masses, infra-red divergences appear on both sides of \cref{eq:matchingirdiv}.
Since the tree-level matching for cubic couplings is trivial,
\begin{equation}
 \tilde{\kappa}_{abc} =  \kappa_{abc}\, ,
\end{equation}
the divergences cancel exactly. Thus, the effective quartic couplings 
$\tilde{\lambda}_{abcd}$ are given by
\begin{align}
\tilde{\lambda}_{abcd}  = &  \lambda_{abcd} - \sum_{x \in \{\phi^H\}} \left(\frac{{\kappa}_{abx} {\kappa}_{xcd}}{M_x^2} + \frac{{\kappa}_{acx} {\kappa}_{xbd}}{M_x^2} + \frac{{\kappa}_{adx} {\kappa}_{xcb}}{M_x^2}  \right) \,.
\end{align}
As already mentioned, the matching at tree level is not sufficient for a precise prediction of the
properties of the scalar sector at the low-energy scale. Thus, one needs to include
loop corrections changing the matching conditions to
\begin{align}
& \tilde{\kappa}_{abc} + \delta \tilde{\kappa}_{abc}(\phi^L,\psi^L) = \kappa_{abc} + \delta \kappa_{abc}(\phi^L,\phi^H,\psi^L,\psi^H)
\end{align}
\begin{align}
    \tilde{\lambda}_{abcd} - \sum_{x \in \{\phi^L\}} \sum_{\rm s,t,u} \frac{\tilde{\kappa}_{abx} \tilde{\kappa}_{xcd}}{m_x^2\to 0} + \delta \tilde{\lambda}_{abcd}(\phi^L,\psi^L) &= 
    \lambda_{abcd} -  \sum_{\rm s,t,u}\left(\sum_{x \in \{\phi^L\}} \frac{{\kappa}_{abx} {\kappa}_{xcd}}{m_x^2\to 0} + \sum_{x \in \{\phi^H\}} \frac{{\kappa}_{abx} {\kappa}_{xcd}}{M_x^2}\right) \nonumber \\[3mm]
    &\quad + \delta \lambda_{abcd}(\phi^L,\phi^H,\psi^L,\psi^H)\, ,
\end{align}
where $\delta X(\phi^L,\phi^H,\psi^L,\psi^H)$, with $X=\lambda,\kappa$, denote the sum of all one-loop
contributions that contain only light fields, mixed heavy and light fields as well
as only heavy fields in the loop. Likewise $\delta \tilde{\lambda}$ can only arise from diagrams involving light fields in the loop since there are no heavy states present in the EFT.
All generic diagrams which can contribute to tree-level 
 and one-loop amplitudes of any renormalisable scalar operator are given in \cref{app:generic}. 
Again, IR divergences caused by light fields are present on both sides which
need to cancel in the matching conditions. 
A detailed discussion on these cancellations is beyond the scope of
this paper but was recently discussed in
Ref.~\cite{Braathen:2018htl}. In
summary, the matching condition can be expressed in terms of IR-finite pieces
\begin{align}
\tilde{\kappa}_{abc}  &= \kappa_{abc} + \overline{\delta} \kappa_{abc}(\phi^L,\phi^H,\psi^L,\psi^H) \\
\tilde{\lambda}_{abcd} &= \lambda_{abcd} -  \sum_{x \in \{\phi^H\}}\sum_{\rm s,t,u}\left(\frac{{\kappa}_{abx} {\kappa}_{xcd}}{M_x^2}\right) + \overline{\delta} \lambda_{abcd}(\phi^L,\phi^H,\psi^L,\psi^H) \, ,
\end{align}
where the one-loop contributions $\overline{\delta} X$ are computed using
modified loop integrals where the IR divergent pieces have been subtracted.
For instance, the scalar two-point integral $B_0$ with vanishing
external momentum (for simplicity we omit the vanishing external momentum in the argument of all loop function) and vanishing masses suffers from a logarithmic IR divergence 
\begin{equation}
B_0(m^2,m^2)|_{m^2\to 0} =  \left. \log \left(\frac{Q^2}{m^2}\right)\right|_{m^2\to 0} \,
\end{equation}
which will necessarily cancel in the matching condition \cref{eq:matchcond}.
Thus, the replacement of the $B_0$ with the modified loop function
\begin{equation}
    \left.  \overline{B_0}(m^2,m^2) \right|_{m^2\to 0} = \left[ B_0(m^2,m^2) - \log \left(\frac{Q^2}{m^2}\right) \right]_{m^2\to 0}
\end{equation}
makes this cancellation manifest without the need to compute the corresponding IR-divergent diagrams in the EFT.
Thus, the calculation of the matching conditions can be performed in a straight-forward way by using the
IR-safe loop functions $\overline{B}_0$, $\overline{B}_1$, $\overline{C}_0$,
$\overline{D}_0$, $\dot{\overline{B}}_0$ and $\dot{\overline{B}}_1$  defined in
\cref{app:loopfunctions}.

\subsection{Renormalisation Scheme}
\label{sec:wave_off}
A simple renormalisation scheme which is applicable to a wide range of models is the \MS/\DR
scheme. Therefore, we are going to stick mainly to this scheme. The only
exception is the treatment of the off-diagonal wave-function renormalisation
(WFR) of the scalar fields. It has been proposed in Ref.~\cite{Bagnaschi:2014rsa} that
these contributions can be dropped by assuming finite counter-terms for some
input parameters. For instance, in the high-scale MSSM one could assume a
counter-term for $\tan\beta$ which exactly cancels the off-diagonal
WFR contributions.
This approach is a more economic calculation and can lead to performance
improvements in the runtime. However, it depends
on the considered model and the chosen input parameters if such a scheme is
possible. Therefore, we provide the possibility to include or
exclude the one-loop contributions from off-diagonal WFR constants during the calculation. 
\\ 
For an appropriate choice of the WFR treatment it is worth to mention
the equivalence between excluding the off-diagonal WFR
constants and the extraction of effective quartic couplings from a pole-mass
matching \cite{Braathen:2018htl,Athron:2016fuq}. Thus, for
the comparison with tools that use a pole-mass matching, the inclusion of
off-diagonal WFR constants should be disabled in the calculation. 

\subsection{Parametrisation of the Results at the Matching Scale}
\label{sec:parametrisation}
\begin{figure}[tb]
\begin{center}
\includegraphics{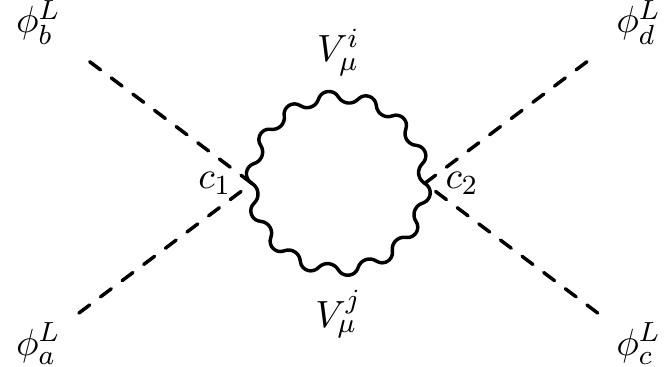}
\end{center}
\caption{One-loop diagram contributing to the shift from \MS--\DR conversion.}
\label{fig:MSDR}
\end{figure}
Using matching conditions to calculate the effective couplings yields solutions that
are functions of the parameters of the UV theory. However, in some cases it might
be better to (at least partially) give their dependence on the EFT parameters. 
This is especially the case for the SM gauge and Yukawa couplings because their values 
are known very precisely. Therefore, one also needs to match these couplings at 
a suitable loop-level. Concerning the matching of the scalar sector, the EFT 
parameters that enter the scalar matching conditions at tree-level need 
to be matched at the one-loop level (and re-inserted into the scalar 
tree-level matching).  For all other parameters, a tree-level matching 
is sufficient as long as we stick to a one-loop matching of the scalar couplings. 
For non-supersymmetric models the scalar parameters which we want to match are free parameters, i.e. in these cases a matching of the SM
parameters at tree-level is always sufficient. This is different for
supersymmetric models because the scalar couplings are related to the other
couplings through $F$- and $D$-terms. We concentrate on the $D$-term
contributions, i.e. the matching of the gauge couplings, because this is the part
important for the matching of scalars that could -- at least in principle --
provide a SM-like Higgs boson. The matching of the gauge couplings is parametrised by
\begin{equation}
 g_i \to g_i + \delta g_i
\end{equation}
and receives two different contributions:
\begin{enumerate}
 \item Thresholds from heavy fields:
 \begin{align}
 \label{eq:dg}
  \delta g_i = \sum_{\phi^H} \frac{1}{16 \pi^2} \frac{C^i(\phi^H)}{12}
   I_2^i(\phi^H) g_i^3 \log\frac{M_{\phi_H}^2}{Q^2} + \sum_{\psi^H}
   \frac{1}{16 \pi^2} \frac{C^i(\psi^H)}{3} I_2^i(\psi^H) g_i^3
   \log\frac{M_{\psi_H}^2}{Q^2} \,, 
 \end{align}
 where $g_i$ is the gauge coupling with respect to the gauge group $i$,
 $I_2^i(x)$ is the Dynkin index of the field $x$ with respect to the gauge
 group $i$, and $C^i(x)$ is a multiplicity factor taking into account the
 charges under non-Abelian gauge groups others than $i$, i.e. in the case of
 the SM gauge group, this counts the colour and isospin multiplicity in the
 loop. 
 \item \MS--\DR conversion: required if an \MS and a \DR renormalised
   quantity are to be matched. This is e.g. the case if non-SUSY models
   are matched onto SUSY ones. There are two different contributions
   which affect the quartic couplings: 
 \begin{itemize}
   \item The finite shifts of the gauge couplings for an $SU(N)$ group are \cite{Martin:1993yx} 
  \begin{equation}
  \label{eq:MSDRshift}
   \delta g_i = \frac{1}{16 \pi^2} g_i^3 \frac{N}{6} \,.
  \end{equation}
  \item Quartic vertices receive an additional shift from \MS--\DR conversion from the diagrams shown in \cref{fig:MSDR}. The amplitude difference of this diagram between the two schemes is 
  \begin{equation}
   \mathcal{M}= c_1 c_2 \,,
  \end{equation}
  where $c_1$ and $c_2$ are the two involved vertices between two
  scalars and two vector bosons. 
 \end{itemize}
\end{enumerate}
The calculation of the two different contributions was implemented in \SARAH and are automatically included in the matching procedure.
\subsection{Above and Below the Matching Scale: Threshold Corrections
  to Fermionic Couplings} 
\label{sec:above}
\begin{figure}[tb]
\begin{center}
\includegraphics{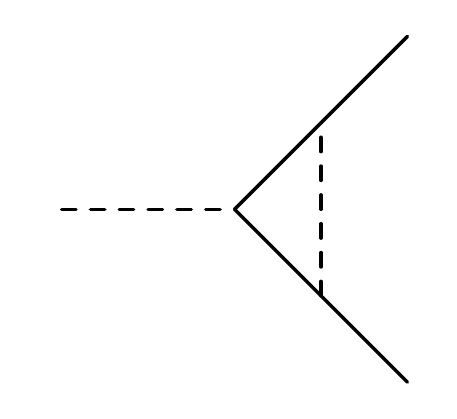}
\quad 
\includegraphics{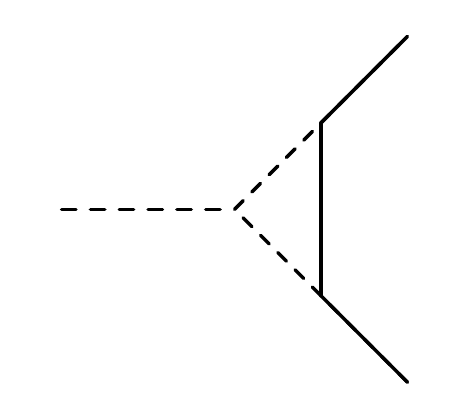}
\quad 
\includegraphics{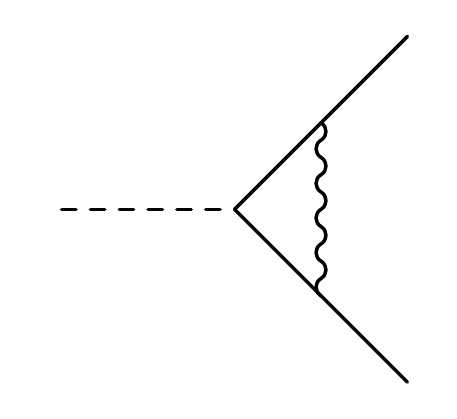}
\end{center}
\caption{Irreducible one-loop diagrams contributing to the threshold
    corrections to scalar-fermion couplings. The diagram on the right only
shows up for \MS--\DR conversion.}
\label{fig:oneloopY}
\end{figure}

So far, we have concentrated on scenarios where the running above the matching
scale can be neglected and the threshold corrections to fermionic couplings
do not play an important role. Of course, there are plenty of situations where it is
necessary to go beyond that. The simplest case is a high-scale SUSY scenario  
which is connected to a common SUSY breaking mechanism like
minimal supergravity (mSugra). Such a SUSY breaking predicts that the masses of
the sparticles are degenerate at the scale of grand unification (GUT), but
not necessarily at the matching scale. 
Thus, finite differences between the running masses are present below
the GUT scale. 
In such cases, one needs to consider the running above the matching scale up to
the GUT scale. Since two-loop renormalisation group equations (RGEs) are
commonly used for that running, it is necessary to include the threshold
corrections to the SM gauge and Yukawa couplings. While the threshold
corrections to the gauge couplings are given by \cref{eq:dg} and
\cref{eq:MSDRshift}, some more work is needed to compute the shifts to the Yukawa
couplings. The general ansatz to calculate these shifts is the same as for the
scalar couplings, i.e. imposing that the $n$-loop amplitudes of corresponding
fields are identical at the matching scale $M_M$. Once again, all IR
divergences must cancel at $M_M$, i.e. one determines the Yukawa couplings
above the threshold scale via 
\begin{equation}
\label{eq:yshift}
Y_{abc} = \tilde{Y}_{abc} - \overline{\delta}Y_{abc} \;,
\end{equation}
where $Y_{abc}$ is for instance a running SM Yukawa coupling and
$\overline{\delta}Y_{abc}$ contains corrections from diagrams containing heavy fields, which
are obtained with IR-safe loop functions, as well as \MS--\DR conversion if
necessary.
\\
If the EFT is not the SM but an extension with additional fermions, also new
Yukawa-like couplings are present below the matching scale. A good example for
such a scenario is for instance split-SUSY with effective
gaugino--Higgsino--Higgs couplings. Of course, the one-loop relation to
calculate these couplings is just given by inverting \cref{eq:yshift}, i.e. 
\begin{equation}
\tilde{Y}_{abc} = Y_{abc} + \overline{\delta}Y_{abc} \;.
\end{equation}
Thus, in a generic approach, both types of Yukawa coupling corrections, 
above (SM-like) and below (BSM-like) the matching scale are obtained simultaneously. Necessary
ingredients are the one-loop diagrams depicted in \cref{fig:oneloopY} together
with the wave-function corrections of the external states.

\section{Implementation in \SARAH and \SPheno}
\label{sec:implementation}
In the last section all necessary ingredients for a matching of two arbitrary
renormalisable scalar sectors at the one-loop level were introduced. In this
section we describe the implementation as well as the usage in the computer
programs \SARAH and \SPheno.

\subsection{General Information about \SARAH and \SPheno}
\SARAH\footnote{\SARAH is available at hepforge: \href{https://sarah.hepforge.org/}{\tt sarah.hepforge.org}} is a
\Mathematica package optimised for an easy, fast and exhaustive study of BSM
models. For a given model, which is defined in form of three
input files, \SARAH derives all tree-level properties, i.e.  mass matrices,
tadpole equations and vertices. Moreover, the analytical calculations of
one-loop self-energies and tadpoles as well as of two-loop renormalisation
group equations (RGEs) are fully automatised in \SARAH based on generic results
given in literature
\cite{Machacek:1983tz,Machacek:1983fi,Machacek:1984zw,Martin:1993zk,Luo:2002ti,Fonseca:2011vn,Goodsell:2012fm,Fonseca:2013bua,Sperling:2013eva,Sperling:2013xqa,Schienbein:2018fsw}.
With version {\tt 3}, \SARAH became the first 'spectrum-generator-generator':
all analytical information derived by \SARAH can be exported to \Fortran code
which provides a fully-fledged spectrum generator based on \SPheno. A \SARAH
generated \SPheno version calculates all masses at the full one-loop level, and
includes the dominant two-loop corrections for neutral scalars
\cite{Goodsell:2014bna,Goodsell:2015ira,Braathen:2017izn}. Beyond that, \SPheno
makes predictions for two- and three-body decays, flavour and precision
observables \cite{Goodsell:2017pdq,Porod:2014xia}, and the EW fine-tuning. In
order to define the properties of the generated \SPheno version, \SARAH needs an
additional input file usually called {\tt SPheno.m}. This input contains the
following information:
\begin{itemize}
 \item The input parameters of the model
 \item The choice for the renormalisation scale
 \item The boundary conditions at the electroweak scale, at the renormalisation scale and at the GUT scale 
 \item Optional: a condition to dynamically determine the GUT scale, e.g. $g_1(m_{\rm GUT})=g_2(m_{\rm GUT})$
 \item A list of particles for which the two- and three body decays should be calculated
\end{itemize}
Since the {\tt SPheno.m} file will be important for the discussion in the
following, we give an example in \cref{app:sphenomssm} how such a file may look
like. For more details, we refer to the manual as well as the \SARAH wiki
page\footnote{\href{http://stauby.de/sarah_wiki/index.php}{\tt
stauby.de/sarah\_wiki/}}. In the following section, we discuss various
aspects that arise in an automatised matching between two models and how they
have been considered through the implementation of two independent approaches.

\subsection{Available Options to Perform the Matching}
\label{sec:matching_overview}
The matching of two scalar sectors can be motivated by a precise investigation
of very different properties of the theories to be matched. The largest 
contributions to threshold corrections often have their origin in one common
sector of the heavy spectrum. It can be of particular interest to track this
origin down in order to learn more about which parts of a given UV-theory are
essential for the predictions in an EFT framework. For this purpose an
analytical evaluation of threshold corrections is preferred. The analytical solutions 
can also easily be ported to other computer programs which is a key feature of
many existing \SARAH routines.
\\

As already discussed, the matching of an 
EFT onto a UV-complete model does not only influence many low-energy observables
but also enters the RGE running and other predictions above the matching scale.
Considering the whole picture of the matching procedure and its numerical
influence in all sectors of the theories to be matched, a numerical
calculation of threshold corrections is preferred because it can easily be embedded
into existing routines of the generated \SPheno code. 
\\

With \SARAH version 4.14.0 we provide two different possibilities to perform
the matching between two arbitrary scalar sectors:
\begin{enumerate}
 \item An analytical calculation within \Mathematica
 \item A fully numerical calculation using only the \SPheno interface
\end{enumerate}
\begin{figure}[tb]
\begin{center}
\includegraphics[width=0.75\linewidth]{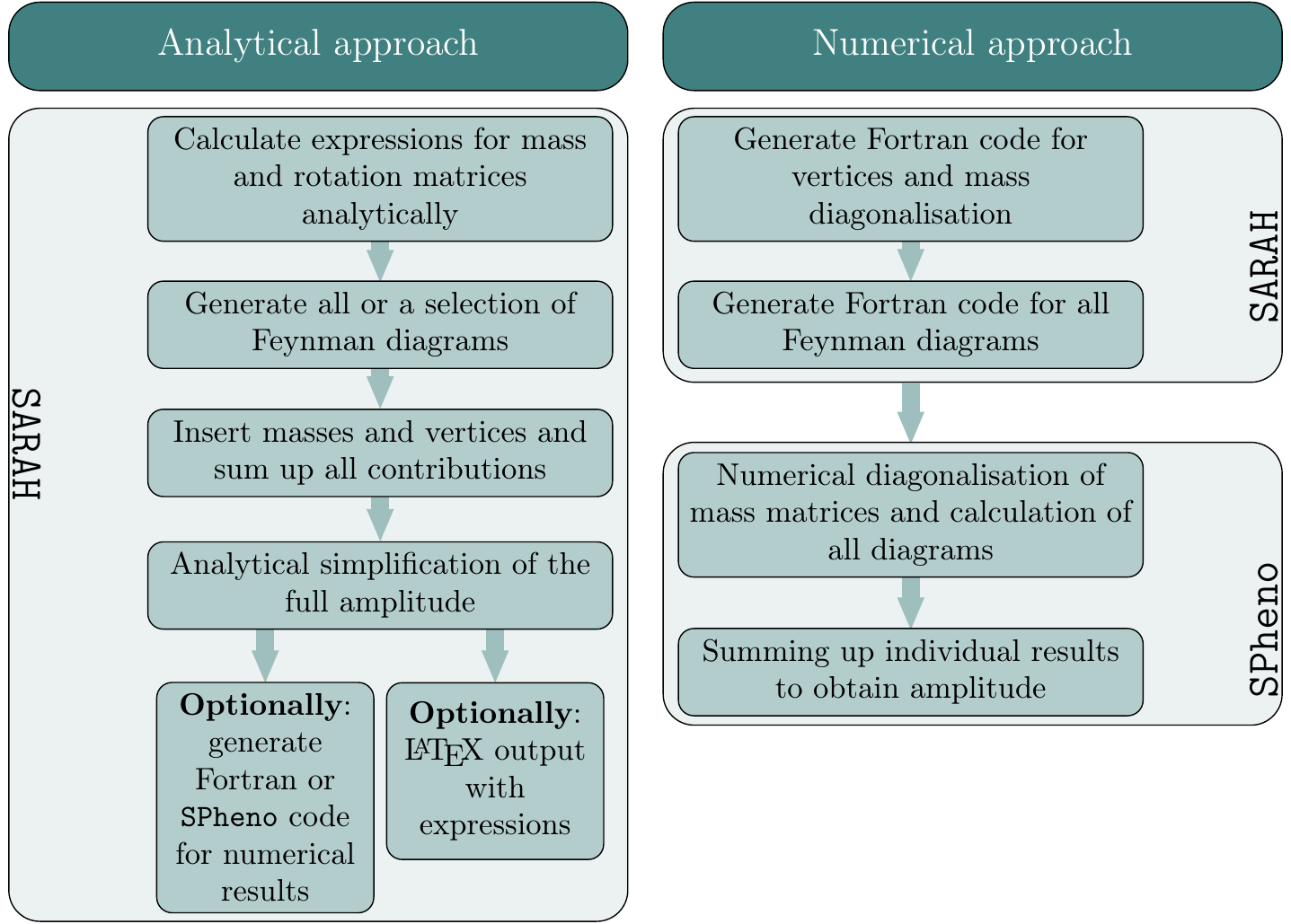}
\end{center}
\caption{Schematic procedure for the two options to calculate matching
  conditions with \SARAH/\SPheno.}
\label{fig:flowchart_approaches}
\end{figure}
It is important to stress that both options are not based on the same routines,
but have been implemented independently. Thus, they offer the possibility to
double check the obtained results. A schematic comparison of the two approaches
is given in \cref{fig:flowchart_approaches}. A summary of the description given here is also available at
the \SARAH wiki page\footnote{\href{http://stauby.de/sarah_wiki/index.php?title=One-Loop_Threshold_Corrections_in_Scalar_Sectors}{\tt http://stauby.de/sarah\_wiki/index.php?title=One-Loop\_Threshold\_Corrections\_in\_Scalar\_Sectors}}.
\\

For the analytical calculation it is necessary that all mass matrices in the
model can be diagonalised analytically. Thus, it is usually necessary to work
with a set of assumptions which  simplify the most general mass matrices in a
given model. In theories with spontaneous symmetry breaking, a high degree of different mixing patterns is
introduced through the presence of VEVs. It has already been argued
that, if these VEVs are responsible for the 
generation of masses in the EFT, i.e. if the low-energy Lagrangian is invariant
under the symmetries broken by these VEVs, a common assumption is to
neglect all small VEVs. 
In addition, flavour violating effects are usually negligible. 
The only exception are scenarios in which large contributions to flavour
violation occur in the new physics sector. 
This could for instance happen in the MSSM with large off-diagonal trilinear
soft-terms which can have a big effect on the Higgs boson mass
\cite{Goodsell:2015yca}. Thus, if any of these assumptions is not justified, it
is necessary to switch to the purely numerical calculation.
\\

Although the focus of the \Mathematica interface is on the derivation of
analytical expressions for the matching conditions, additional routines have
been implemented to make these results easily usable in numerical calculations. 
This has the advantage that the obtained code for numerical evaluations can be much faster
than the fully numerical interface because many simplifications can be
performed on the analytical level. In addition, the obtained results can be
exported into \LaTeX\ files which makes a evaluation of the expressions in a
human readable format possible.
On the other hand, the fully numerical implementation has several advantages: (i) the
RGE running above the matching scale can be performed, (ii) shifts to fermionic
couplings can be included, (iii) several EFTs appearing in models with more
than one matching scale can be automatically linked (iv) flavour violating effects can be included.
\\

Before describing the user interface of the new routines, we want to comment 
on a few subtleties to provide a better understanding on the importance of
certain user inputs.
\begin{enumerate}
\item {\bf Model files}: in principle, one can set up specific model files for
    the UV theory where for instance EW VEVs are dropped from the very
    beginning. However, this complicates further studies of the UV theory.
    Thus, we are going to work in the following with the default
    model files delivered with \SARAH. For instance, we use the MSSM
    implementation which includes EW VEVs and apply the simplifying assumption
    to neglect these VEVs during the matching procedure. However, the
    considered EFT may require the development of further model files. For
    instance, various split-SUSY models that contain
    only the fermionic degrees of freedom of their corresponding SUSY models
    already have been implemented in the new \SARAH version.

 \item {\bf Normalisation of couplings}: in many models studied in literature, the coefficients in front of the scalar couplings are often chosen differently from \cref{eq:LagUV}. For instance, a common convention for the SM Lagrangian reads
 \begin{equation}
     \mathcal{L}_{SM} = \left(D_\mu H\right)\left( D^\mu H\right)^\dagger + \sum_f \bar{f} \slashed{D} f - \left(\frac12 \lambda_{SM} |H|^4 + \mu^2 |H|^2 \right) -  \left(Y_d H^\dagger \bar{d} q + Y_e H^\dagger \bar{e} l - Y_u H \bar{u}q + \text{h.c.} \right)
 \end{equation}
 Thus, after replacing $H^0 \to \frac{1}{\sqrt{2}} \left(h + i G + v\right)$ the vertex in \cref{eq:LagUV} between four Higgs fields $h$ is
\begin{equation}
    \tilde{\kappa}_{hhhh}= - 3 \lambda_{SM} \, .
\end{equation}
Therefore, the correct matching condition to calculate $\lambda_{SM}$ becomes 
\begin{equation}
    \label{eq:effnorm}
    \lambda_{SM} = -\frac{1}{3} \left( \kappa_{hhhh} + \bar{\delta} \kappa_{hhhh} + \sum_{s,t,u}\, \sum_{x \in \{\phi^H\}} \frac{ \left( \kappa_{hhx} \right)^2}{M_x^2} \right)\, ,
\end{equation}
where $\kappa_{i}$ denotes tree-level vertices in the UV theory while $\bar{\delta}$ are the corresponding one-loop shifts.
The relative normalizations between operators in the considered UV and
the effective theory, such as for example the factor
$-\nicefrac{1}{3}$ in \cref{eq:effnorm}, have to be provided by the
user.

\item {\bf Superposition of fields}: when matching a scalar sector involving
    multiple (light) scalar fields with identical quantum numbers, often linear combinations of external fields contribute to the matching
    of different parameters. For instance, consider
    the couplings $\lambda_4$ and $\lambda_5$ in a THDM:
\begin{equation}
\mathcal{L}_{THDM} = \dots - \lambda_4 |H_1^\dagger H_2|^2 - \frac12\left(\lambda_5 (H_1^\dagger H_2)^2 + \text{h.c.} \right) \, ,
\end{equation}
where the two $SU(2)$ doublets $H_1$ and $H_2$ have the same hypercharge. 
We find that any vertex involving $\lambda_4$ receives also contributions from
$\lambda_5$ and vice versa. For instance consider the couplings 
\begin{align}
\kappa_{h_1 h_2 H_1^+ H_2^-} = -\frac12 \left(\lambda_4 + \lambda_5 \right) \, ,\\
\kappa_{h_1 A_2 H_1^+ H_2^-} = \frac12 i \left(\lambda_4 - \lambda_5 \right) \, ,
\end{align}
after splitting the two doublets into their charged ($H^\pm_{1,2}$), CP-even
($h_{1,2}$) and CP-odd ($A_{1,2}$) components (note that the gauge eigenstates introduced here also correspond to the mass eigenstates as we assume vanishing VEVs). 
For simplicity, we assume real parameters . Thus, to obtain the matching
conditions for $\lambda_4$ and $\lambda_5$ separately, it is necessary to
calculate  the superpositions
\begin{align}
    \lambda_4 = -\left(\mathcal{M}(h_1 h_2 H_1^+ H_2^-)+i \mathcal{M}(h_1 A_2 H_1^+ H_2^-)\right) \, , \label{eq:match2hdm1}\\
\lambda_5 = -\left(\mathcal{M}(h_1 h_2 H_1^+ H_2^-)-i \mathcal{M}(h_1 A_2 H_1^+ H_2^-)\right)\, .\label{eq:match2hdm2}
\end{align}
These conditions are user input as well.
\end{enumerate}

\subsection{Analytical Approach}
\label{sec:ana_approach}
In order to use the \Mathematica interface to obtain analytical expressions for
the matching conditions, one needs to initialize a \Mathematica kernel, load
\SARAH and start the considered high-scale model. This can be done by opening a
new \Mathematica notebook and entering the commands
\begin{MIN}
<<SARAH.m
Start["<Model>"]
\end{MIN}
where  \lstinline{<Model>} could be for instance \lstinline{MSSM} or
\lstinline{NMSSM}. In the next step, there are two possibilities to obtain the
matching conditions analytically:
\begin{enumerate}
 \item one can calculate individual effective couplings in an interactive mode or
 \item use a \textit{batch mode} to calculate several matching conditions at
     once and to optionally obtain \LaTeX\ , \Fortran and \SPheno outputs.
\end{enumerate}
We are going to give details about both options which are based on the new command 
\begin{MIN}[numbers=none]
InitMatching[Options]
\end{MIN}
where possible options are
\begin{itemize}
 \item  \lstinline{Parametrisation -> $LIST}
     \begin{itemize}
         \item Default: \lstinline!{}!
         \item Description: list of specific parametrisations of selected model parameters
         \item Example: \lstinline!{vu -> v Sin[ArcTan[TanBeta]], ...}!
     \end{itemize}
 \item \lstinline{Assumptions -> $LIST}
     \begin{itemize}
     \item Default: \lstinline!{}!
     \item Description: list of assumptions for parameters in the model in order to simplify the expressions
     \item Example: \lstinline!{TanBeta>0}!
     \end{itemize}
 \item \lstinline{SolveTadpoles -> $LIST}
     \begin{itemize}
         \item Default: \lstinline!{}!
         \item Description: list of parameters which are obtained by the tadpole equations
         \item Example: \lstinline!{mHu2,mHd2}!
     \end{itemize}
 \item \lstinline{ReadLists -> $BOOL}
     \begin{itemize}
     \item Default: \lstinline{False}
     \item Description: if set to {\lstinline|True|}, the calculation of
         vertices is skipped, but results stored in a previous session are
         used. This can be a significant performance boost.
     \end{itemize}
 \item \lstinline{InputFile -> $FileName}
     \begin{itemize}
     \item Default: \lstinline{False}
     \item Description: can be used to define an input file containing all necessary information
     \end{itemize}
\end{itemize}
A short description of this command can be obtained within a \SARAH
session by invoking the command \lstinline{?InitMatching}.
\\

If the interactive mode is demanded, the option \lstinline{InputFile} has to be
omitted while values for \lstinline{Parametrisation}, \lstinline{Assumptions}
and \lstinline{SolveTadpoles} should be provided to allow for an analytical
diagonalization of all mass matrices. The usage of the batch mode requires only the 
option \lstinline{InputFile} and serves a high reproducibility of the obtained results
by providing only one single input file.\\

The provided assumptions and parametrisations are used to calculate analytical
expressions for all masses
and rotation matrices. If this is not possible, because \Mathematica 
cannot diagonalize the mass matrices analytically (using the build-in
functions \lstinline{Eigensystem} and \lstinline{SingularValueDecomposition}),
one can either use the purely numerical interface explained in \cref{sec:num_approach} 
or choose appropriate simplifying assumptions.

\subsubsection{Interactive Mode: Calculating Individual Matching Conditions}
Initializing the matching routines using the \lstinline{InitMatching} function
with the options described in the previous paragraph while not specifying the
option \lstinline{InputFile} 
enables the interactive mode. The necessary vertices of the high-scale theory 
are calculated or loaded from a previous session and the masses/rotation matrices are
derived. However, no further calculations are performed at this point.
\\

\textbf{\underline{Example initialization: }} consider a
high-scale MSSM scenario where all SUSY particles have a degenerate mass
\lstinline{MSUSY} while only the SM Higgs remains light. A possible parametrisation may look like
\begin{MIN}[firstnumber=3]
MyParametrisation = {|\newline
|vd -> v Cos[ArcTan[TanBeta]],|\newline
|vu -> v Sin[ArcTan[TanBeta]],|\newline
|v -> epsUV,|\newline
|g1 -> g1Q, |\newline
|g2 -> g2Q,|\newline
|g3 -> g3Q, |\newline
|Yu[a_,b_] :> 1/Sin[ArcTan[TanBeta]]*Delta[3,b]Delta[a,3]YuQ[a,b],|\newline
|Yd[a_,b_] :> 1/Cos[ArcTan[TanBeta]]*Delta[3,b]Delta[a,3]YdQ[a,b],|\newline
|Ye[a_,b_] :> 1/Cos[ArcTan[TanBeta]]*Delta[3,b]Delta[a,3]YeQ[a,b],|\newline
|T[Yu][a__] :> Delta[a] At Yu[a],|\newline
|T[Yd][a__] :> Delta[a] Ab Yd[a],|\newline
|T[Ye][a__] :> Delta[a] Ae Ye[a],|\newline
|mq2[a__] :> Delta[a] MSUSY^2,|\newline
|mu2[a__] :> Delta[a] MSUSY^2,|\newline
|md2[a__] :> Delta[a] MSUSY^2,|\newline
|me2[a__] :> Delta[a] MSUSY^2,|\newline
|ml2[a__] :> Delta[a] MSUSY^2,|\newline
|conj[x_] -> x, |\newline
|MassB -> MSUSY,|\newline
|MassWB -> MSUSY,|\newline
|MassG -> MSUSY, |\newline
\textbackslash|[Mu] -> MSUSY, |\newline
|B[|\textbackslash|[Mu]] -> MSUSY^2 |\newline
|};
\end{MIN}
Note that the symbols \lstinline{MSUSY}, \lstinline{At}, \lstinline{Ab},
\lstinline{Ae} and \lstinline{TanBeta} are not defined in the MSSM model file. 
Thus, additional information about these symbols must be provided using the
\lstinline{Assumptions} option, otherwise they are assumed to be 
arbitrary complex numbers. The initialization is invoked by
\begin{MIN}[firstnumber=4]
InitMatching[Parametrisation->MyParametrisation, |\newline| SolveTadpoles->{mHd2,mHu2}, |\newline| Assumptions -> {TanBeta > 0, MSUSY > 0, At>0, Ab>0, Ae>0}]
\end{MIN}
There are a few important comments concerning the parametrisation which we have used in this example:
\begin{itemize}
    \item The symbol \lstinline{epsUV} is used to indicate dimensionful
        parameters \lstinline{X} which are to be neglected in the UV theory. One should
        always use this parameter instead of the simpler rule \lstinline{X->0}
        to avoid problems caused by a division by 0.
 \item It is recommended to express all matching conditions in terms of the running
     parameters of the effective theory, see \cref{sec:parametrisation}.
     Therefore, we express the MSSM gauge and Yukawa couplings by the SM ones
     using the suffix \lstinline{Q} which marks the running parameters (instead of
     \lstinline{g1} we e.g. specify it to be \lstinline{g1Q}). For these
     parameters, only the tree-level matching conditions are required.
     The one-loop matching conditions for the gauge couplings, discussed in
     \cref{sec:parametrisation}, are automatically derived.
 \item \lstinline{Delta[a,b]} is the \SARAH internal symbol for the Kronecker
     delta $\delta_{ab}$. We use it here to include only contributions from
     third generation Yukawa couplings, and to force diagonal soft masses for
     the sfermions. 
 \item In order to simplify the analytical calculation, we assume that all
     parameters are real. This is translated by \lstinline{conj[x_]->x}. The
     object \lstinline{conj} is the \SARAH internal command for complex conjugation.
\end{itemize}
When all calculations are finished, it is possible to validate if the obtained
mass spectrum at the matching scale is as expected 
\begin{MIN}[firstnumber=5]
?M
\end{MIN}
which yields the result
\begin{MOUT}[firstnumber=5]
|\hspace{7mm}|M[Ah]={0,MSUSY}|\newline
|M[Cha]={MSUSY,MSUSY}|\newline
|M[Chi]={-MSUSY,MSUSY,MSUSY,MSUSY}|\newline
|M[Fd]={0,0,0}|\newline
|M[Fe]={0,0,0}|\newline
|M[Fu]={0,0,0}|\newline
|M[Glu]=MSUSY|\newline
|M[hh]={0,MSUSY}|\newline
|M[Hpm]={0,MSUSY}|\newline
|M[Sd]={MSUSY,MSUSY,MSUSY,MSUSY,MSUSY,MSUSY}|\newline
|M[Se]={MSUSY,MSUSY,MSUSY,MSUSY,MSUSY,MSUSY}|\newline
|M[Su]={MSUSY,MSUSY,MSUSY,MSUSY,MSUSY,MSUSY}|\newline
|M[Sv]={MSUSY,MSUSY,MSUSY}
\end{MOUT}
As expected, the spectrum at the matching scale contains one massless CP-even
Higgs boson which corresponds to the SM-like Higgs boson. Also all SM-like
fermions remain massless while the heavy fields are degenerate in
the mass parameter \lstinline{MSUSY}. 
\\
The rotation matrices are stored in the array \lstinline{ReplacementRotationMatrices} and read in our example
\begin{MOUT}[numbers=none]
{ZD[a__Integer] -> {{1, 0, 0, 0, 0, 0}, {0, 1, 0, 0, 0, 0}, {0, 0, 1, 0, 0, 0}, {0, 0, 0, 1, 0, 0}, {0, 0, 0, 0, 1, 0}, {0, 0, 0, 0, 0, 1}}[[a]], 
...
ZH[a__Integer] -> {{1/Sqrt[1 + TanBeta^2], 1/Sqrt[1 + TanBeta^(-2)]}, {-(TanBeta/Sqrt[1 + TanBeta^2]), 1/Sqrt[1 + TanBeta^2]}}[[a]], 
ZA[a__Integer] -> {{-(1/Sqrt[1 + TanBeta^2]), 1/Sqrt[1 + TanBeta^(-2)]}, {TanBeta/Sqrt[1 + TanBeta^2], 1/Sqrt[1 + TanBeta^2]}}[[a]], 
ZP[a__Integer] -> {{-(1/Sqrt[1 + TanBeta^2]), 1/Sqrt[1 + TanBeta^(-2)]}, {TanBeta/Sqrt[1 + TanBeta^2], 1/Sqrt[1 + TanBeta^2]}}[[a]], 
ZN[a__Integer] -> {{0, 0, 1/Sqrt[2], 1/Sqrt[2]}, {0, 0, -(1/Sqrt[2]), 1/Sqrt[2]}, {0, 1, 0, 0}, {1, 0, 0, 0}}[[a]],
...}
\end{MOUT}

Let us now continue with the description of the analytical interface.
After the successful initialization and calculation of all mass and rotation matrices,
one can compute the leading order (LO) and next-to leading order (NLO) corrections to an amplitude
with the external fields given in the list \lstinline{fieldlist}
\begin{MIN}[numbers=none]
    EFTcoupLO[fieldlist, Options]
    EFTcoupNLO[fieldlist, Options]
\end{MIN}
The \lstinline{fieldslist} can contain two, three or four scalar fields
including their generation indices to obtain 
effective mass parameters and cubic or quartic couplings. Note that the matching
of effective mass parameters is only demanded if no spontaneous symmetry
breaking occurs in this sector of the theory.
The possible options for the function \lstinline{EFTcoupNLO} are

\begin{itemize}
    \item \lstinline{Topologies -> $LIST}
        \begin{itemize}
            \item Default: \lstinline!{}!
            \item Description: list of topologies to include into the
                calculation. If empty, all topologies are used. Topologies are
                denoted as in \cref{app:generic}.
            \item Example: \lstinline|{B[4][1],B[4][2][1], B[4][2][2]}| or equivalently
                \lstinline|{B[4]}|.
        \end{itemize}
    \item \lstinline{ExcludeTopologies -> $LIST}
        \begin{itemize}
            \item Default: \lstinline!{OffdiagonalWFRs}!
            \item Description: list of topologies to be excluded from the
            calculation. The filtering of \lstinline{ExcludeTopologies} is also
        applied on the topology groups given in the \lstinline{Topologies}
    option, e.g. if \lstinline!{B[4]}! is given in the \lstinline{Topologies}
list but \lstinline{B[4][2][2]} in the \lstinline{ExcludeTopologies} list,
then only \lstinline{B[4][1]} and \lstinline{B[4][2][1]} are computed.
            \item Example: \lstinline|{OffdiagonalWFRs, DiagonalWFRS}| to exclude all contributions on external legs.
        \end{itemize}
    \item \lstinline{ExcludeFields -> $LIST}
        \begin{itemize}
            \item Default: \lstinline!{}!
            \item Description: list of fields to be excluded when appearing
                as internal fields.
            \item Example: \lstinline!{Cha,Chi}! e.g. to exclude
                electroweakinos within a split SUSY scenario.
        \end{itemize}
    \item \lstinline{InternalPatterns -> $LIST}
        \begin{itemize}
            \item Default: \lstinline!{}!
            \item Description: compute only diagrams with certain internal field-type patterns. For an empty list all patterns are computed. 
            \item Example: \lstinline!{S,SS,SSS,SSSS}! computes
              corrections from heavy scalars only while
              \lstinline!{FF}! computes diagrams that contain exactly
              two internal fermions. 
        \end{itemize}
    \item \lstinline{GaugeThresholds->$BOOL}
        \begin{itemize}
            \item Default: \lstinline{True}
            \item Description: whether to include the contributions from one-loop gauge
                coupling thresholds to the tree-level amplitude or not
        \end{itemize}
    \item \lstinline{ShiftMSDR-> 0/1/2/Automatic}
        \begin{itemize}
            \item Default: \lstinline{Automatic}
            \item Description: whether to include the $\MS-\DR$ conversion
                factors. 0: no, 1: inclusive, 2: exclusive,
                \lstinline{Automatic}: decide between 1 and 0 depending on the type of
                considered model (SUSY or non-SUSY). \lstinline{exclusive} means that only the conversion factor is calculated while \lstinline{inclusive} gives the full result plus conversion factor (default for SUSY models).
        \end{itemize}
    \item \lstinline{Debug -> $BOOL}
        \begin{itemize}
            \item Default: \lstinline{False}
            \item Description: multiplies each amplitude with a debug
              variable marking its topology and field insertion
            \item Example: the term \lstinline{debug[C[4][1]][hh[2],hh[2],hh[2]]} may
                be multiplied with the expression of the amplitude of the triangle diagram (\lstinline{C[4][1]}, see \cref{eq:triangle}) with
                three heavy internal Higgs bosons (\lstinline{hh[2],hh[2],hh[2]}).
        \end{itemize}
    \item \lstinline{SimplifyResults -> $BOOL}
        \begin{itemize}
        \item Default: \lstinline{True}
        \item Description: whether to simplify the results using the
          given assumptions or not.
        \end{itemize}
    \item \lstinline{LoopReplace -> $FUNCTION}
        \begin{itemize}
        \item Default: \lstinline{AnalyticLoopFunctions}
        \item Description: the amplitudes contain loop functions in the \textsc{FormCalc}
            notation  (e.g. a $B_0(0,m_1^2,m_2^2)$ function is denoted by
            \lstinline{B0i[bb0,0,m1^2,m2^2]}). The function
            \lstinline{AnalyticLoopFunctions} replaces them with the IR-save
            loop functions defined in \cref{app:loopfunctions}. However, for a
            better readability one may set this to the \lstinline{Identity}
            function.
		\item Example: \lstinline{Identity}
        \end{itemize}
\end{itemize}
To view a short description of the options within a \SARAH session one can
invoke the commands \lstinline{?EFTcoupNLO} and
\lstinline{Options[EFTcoupNLO]}. The function \lstinline{EFTcoupLO} only
provides the options \lstinline{SimplifyResults}, \lstinline{Debug} and
\lstinline{ExcludeFields}.
\\
\\
\textbf{\underline{Example calculation:}} proceeding with the high-scale MSSM
example i.e. the MSSM $\to$ SM matching we can use the introduced functions 
to calculate the expressions for the effective quartic coupling $\lambda_{SM}$ 
of the SM Higgs boson at the matching scale. The tree-level matching condition 
is calculated as follows
\begin{MIN}[firstnumber=7]
-1/3 EFTcoupLO[{hh[1],hh[1],hh[1],hh[1]}]
\end{MIN}
the output reads
\begin{MOUT}[firstnumber=7]
(((g1Q^2 + g2Q^2) (-1 + TanBeta^2)^2)/(4 (1 + TanBeta^2)^2))
\end{MOUT}
where the number in the square brackets denotes the generation index of the
Higgs field \lstinline{hh}. Note the coefficient $-\frac13$, which we have included to get the value for
$\lambda_{SM}$ as explained at the end of \cref{sec:matching_overview}. Thus,
we found at leading order
\begin{equation}
    \lambda^{(\rm LO)}_{SM} = \frac14 (g_1^2+g_2^2) \frac{(\tan^2\beta-1)^2}{(\tan^2\beta+1)^2} \,.
\end{equation}
The full expression at the one-loop order is rather lengthy. Therefore, we make a few
approximations and include only the terms involving the top quark Yukawa
coupling. This can be achieved by setting all other couplings to zero. The
command
\begin{MIN}[firstnumber=8]
-1/3 EFTcoupNLO[{hh[1],hh[1],hh[1],hh[1]}]//.{g1Q->0, g2Q->0, YeQ[3,3]->0, YdQ[3,3]->0, At -> (Xt + MSUSY/TanBeta)}
\end{MIN}
where we have introduced the stop mixing parameter $X_t=A_t - \mu
\tan^{-1}\beta$, yields 

\begin{MOUT}[firstnumber=8]
((12 MSUSY^2 Xt^2 - Xt^4 -12 MSUSY^4 Log[UVscaleQ^2/MSUSY^2]) YuQ[3,3]^4)/(32 MSUSY^4 Pi^2)
\end{MOUT}
where the symbol \lstinline{UVscaleQ} is the name for the renormalisation scale
used in the loop functions. Note, because of the assumption $g_i \to 0$ this corresponds only to the
leading one-loop shift but not to the full NLO expression (including the
tree-level contributions), i.e. we found 
\begin{equation}
    \delta \lambda^{(\rm NLO)}_{SM} = \frac{Y_t^4}{32 \pi^2}\left( \frac{12 X_t^2}{M_{\rm SUSY}^2} - \frac{X_t^4}{M_{\rm SUSY}^4}  \right) - \frac{3 Y_t^4}{8\pi^2} \log \frac{Q_{match}^2}{M_{\rm SUSY}^2}  \, ,
\end{equation}
which is the well-known leading one-loop shift maximized for
$X_t=\sqrt{6} M_{\rm SUSY}$. 
\\
\\
\textbf{\underline{Advanced Examples:}} the root directory of the new \SARAH version includes the file 
\begin{center}
    \lstinline{Example_Matching.nb}
\end{center}
which contains already evaluated Cells that describe the example usage of all possible Options of \lstinline{EFTcoupNLO} 
(e.g. the selection of specific topologies or debugging) within the high-scale MSSM.
%
\subsubsection{Batch Mode}
\label{sec:math_batch}
The complexity of the calculation requires a high degree of reproducibility of
the results. 
For this purpose it is possible to write input files that contain all necessary
information for the matching to a given EFT model. This includes all
information already discussed in the interactive mode. In addition,
the correspondence between effective couplings in the low-energy model and amplitudes in
the UV model, as it was demonstrated for the THDM matching, have to be
defined.\\

The \textit{batch mode} is invoked during the initialisation by specifying the
input file \lstinline{<FileName>} located in the directory of the loaded \SARAH
model
\begin{MIN}[numbers=none]
InitMatching[InputFile-><FileName>]
\end{MIN}
The mandatory content of the input file is
\begin{lstlisting}[language=SLHA,title=Matching.m]
$NameUV= "STRING" ;
$ParametrisationUV = { ... };
$AssumptionsMatching={ ... }
$SolveTadpolesUV = { ... };
$MatchingConditions = { 
   {Parameter1, coefficient field1.field2.field3.field4},
   ...
   {Parameter2, coefficient field1.field2.field3}
  }
\end{lstlisting}
The purpose of the different keywords is
\begin{itemize}
 \item \lstinline{$NameUV}: defines a name for the current setup. This also
     determines the name of the output directory in which the results are saved
     into as well as the file name of the \SPheno binary.
 \item \lstinline{$ParametrisationUV}: the parametrisation in the UV. This is
     equivalent to option \lstinline{Parametrisation} when running
     \lstinline{InitMatching} without an input file.
 \item \lstinline{$SimplificationsMatching}: a list of simplifications which
     are only applied at the matching scale. 
 \item \lstinline{$AssumptionsMatching}: a list of assumptions at the matching
     scale equivalent to the \lstinline{Assumptions} option when running
     \lstinline{InitMatching} without input file.
 \item \lstinline{$SolveTadpolesUV}: the equivalent to the option
     \lstinline{SolveTadpoles} of \lstinline{InitMatching}.  
 \item \lstinline{$MatchingConditions}: a list of matching conditions which
     relates a parameter in the EFT to amplitudes in the high-scale model
     containing light external fields only, similar to \cref{eq:match2hdm1,eq:match2hdm2}. 
\end{itemize}
Up to \lstinline{$NameUV} and \lstinline{$MatchingConditions} this is the same
information which is otherwise passed to \lstinline{InitMatching} and
\lstinline{EFTcoupLO}/\lstinline{EFTcoupNLO} in the interactive mode. In
addition, one can define options to control the generation of \LaTeX\ or \SPheno output.
This is described in more detail below. First, consider an input file example
which defines a high-scale SUSY scenario
\begin{lstlisting}[language=SLHA,title=Matching\_SimpleHighScaleSUSY.m]
$NameUV="SimpleHighScaleSUSY";

$ParametrisationUV = {
   vd -> v Cos[ArcTan[TanBeta]],
   vu -> v Sin[ArcTan[TanBeta]],
   v -> epsUV,
   ... (* as in the interactive mode *)
   \[Mu]->MSUSY
  };
   
$SimplificationsMatching={
 UVscaleQ->MSUSY,
 conj[x_] :> x
};
   
$AssumptionsMatching={
 TanBeta>0, 
 MSUSY>0
}

$SolveTadpolesUV = {mHd2, mHu2};

$MatchingConditions = {
   \[Lambda] -> -1/3 hh[1].hh[1].hh[1].hh[1]
   };
\end{lstlisting}
Here, we skipped most of the lines for \lstinline{$ParametrisationUV} because they are
similar to the definition of \lstinline{MyParametrisation} in the
last subsection. For simplicity, we set here all trilinear sfermion couplings
as well as the matching scale \lstinline{UVscaleQ} equal to \lstinline{MSUSY}.
\\

If the option \lstinline{InputFile->"Matching_SimpleHighScaleSUSY.m"} is given to
\lstinline{InitMatching}, \SARAH will calculate all matching conditions defined
in \lstinline{$MatchingConditions}. The information is stored in the arrays
\begin{MIN}[numbers=none]
SA`MatchingResultsLO
SA`MatchingResultsNLO
\end{MIN}
and is also written to the destination directory 
\begin{center}
{\tt \$SARAH\_Directory/Output/\$Model/EWSB/Matching/\$NameUV} 
\end{center}
Thus, one can work with the results within other \Mathematica sessions as well.

\paragraph{\LaTeX\  Output} One can use the batch mode to obtain \LaTeX\ files
which give information about calculated masses, rotation matrices and
 matching conditions in a human readable format. In order to
produce this output, the input file must contain additional information which
maps the additional symbols onto \LaTeX\ symbols
\begin{lstlisting}[firstnumber=11,language=SLHA,title=Matching.m]
$EFTcouplingsToTeX = True;
$AdditionalTeXsymbols={ ... }; 
\end{lstlisting}
The meaning of these lines is
\begin{itemize}
 \item \lstinline{$EFTcouplingsToTeX}: if set to \lstinline{True}, all
     information obtained during the matching is exported into a \LaTeX\ file
     ready to be compiled by standard \LaTeX\ compilers.
 \item \lstinline{$AdditionalTeXsymbols}: a list containing replacement
     rules that define the correspondence between \LaTeX\ and \Mathematica\
     expressions which are for instance used in the defined parametrisation.
     This will improve the readability of the \LaTeX\ document significantly.
\end{itemize}
Thus, for our chosen example, the entries might read
\begin{lstlisting}[firstnumber=26,language=SLHA,title=Matching\_SimpleHighScaleSUSY.m]
$EFTcouplingsToTeX=True;
$AdditionalTeXsymbols={
 {\[Lambda], "\\lambda"},
 {TanBeta, "t_{\\beta}"},
 {MSUSY, "M_{\\text{SUSY}}"},
 {g1Q, "g_1^{\\rm MS}"},
 {g2Q, "g_2^{\\rm MS}"},
 {g23, "g_3^{\\rm MS}"},
 {YuQ[3,3], "Y_t^{\\rm MS}"},
 {YdQ[3,3], "Y_b^{\\rm MS}"},
 {YeQ[3,3], "Y_\\tau^{\\rm MS}"}
};
\end{lstlisting}
where the additional backslash is a necessary escape character.
The \LaTeX\ files are saved in the same directory
\lstinline{$SARAH_Directory/Output/$Model/EWSB/Matching/$NameUV} as the other
outputs. 

\paragraph{\SPheno output}
With little effort, it is also possible to generate a \SPheno version which
includes the analytical matching conditions to be used within an iterative
running between the matching and the EW scale. 
In order to do so, two steps are necessary:
\begin{enumerate}
 \item Export the \Mathematica expressions into \Fortran code and write a corresponding {\tt SPheno.m} file
 \item Run the EFT model using this {\tt SPheno.m} 
\end{enumerate}
The first step is again steered through the input file of \lstinline{InitMatching} by adding the following information
\begin{lstlisting}[firstnumber=38,language=SLHA,title=Matching\_SimpleHighScaleSUSY.m]
$ExportToSPheno=True;
$SPhenoEFTmodel="SM";
$SPhenoMINPAR={
 {1, MSUSY},
 {2, TanBeta}};
$SPhenoBoundaryHighScale={};
$SPhenoBoundaryRenScale={};
$SPhenoTadpoles={mu2};
$SPhenoMatchingScale={MSUSY};
$SPhenoRenScale=173.;
$SPhenoMatchingEWSB=Default[OHDM];
\end{lstlisting}
The export into \SPheno routines is enabled with the first line. This option
is sufficient to obtain \Fortran routines for all matching conditions at the
one-loop level. All other information must be given to automatically 
generate a suitable \lstinline{SPheno.m} for the EFT model. 
Most variables have a 1:1 correspondence to the standard variables (without
the \lstinline{$SPheno} prefix) used in 
\lstinline{SPheno.m} files discussed in \cref{app:sphenomssm}. 
The new option is \lstinline{$SPhenoMatchingScale} which defines at which scale
the matching should be performed. 
\\
Running \lstinline{InitMatching} with an input file containing these lines, produces two outputs:
\begin{itemize}
\item The file \lstinline{EFTcoupling1_SPhenoEFT_MSSM_SimpleHighScaleSUSY.f90},
    located in the output directory of the MSSM model, 
    which contains the matching conditions in \Fortran format
\begin{lstlisting}[language=Fortran90]
Real(dp) Function EFTcoupling1(g1Q,g2Q,TanBeta,YdQ,YeQ,YuQ,UVscaleQ)
Implicit None
Complex(dp),Intent(in) :: YdQ(:,:),YeQ(:,:),YuQ(:,:) 
Real(dp),Intent(in) :: g1Q,g2Q,TanBeta,UVscaleQ

EFTcoupling1=-(-2*g2Q**4*TanBeta**4 + ... &
	& )/(192._dp*Pi**2*TanBeta**4*(1 + TanBeta**2)**2)

End Function EFTcoupling1
\end{lstlisting}
where most of the terms in the sum have been omitted as they are not
important for the discussion. \\

\item a \Mathematica file named \lstinline{SPhenoEFT_MSSM_SimpleHighScaleSUSY.m} which is located in the
    model directory of the SM. This file may look like
\begin{lstlisting}[language=SLHA]
ModelName = "SimpleHighScaleSUSY";
OnlyLowEnergySPheno = False;
MINPAR={{1, MSUSY}, {2, TanBeta}}
ParametersToSolveTadpoles={mu2}
UseParameterAsGUTscale = {{MSUSY}};

RenormalizationScaleFirstGuess = 173.^2;
RenormalizationScale = 173.;

DEFINITION[MatchingConditions]= Default[OHDM];
BoundaryHighScale = {};
BoundaryRenScale = {};

BoundaryMatchingUV = {
{\[Lambda],EFTcoupling1[g1,g2,TanBeta,Yd,Ye,Yu,mGUT]}
};

ListDecayParticles = Automatic;
ListDecayParticles3B = Automatic;

RealParameters = {MSUSY, TanBeta};

SelfDefinedFunctions = {
  ReadString["$SARAH_Directory/Output/MSSM/EWSB/Matching/SimpleHighScaleSUSY/\
  EFTcoupling1_SPhenoEFT_MSSM_SimpleHighScaleSUSY.f90"]
};
\end{lstlisting}
One can see that this file contains the information given to
\lstinline{InitMatching} (line 1-12). In addition, the information about the matching and
the corresponding \Fortran routines (using parameter without the \lstinline{Q} prefix) have been automatically added by \SARAH (line 14-16 and 21-26). 
\end{itemize}
The second step to generate a numerical code that includes the computed matching conditions 
is to run a new \Mathematica kernel and call the \SARAH routine \lstinline{MakeSPheno} using the
generated \lstinline{SPheno.m}, i.e.
\begin{MIN}
<<SARAH.m`
Start["SM"];
MakeSPheno[InputFile->"SPhenoEFT_MSSM_HighScaleSUSY.m"]
\end{MIN}
This generates all necessary \Fortran routines for the high-scale SUSY
implementation. The code is compiled in the same way as other \SARAH generated
\SPheno modules:
\begin{enumerate}
 \item Copy the \SARAH output to a new sub-directory of your \SPheno
     installation\footnote{\SPheno can be downloaded from \href{https://spheno.hepforge.org/}{\tt spheno.hepforge.org}}
 \item Copy the code to a new \SPheno sub-directory 
 \begin{lstlisting}[style=terminal]
 > cp -r $SARAH_Directory/Output/SM/EWSB/SPheno $SPheno_Directory/SimpleHighScaleSUSY/
 \end{lstlisting}
\item Compile the code 
\begin{lstlisting}[style=terminal,firstnumber=2]
 > cd $SPheno_Directory
 > make Model=SimpleHighScaleSUSY
\end{lstlisting}
\item Run \SPheno 
\begin{lstlisting}[style=terminal,firstnumber=4]
 > ./bin/SPhenoSimpleHighScaleSUSY
\end{lstlisting}
\end{enumerate}
For the last step, a Les Houches input file \cite{Allanach:2008qq} must be
provided which includes the numerical values for $M_{\rm SUSY}$ and $\tan\beta$
as well as settings for \SPheno. \SARAH generates also a template for such a
file which is located in 
\begin{center}
    \lstinline{$SARAH_Directory/Output/SM/EWSB/SPheno/Input_Files/}.
\end{center}
The actual behaviour of the compiled \SPheno code is described and compared with
the fully numerical approach in the next section.
\subsubsection{Matching at two scales}
\label{sec:two_scales_analytical}
\begin{figure}[tb]
\begin{center}
\includegraphics[width=\linewidth]{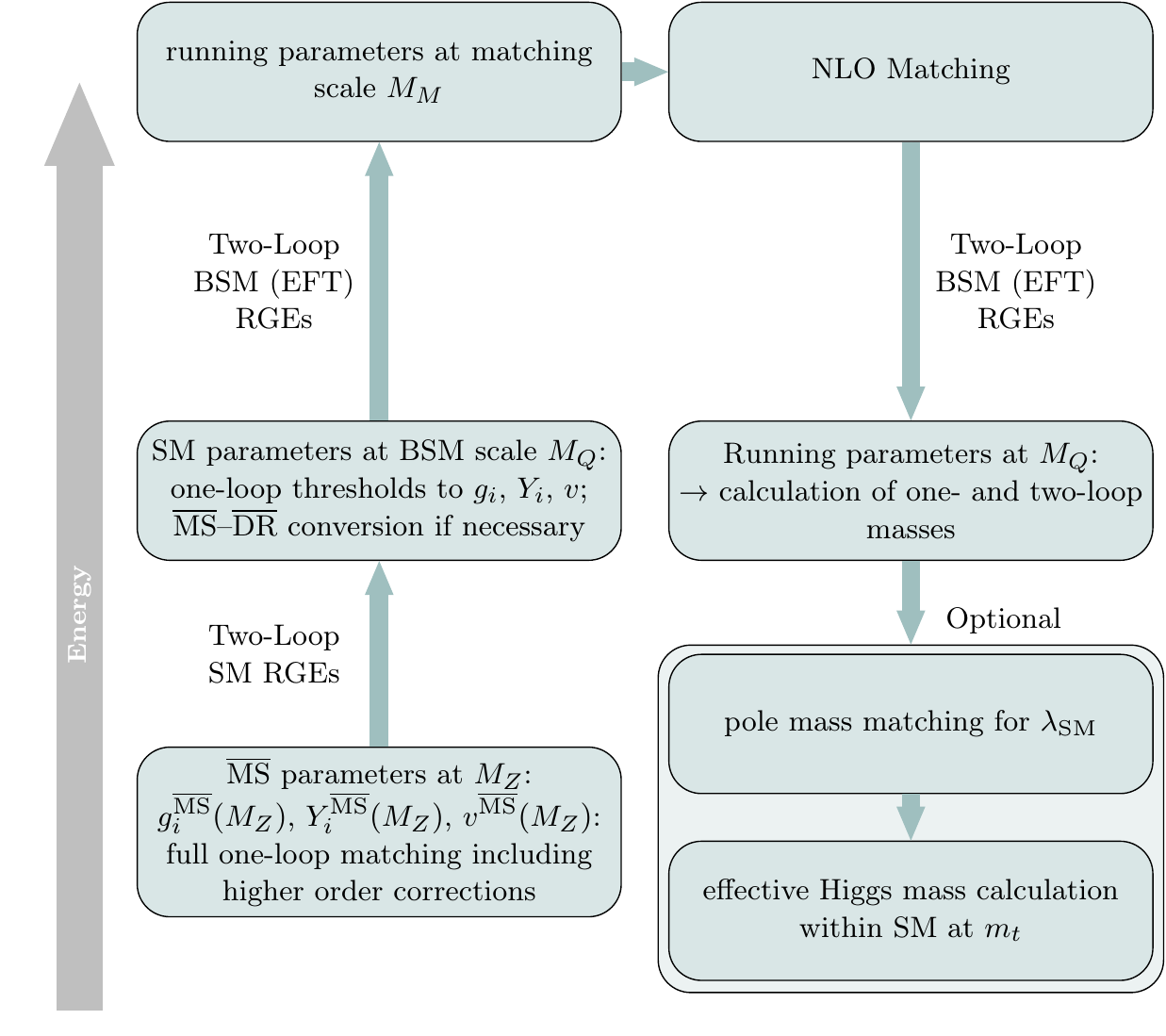}
\end{center}
\caption{Schematic procedure of the numerical evaluation of a parameter point
    with \SPheno when using the analytical setup to calculate matching
    conditions and implement them in \SPheno}. 
    \label{fig:flowchart_spheno_analytical}
\end{figure}

The analytical matching procedure discussed so far supports the derivation of
effective scalar couplings from a high-scale theory at a single matching scale.
Thus, towers of effective theories where the different sets of RGEs are needed
between the different matching scales are not a priori possible in this approach.
 On the other side, \SPheno always provides the possibility to perform a pole-mass matching
between a given BSM model and the SM as described in detail in
Ref.~\cite{Staub:2017jnp}. Thus, the functionality can be used to obtain
precise prediction for scenarios like 
\begin{eqnarray*}
    \text{BSM}_2 \quad \xrightarrow{\text{analytical matching}} \quad \text{BSM}_1 \quad \xrightarrow{\text{pole-mass matching}} \quad \text{SM} 
\end{eqnarray*}
where large scale separations between the two BSM models as well as the SM
exist. This is for instance the case for split-SUSY where the electroweakinos
are in the multi-TeV range. Thus, such scenarios are already fully covered. An
even more general implementation to allow for an arbitrary number of matching
scales and an RGE running in-between is only possible with the numerical
approach which we discuss next. A schematic overview about the numerical
evaluation of a parameter point when using the analytical calculation of
matching conditions is shown in \cref{fig:flowchart_spheno_analytical}.

\subsection{Numerical Approach}
\label{sec:num_approach}
The second option to generate a \SPheno version for an effective model
including the matching conditions to a UV theory is to set up a suitable {\tt
SPheno.m} for the EFT from the very beginning. This file must include the following
information in addition to the standard information which is usually defined in
the {\tt SPheno.m} files, see \cref{app:sphenomssm}:
\begin{lstlisting}[language=SLHA,title=SPheno.m]
MatchingToModel= { $MODELNAME };
MatchingScale= { $LIST_OF_PARAMETERS };

IncludeParticlesInThresholds={ $LIST_OF_FIELDS };

AssumptionsMatchingScale={  $LIST };
BoundaryMatchingScaleUp={ $LIST };
BoundaryMatchingScaleDown={ $LIST };

ParametersToSolveTadpoleMatchingScale={ $LIST_OF_PARAMETERS};
\end{lstlisting}
Note, this ansatz is not restricted to a single matching scale. Therefore, all entries are arrays of the dimension of the number of matching scales. 
The purpose of the different entries is
\begin{enumerate}
 \item {\tt MatchingToModel} is used to define the UV model(s), i.e. the model directory in \SARAH.
 \item {\tt IncludeParticlesInThresholds} defines the list of particles which are included in the loop calculations.
 \item {\tt AssumptionsMatchingScale} is used to define simplifying assumptions at the matching scale. A common choice is to neglect the contributions from EW VEVs or other small parameters.
 \item {\tt BoundaryMatchingScaleUp} defines the boundary conditions to relate the parameters of the UV theory to the running parameters of the EFT when the RGEs run from low to high scales.
 \item {\tt BoundaryMatchingScaleDown} defines the boundary conditions to relate the parameters of the UV theory to the running parameters of the EFT when the RGEs run from high to low scales.
 \item {\tt ParametersToSolveTadpoleMatchingScale} defines the parameters that are fixed by the tadpole equations in the full theory.
\end{enumerate}
Also one-loop matching conditions for fermionic interactions are available in the numerical approach. The full one-loop coupling is also indicated by using {\tt EFTcoupNLO}, i.e. 
\begin{lstlisting}[language=SLHA]
EFTcoupNLO[F1.F2.S][PL] 
EFTcoupNLO[F1.F2.S][PR]
\end{lstlisting}
where {\tt F1} and {\tt F2} are the involved fermions and {\tt S} is the
involved scalar. Yukawa-like interactions are chiral couplings. Therefore, the
main difference to purely scalar couplings is the second argument containing
{\tt PL}/{\tt PR} (for $P_{L,R} = \frac12 (1 \pm \gamma_5)$) to define which
part of the coupling is meant. Moreover, the keyword {\tt ShiftCoupNLO} can be
used just to obtain the one-loop shift to a coupling, e.g.  
\begin{lstlisting}[language=SLHA]
ShiftCoupNLO[F1.F2.S][PL] 
ShiftCoupNLO[F1.F2.S][PR]
\end{lstlisting}
Several examples for the usage of these options are given below. 

\subsubsection{One Matching Scale Without RGE Running Above}
We start again with the simplest example of high-scale SUSY without any RGE running above the matching scale. Thus, the produced \SPheno code will generate the same results as 
the one with the analytical approach in the last section. In order to set up a
high-scale SUSY version with degenerate SUSY masses at the matching scale, the
corresponding lines in the {\tt SPheno.m} located in the model directory of the EFT, i.e. \lstinline{Models/HighScaleSUSY/MSSM/SPheno.m}, must read
\begin{lstlisting}[language=SLHA,title=SPheno.m]
MatchingToModel= {"MSSM"};
MatchingScale = {m0};

UseParameterAsGUTscale = {m0};

IncludeParticlesInThresholds={
 {hh,Ah,Hpm,Su,Sd,Se,Sv,Chi,Cha}
};

AssumptionsHighScale={
 {
  {vd,epsUV*Cos[ArcTan[TanBeta]]}, 
  {vu,epsUV*Sin[ArcTan[TanBeta]]}
 }
};

BoundaryMatchingScaleUp={
{
  {Yu, Sqrt[1+TanBeta^2]/TanBeta*Yu},
  {Yd, Sqrt[1+TanBeta^2]*Yd},
  {Ye, Sqrt[1+TanBeta^2]*Ye}
 }
};

BoundaryMatchingScaleDown={
{
  {\[Lambda], -1/3 EFTcoupNLO[hh[1].hh[1].hh[1].hh[1]]}
 }
};

ParametersToSolveTadpoleMatchingScale={
 {mHd2,mHu2}
};
\end{lstlisting}
Note that for simple high-scale theories, without additional light fields,  
the \lstinline{SPheno.m} could also be stored in the SM model directory as the 
two models are technically the same. The newly introduced models are described 
in \cref{sec:moddesc}.
\\ 

The definitions are very similar to the analytical approach: the symbol 
\lstinline{epsUV} again has been used to neglect specific parameters at 
the matching scale. An important difference is that we have not singled 
out the contributions from only third generation Yukawas because this 
would not give any performance improvement for the numerical calculation. 
Note, that it is also not necessary to define the matching for the Yukawas 
when running down.  Moreover, we have used the option
to define the scale where the RGE running should stop as function of an input parameter
(\lstinline!UseParameterAsGUTscale = {m0}!)\footnote{The naming of this keyword,
which was originally introduced for other purposes, might be misleading because
the chosen scale need not be connected to any GUT theory.}. Thus, \SPheno will
run the RGEs only to that scale and evaluate the SUSY boundary conditions.
\\

The process to generate the \SPheno output and to compile the \Fortran code is
identical to the final steps for the analytical approach:
\begin{enumerate}
 \item Run {\tt MakeSPheno} of \SARAH with the new input file
 \begin{MIN}
<<SARAH.m
Start["HighScaleSUSY/MSSM"]
MakeSPheno[]
\end{MIN}
\item Copy the files and compile \SPheno
 \begin{lstlisting}[style=terminal]
 > cp -r $SARAH_Directory/Output/SM/EWSB/SPheno $SPheno_Directory/HighScaleSUSY/
 > cd $SPheno_Directory
 > make Model=HighScaleSUSY
\end{lstlisting}
\end{enumerate}

For a \SPheno version generated in that way, two additional flags are available
in the Les Houches input file to have some control over the calculations:
\begin{lstlisting}[language=SLHA,title=LesHouches.in]
Block SPhenoInput   # SPheno specific input 
 201 1 # Turn on/off one-loop contributions in the matching 
 202 0 # Turn on/off contributions from off-diagonal wave-function renormalisation
\end{lstlisting}
Thus, these flags can be used to:
\begin{enumerate}
 \item[{\tt 201}] Turn on/off all one-loop contributions to the matching. By
     default, they are turned on. This might be helpful to check the size and
     importance of the one-loop corrections.
 \item[{\tt 202}] Turn on/off the contributions from the off-diagonal
     wave-function renormalisation. By default, they are turned off. See
     \cref{sec:wave_off} for more details. 
\end{enumerate}

\subsubsection{Running Above the Matching Scale}
We can modify the last example easily to include also the running above the
matching scale. This might be for instance necessary if one wants to apply the
SUSY boundary conditions at the scale where the gauge couplings do unify but
not at the matching scale. In order to do so, one needs to remove {\tt
UseParameterAsGUTscale = \{m0\}} from the last example and put instead
\begin{lstlisting}[language=SLHA,title=SPheno.m.GUT,firstnumber=4]
ConditionGUTscale = g1 == g2;
\end{lstlisting}
Thus, \SPheno stops the running once the condition $g_1(Q)=g_2(Q)$ is fulfilled.\\
In addition, the matching conditions for the Yukawas are changed to
\begin{lstlisting}[language=SLHA,title=SPheno.m.GUT,firstnumber=17]
BoundaryMatchingScaleUp={
 {
  {Yu, Sqrt[1+TanBeta^2]/TanBeta*Yu - Sqrt[2]*ShiftCoupNLO[bar[Fu].Fu.hh[1]][PL]/ZH[1,2]},
  {Yd, Sqrt[1+TanBeta^2]*Yd - Sqrt[2]*ShiftCoupNLO[bar[Fd].Fd.hh[1]][PL]/ZH[1,1]},
  {Ye, Sqrt[1+TanBeta^2]*Ye - Sqrt[2]*ShiftCoupNLO[bar[Fe].Fe.hh[1]][PL]/ZH[1,1]}
 }
};
\end{lstlisting}
The need for the normalization onto the tree-level rotation matrix elements
\lstinline{ZH} is described in the next section.
In that way, we can include the one-loop shifts to all Yukawa couplings which
are necessary to have a consistent RGE running with two-loop SUSY RGEs between
the matching and GUT scale, see also the discussion in \cref{sec:above}. Note,
we did not consider any generation indices for the involved fermions, i.e.
the result of {\tt ShiftCoupNLO} is a $3\times 3$ matrix. If one wants to safe
program run-time it is possible to consider the one-loop shifts to the top Yukawa
couplings only. 
\begin{lstlisting}[language=SLHA,title=SPheno.m.GUT,firstnumber=18]
{...,
 {Yu, Sqrt[1+TanBeta^2]/TanBeta*Yu},
 {Yu[3,3], Yu[3,3] - Sqrt[2]*ShiftCoupNLO[bar[Fu[3]].Fu[3].hh[1]][PL]/ZH[1,2]},
 ...
}
\end{lstlisting}
Moreover, the shifts for the gauge couplings are applied automatically. 

\subsubsection{Several Matching Scales}
\label{sec:multiscale}
With the above settings one can now implement an arbitrary number of matching
scales. However, as we have noted already in \cref{sec:two_scales_analytical},
the pole-mass matching to the SM is automatically included in the \SPheno
output. Thus, if a second matching scale, which is not too far away from the EW
scale, is needed, one can simply rely on that. However, if more than two
matching scales are needed, or if the matching to the SM should take place at
such a high scale where the pole-mass matching might suffer from numerical
problems\footnote{We elaborate a bit on that issue in \cref{sec:HSMSSM}.}, one
can now start to build up towers of EFTs by defining more matching scales in
{\tt SPheno.m}. For instance, the full input to define the tower
\begin{center}
 SM \quad $\to$ \quad THDM \quad $\to$ \quad   THDM + electroweakinos \quad $\to$ \quad MSSM
\end{center}
is given in \cref{app:3scales}. In this example we also make use of the
functionality to calculate new fermionic couplings at the one-loop level below
a matching scale:
\begin{lstlisting}[language=SLHA,title=SPheno.m,firstnumber=142]
BoundaryMatchingScaleDown={
...
 {g1u,-2*(EFTcoupNLO[Chi[fB].Chi[FHu0].hh[2]][PL])},
 {g2u,2*(EFTcoupNLO[Chi[fW0].Chi[FHu0].hh[2]][PL])}, 
 {g1d,2*(EFTcoupNLO[Chi[fB].Chi[FHd0].hh[1]][PL])},
 {g2d,-2*(EFTcoupNLO[Chi[fW0].Chi[FHd0].hh[1]][PL])},
...
}
\end{lstlisting}
Here, $g_{1,2}^{u,d}$ are the split-SUSY couplings between the Higgs boson and a
Higgsino-Gaugino pair, see e.g. Ref. \cite{Bagnaschi:2014rsa}. We include these corrections by considering the one-loop
amplitude between the Higgs boson and a pair of neutralinos. In this example we have
also used another feature: we have not explicitly defined the generation
indices of the involved neutralinos. The reason for this is: even if the
neutralino mass matrix contains only zero's under the given approximations
($\mu,M_i \ll M_{\rm SUSY}$), it is not clear how the mass eigenstates are
ordered in the numerical run. Therefore, we have used the name of the gauge
eigenstates. By doing that, \SPheno checks during the numerical evaluation
which of the mass eigenstates has the biggest contribution of the given gauge
eigenstate. Of course, if the rotation matrix for the neutralinos is not
equivalent to the unit matrix, i.e. if some mixing appears for instance because
of effects of non-vanishing $\mu$, one needs to define 
\begin{lstlisting}[language=SLHA,title=SPheno.m,firstnumber=142]
BoundaryMatchingScaleDown={
...
 {g1u,-2*(EFTcoupNLO[Chi[fB].Chi[FHu0].hh[2]][PL]/(ZN[fB,1]*ZN[FHu0,4]))},
 ...
...
}
\end{lstlisting}
Thus, the rotation to mass eigenstates, which should take place just at the
weak scale, is divided out.

\subsubsection{Summary}
\begin{figure}[tb]
\begin{center}
\includegraphics[width=\linewidth]{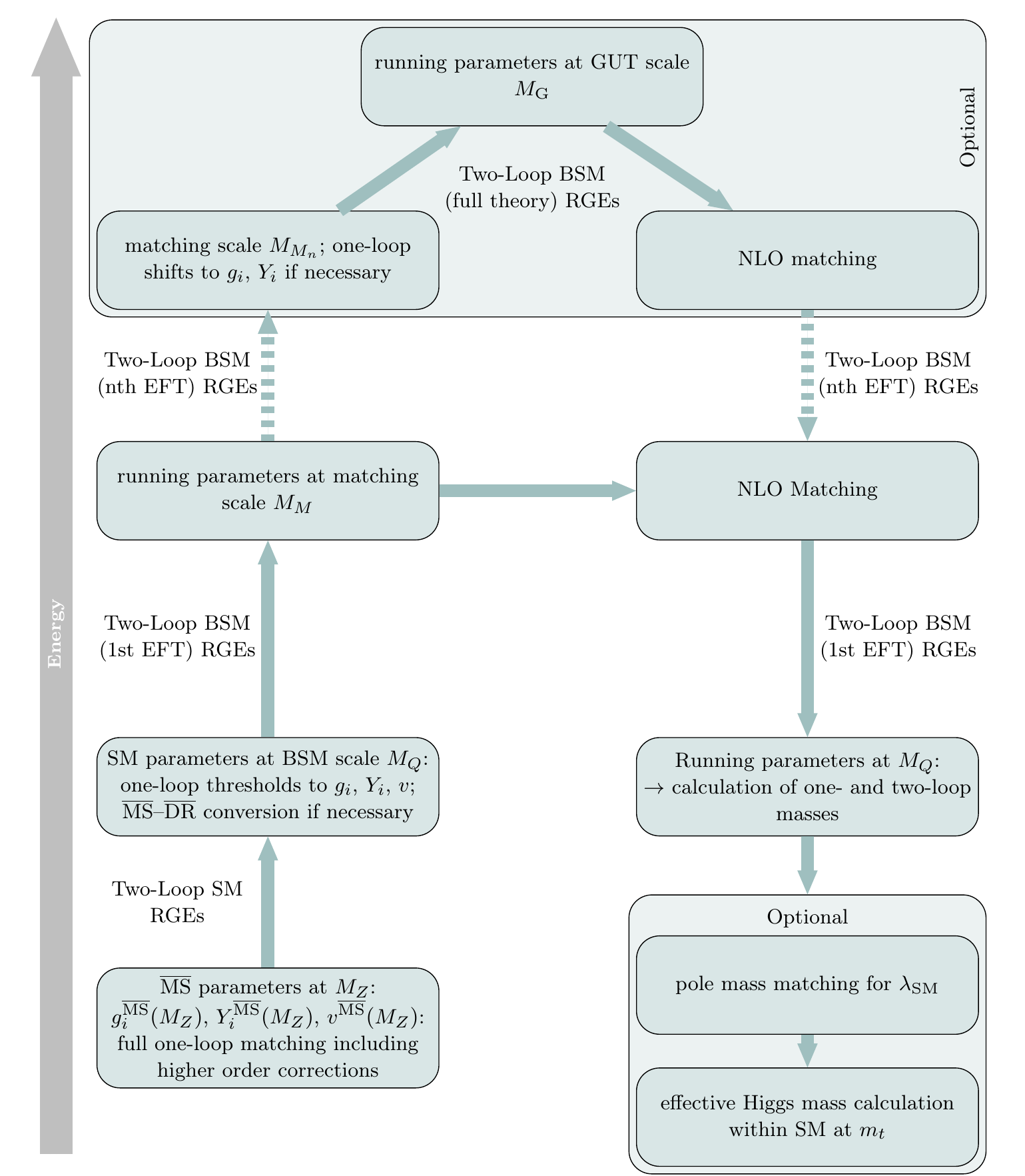}
\end{center}
\caption{Schematic procedure of the numerical evaluation of a parameter point
    with \SPheno when using the numerical setup to calculate matching
conditions and implement them in \SPheno.}
\label{fig:flowchart_spheno_numerical}
\end{figure}

A summary of the numerical evaluation of a parameter point with \SPheno which
includes several matching scales $M_{M_n}$ and optionally also the running to
the GUT scale $M_G$ is given in \cref{fig:flowchart_spheno_numerical}.

\subsection{Included Models and Input Files in \SARAH}
\label{sec:moddesc}
\begin{figure}
\begin{center}
\includegraphics[width=0.8\linewidth]{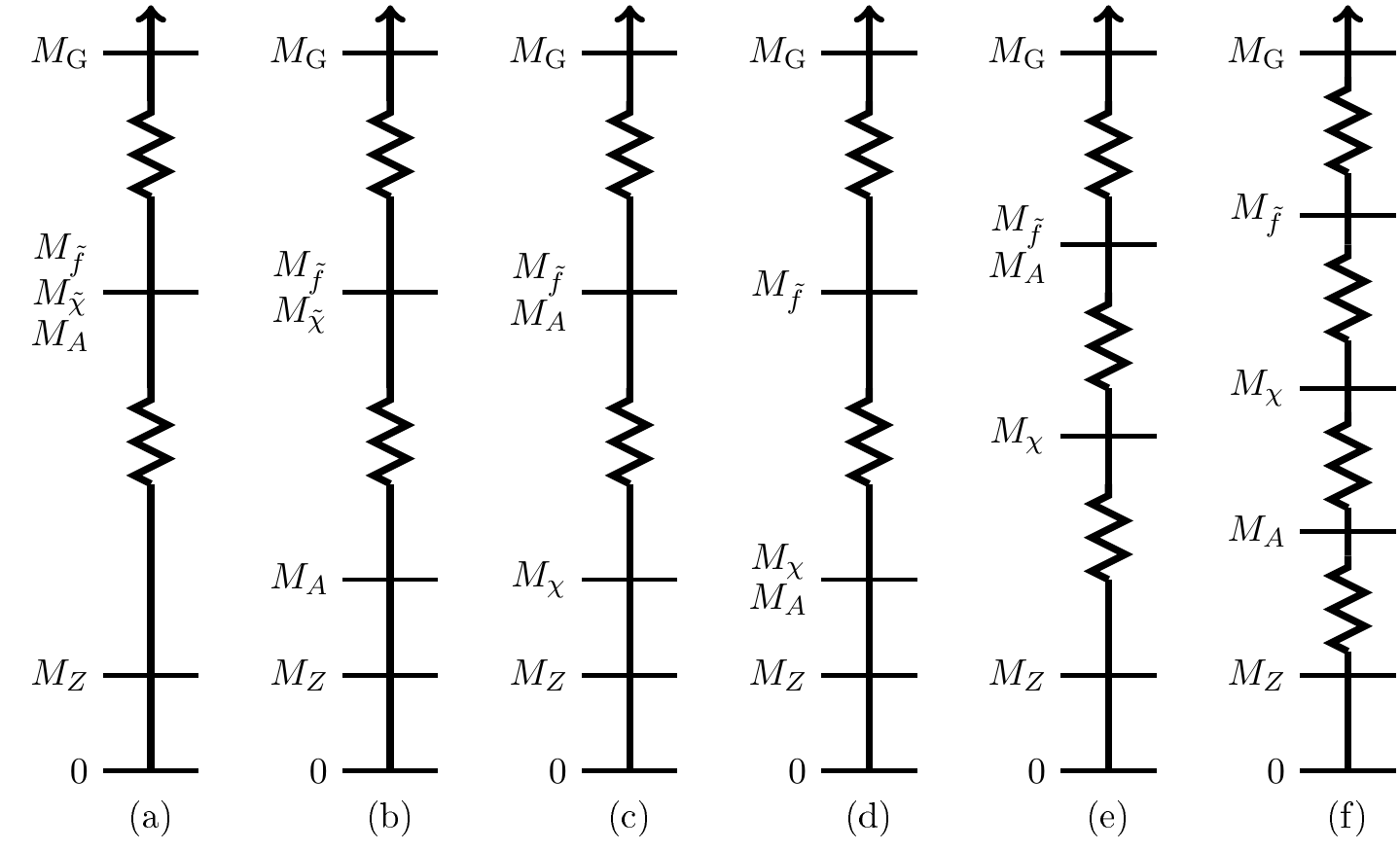}
\end{center}
\caption{Hierarchies for the MSSM so far included in \SARAH. For the NMSSM
    versions similar to (a) and (c) exist as well. The zigzag line represent a
large energy gap which is bridged with two-loop RGEs. The fields of the different mass symbols are explained in \cref{sec:results}.}
\label{fig:hierarchies}
\end{figure}

Several models which make use of the new functionality 
have already been implemented and are part of the publicly available \SARAH version. 
All hierarchies considered for the MSSM so far are summarised in \cref{fig:hierarchies}. Also
for the NMSSM with very heavy particles two models exist: the high-scale NMSSM,
where all SUSY fields are integrated out and a split-NMSSM, where the singlet
and the SUSY fermions are kept. 
The names of the new models that make use of the numerical approach are listed
in \cref{tab:models}. Also for the analytical approach several input files are
now included in \SARAH. Those are summarised in \cref{tab:matching.m}. Based on
these examples and by the explanations in this section, it is now 
straight-forward for the users to implement their own scenarios. 
\begin{table}[tb]
\begin{center}
\begin{tabular}{cccc}
\hline
 Model Name & EFT & UV model(s) & hierarchy \\
\hline 
{\tt HighScaleSUSY/MSSM}  & SM & MSSM  & (a)\\
{\tt HighScaleSUSY/NMSSM}  & SM & NMSSM & (a) \\
{\tt HighScaleSUSY/MSSMlowMA}  & THDM & MSSM & (b) \\
{\tt SplitSUSY/MSSM}  & SM+EWkinos & MSSM & (c) \\
{\tt SplitSUSY/NMSSM}  & SM+singlet+EWkinos & SMSSM & $\sim$(c) \\
{\tt SplitSUSY/MSSMlowMA}  & THDM+EWkinos & MSSM & (d) \\
{\tt SplitSUSY/MSSM\_2scale}  & SM & MSSM $\to$ SM+EWkinos & (e) \\
{\tt SplitSUSY/MSSM\_3scale}  & SM & MSSM $\to$ THDM+EWkinos $\to$ THDM & (f) \\
\hline 
\end{tabular}
\end{center}
\caption{The names of the new models which are part of \SARAH 4.14.0. The
    hierarchy in the last column refers to \cref{fig:hierarchies}. For the
    split-NMSSM also a light singlet is present, i.e. the hierarchy is similar
to $(c)$, but not identical.}
\label{tab:models}
\end{table}
\begin{table}[tb]
\begin{center}
\begin{tabular}{cccc}
\hline
 File Name & EFT & UV model & hierarchy \\
\hline 
{\tt MSSM/Matching\_HighScaleSUSY.m}  & SM & MSSM & (a) \\
{\tt NMSSM/Matching\_HighScaleSUSY.m}  & SM & NMSSM & (a) \\
{\tt MSSM/Matching\_SplitSUSY.m}  & SM+EWkinos & NMSSM & $\sim$(c) \\
{\tt MSSM/Matching\_THDM.m}  & THDM & MSSM & (b) \\
{\tt SMSSM/Matching\_SplitSUSY.m}  & SM+singlet+EWkinos  & SMSSM & (c) \\
\hline 
\end{tabular}
\end{center}
\caption{Input files for the analytical approach which are now delivered with
    \SARAH. The hierarchy in the last column refers to \cref{fig:hierarchies}.
    For the split-NMSSM also a light singlet is present, i.e. the hierarchy is
similar to $(c)$, but not identical. }
\label{tab:matching.m}
\end{table}

\section{Examples, Self-Consistency Checks and Comparisons with Other Codes}
\label{sec:results}
The following section describes realistic examples of practical applications of the presented framework.
We consider different high-scale SUSY scenarios which were
already studied intensively in literature. In particular comparisons between 
predictions for the SM Higgs boson mass derived with our generic setup against dedicated
tools and calculations are made. In this context, we demonstrate also
the perfect agreement between the two available options to use \SARAH/\SPheno
for numerical studies. Finally, we also show that one can easily obtain precise
results for other high-scale extensions for which no other tool existed so far.

\subsection{Low-Energy Limits of the MSSM}
In the introduction it was already mentioned that SUSY models with a SUSY
breaking scale well above the electroweak scale became more popular in the
recent years. While in these scenarios the direct observation of SUSY states
is difficult or even impossible, these models are severely constrained by the
Higgs boson mass measurements. For instance, if the masses of all superpartners are
degenerate, the highest possible SUSY breaking scale in the MSSM is about
$10^{10}$~GeV \cite{Bagnaschi:2014rsa}. For higher scales, the predicted $m_h$
always becomes too large. Since the Higgs boson mass in these models is the crucial
observable, a precise calculation is mandatory and specialised codes have been
developed to get reliable predictions. We are going to consider three different
cases: (i) split-SUSY in which all SUSY scalars are very heavy, but
electroweakinos might stay moderately light, (ii) high-scale SUSY in which all
SUSY masses and the additional Higgs boson masses are large and degenerate,  (iii)
high-scale SUSY with a second light(ish) Higgs doublet. In all three cases we
work with the following reduced set of input parameters
\begin{equation}
    m_0,\, M_\chi,\, M_A,\, A_0\,\,\, \text{and}\,\,\, \tan\beta \,, 
\end{equation}
with
\begin{align}
m_{\tilde{u}}^2 = m_{\tilde{d}}^2 = m_{\tilde{q}}^2 = m_{\tilde{l}}^2 =m_{\tilde{e}}^2 \equiv {\bf 1} m_0^2\, , \\
M_1 = M_2 = M_3 = \mu \equiv M_\chi\, , \\
B_\mu \equiv M_A^2 \frac{\tan\beta}{1+\tan^2\beta}\, , \\
T_i \equiv A_0 Y_i \hspace{0.6cm} \text{and} \hspace{0.6cm} i=\{e,d,u\} \, .
\end{align}
Here, $m^2_{\tilde{f}}$ are the soft masses squared for all chiral superfields,
$M_A$ is the mass of the heavy Higgs doublet, $M_i$ are the soft gaugino
masses, $\mu$ is the Higgsino mass term in the superpotential, and $B_\mu$,
$T_i$ are the soft-breaking equivalents of the $\mu$-term and the Yukawa
couplings in the superpotential.

\subsubsection{Split-SUSY: MSSM $\to$ SM \& Electroweakinos \& Gluinos}
\begin{figure}[htb]
\includegraphics[width=0.49\linewidth]{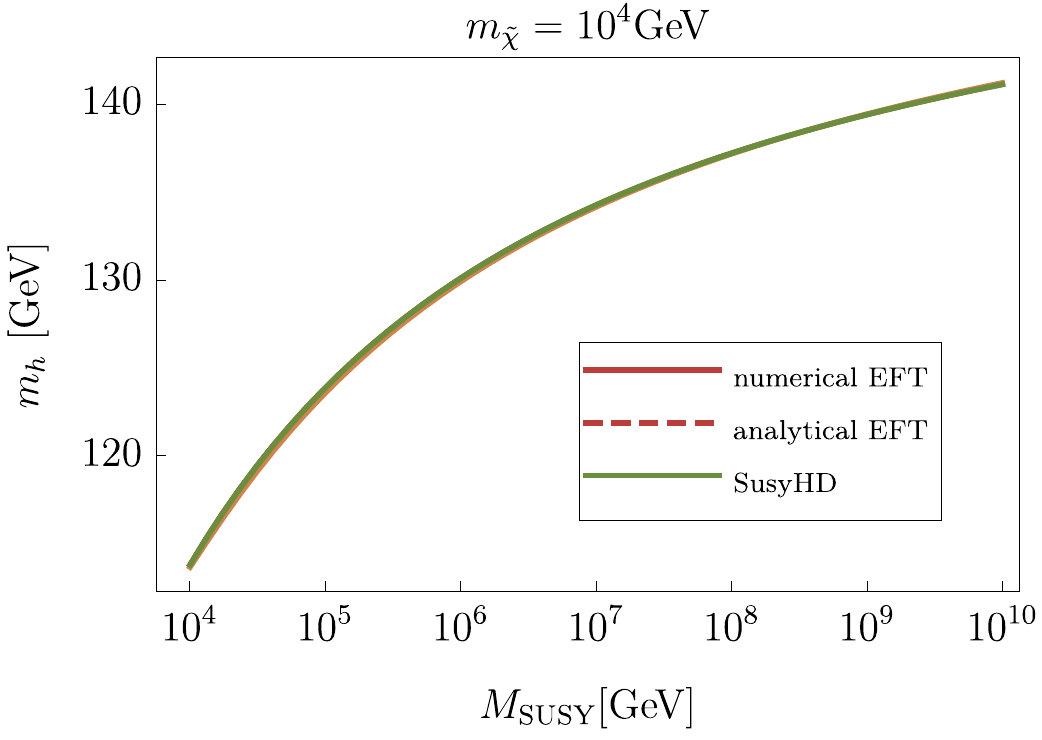} \hfill 
\includegraphics[width=0.49\linewidth]{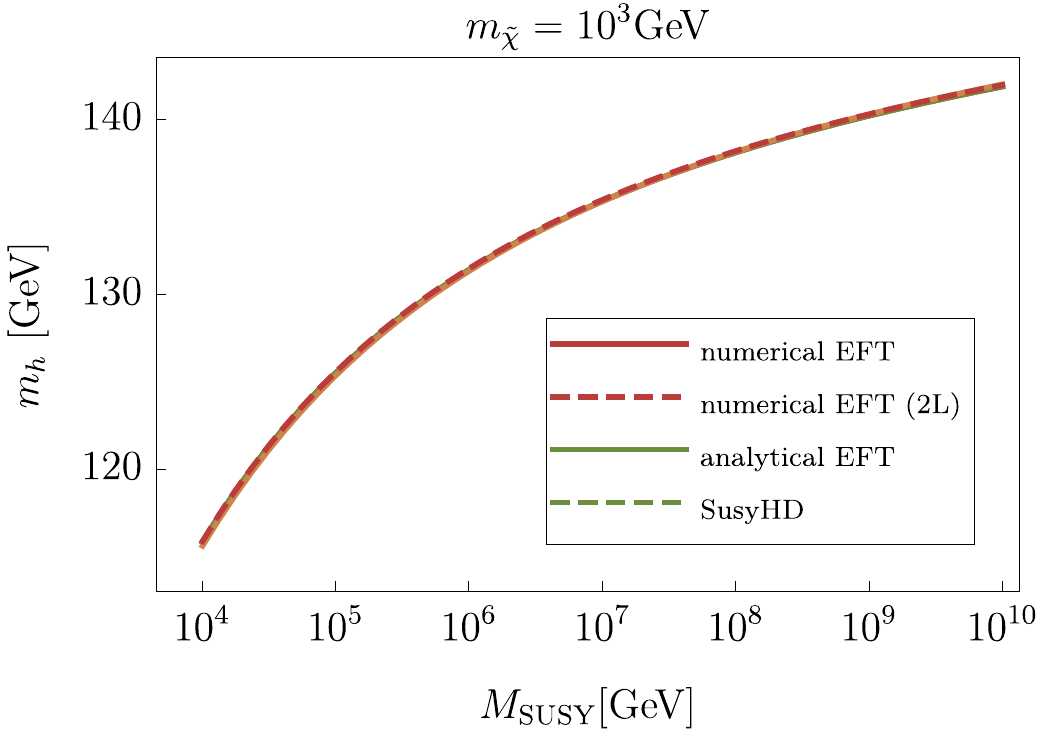} \\ 
\includegraphics[width=0.49\linewidth]{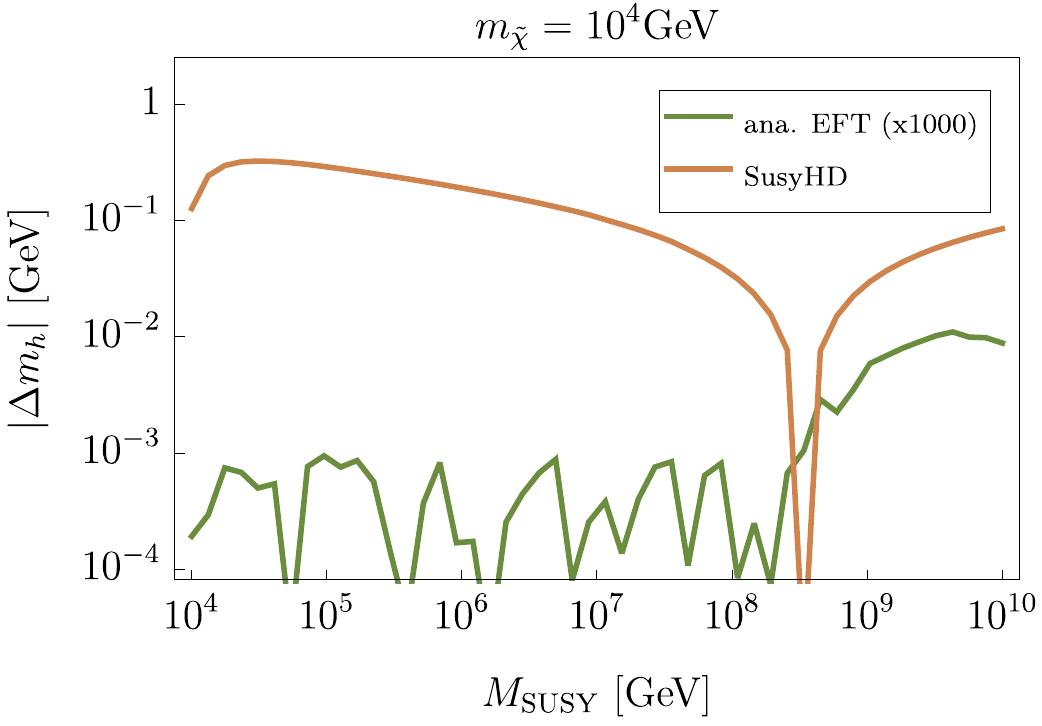} \hfill 
\includegraphics[width=0.49\linewidth]{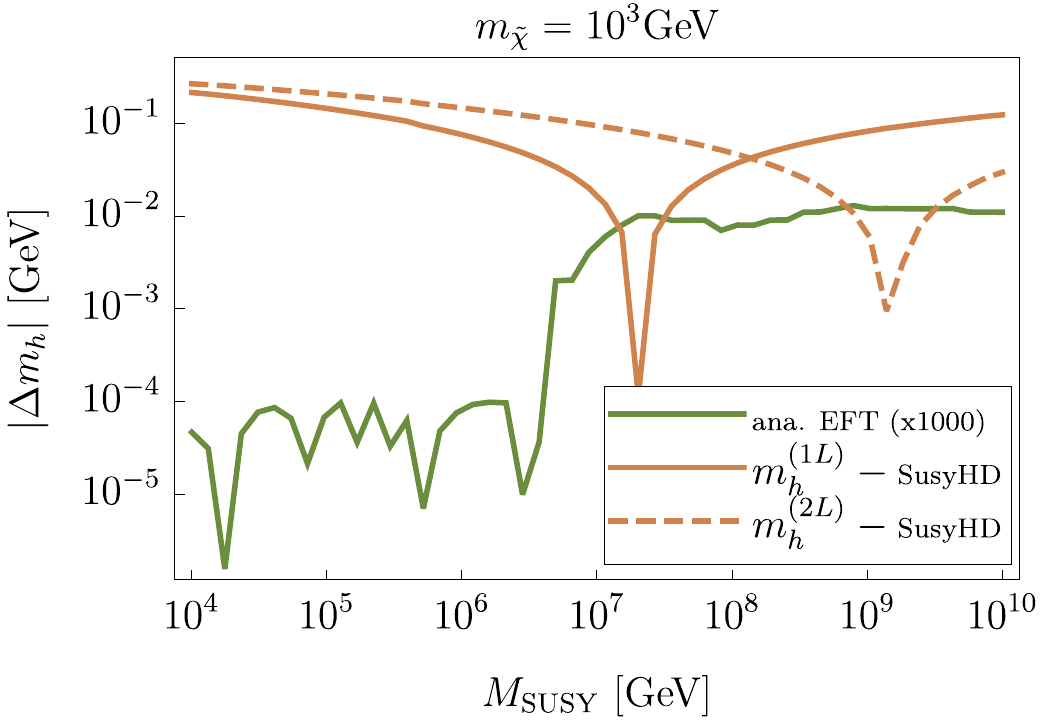}  
\caption{Prediction for the SM Higgs boson mass $m_h$ in split-SUSY as function of
    $M_{\rm SUSY}$, which is the mass scale of all SUSY scalars, and for two
    different values of $M_\chi$, which is the mass scale of all SUSY fermions.
    In addition, we used here $\tan\beta=3$. The blue and purple lines show the
    results of \SPheno using the analytical or numerical approach to obtain the
    matching conditions at $M_{\rm SUSY}$, while the green line corresponds to
    the results of {\tt SusyHD}. For \SPheno and $M_\chi=1$~TeV, we show also
    the result when turning on the two-loop corrections at the low scale. The
    first row shows the calculated values for $m_h$, while the second give the
    differences between the calculations compared to the numerical approach.}
\label{fig:split}
\end{figure}
Split-SUSY with very heavy SUSY scalars but significantly lighter SUSY fermions
keeps most of the nice SUSY properties like gauge coupling unification and
provides a viable dark matter candidate. In this setup, the full MSSM is
matched to the SM extended by additional fermions. The Lagrangian of the
effective theory reads
\begin{align}
\lagr_{\rm EFT} = & \lagr_{\rm SM} - 
\left(\frac12 M_3 \tilde{g}^\alpha \tilde{g}^\alpha + \frac12 M_2 \tilde{W}^a \tilde{W}^a + \frac12 M_B \tilde{B}\tilde{B} + \mu \tilde{H}_u^T \epsilon \tilde{H}_d \, + \, \text{h.c.}  \right) \nonumber \\
& -\left[\frac{1}{\sqrt{2}} H^\dagger \left(\tilde{g}_{2u}  \sigma^a \tilde{W}^a + \tilde{g}_{1u} \tilde{B} \right) \tilde{H}_u +   
\frac{1}{\sqrt{2}} H^T \epsilon \left(-\tilde{g}_{2d}  \sigma^a \tilde{W}^a + \tilde{g}_{1d} \tilde{B} \right) \tilde{H}_d \, + \, \text{h.c.} \right] \,,
\label{EQ:LagSplit}
\end{align}
where the Yukawa couplings $g_{1,2}^{u,d}$ are as in the example of \cref{sec:multiscale}, $\sigma^a$ are the pauli matrices and $\alpha=1,...,8$.
In order to calculate the Higgs boson mass in this model, the common approach is to
(i) decouple the SUSY scalars at the scale $M_{\rm SUSY}$ and calculate
$\lambda_{SM}(M_{\rm SUSY})$ including important higher-order corrections, (ii)
run the split-SUSY RGEs to the scale $M_\chi$ of the remaining SUSY states and
calculate the shift in $\lambda_{SM}(M_\chi)$, (iii) run the SM RGEs to $m_t$ and
calculate $m_h(m_t)$ at the two-loop level. The full results for the one-loop
matching conditions at $M_{\rm SUSY}$ and $M_\chi$ were given in
Ref.~\cite{Bagnaschi:2014rsa}. Also the dominant two-loop corrections to
$\lambda_{SM}$ of order $O(\alpha_t^2\alpha_s)$ have been included in this
reference. These results were implemented into the code {\tt SusyHD}
\cite{Vega:2015fna} and also {\tt FlexibleSUSY}
\cite{Athron:2014yba,Athron:2017fvs} uses the matching conditions from
literature. \\

We have compared the analytical expressions of Ref.~\cite{Bagnaschi:2014rsa}
for the one-loop thresholds with the results of \SARAH and found perfect
agreement. Thus, we can immediately go to the discussion of the comparison of
the numerical results of \SPheno and {\tt SusyHD}. Even if the expressions for
the thresholds agree, there are many other ingredients which enter the Higgs
mass prediction. Most importantly, the determination of the top Yukawa coupling
which affects all comparisons shown here. Also higher-order corrections  for
high-scale SUSY scenarios are implemented to some extent in other codes which
are not (yet) available in our generic setup. The corresponding model in \SARAH
which we have set up for this scenario is
\begin{center}
{\tt SplitSUSY/MSSM} 
\end{center}
We show in \cref{fig:split} the calculated Higgs boson mass by {\tt SusyHD}\footnote{During this comparison we found a bug in the two-loop RGEs of
$\lambda_{SM}$ for split-SUSY as implemented in {\tt SusyHD}. The contribution
$\frac{21}{2} \tilde{g}_{2d}^2
\tilde{g}_{2u}(\tilde{g}_{1d}^2+\tilde{g}_{1u}^2)$ misses one power of
$\tilde{g}_{2u}$. We fixed that and in all following results the patched
version of {\tt SusyHD} is used.} and \SPheno as function of $M_{\rm SUSY}$ for
two different choices of $M_\chi$. First of all, one can see that the overall
agreement is very good between all calculations: for the two calculations
implemented in \SARAH/\SPheno we find agreement up to the numerical precision,
while the biggest difference between \SPheno and {\tt SusyHD} is well below one
GeV for all considered values of $M_{\rm SUSY}$. \\

For the case of electroweakino masses of 1~TeV we show also the \SPheno result
when using a two-loop fixed-order calculation in the EFT. We see, in agreement
with a previous study in Ref.~\cite{Braathen:2017izn}, that the two-loop
contributions of the additional fermions have only a mild effect on the SM-like
Higgs boson mass. 

\subsubsection{High-scale SUSY: MSSM $\to$ SM}
\label{sec:HSMSSM}
\begin{figure}[tb]
\includegraphics[width=0.49\linewidth]{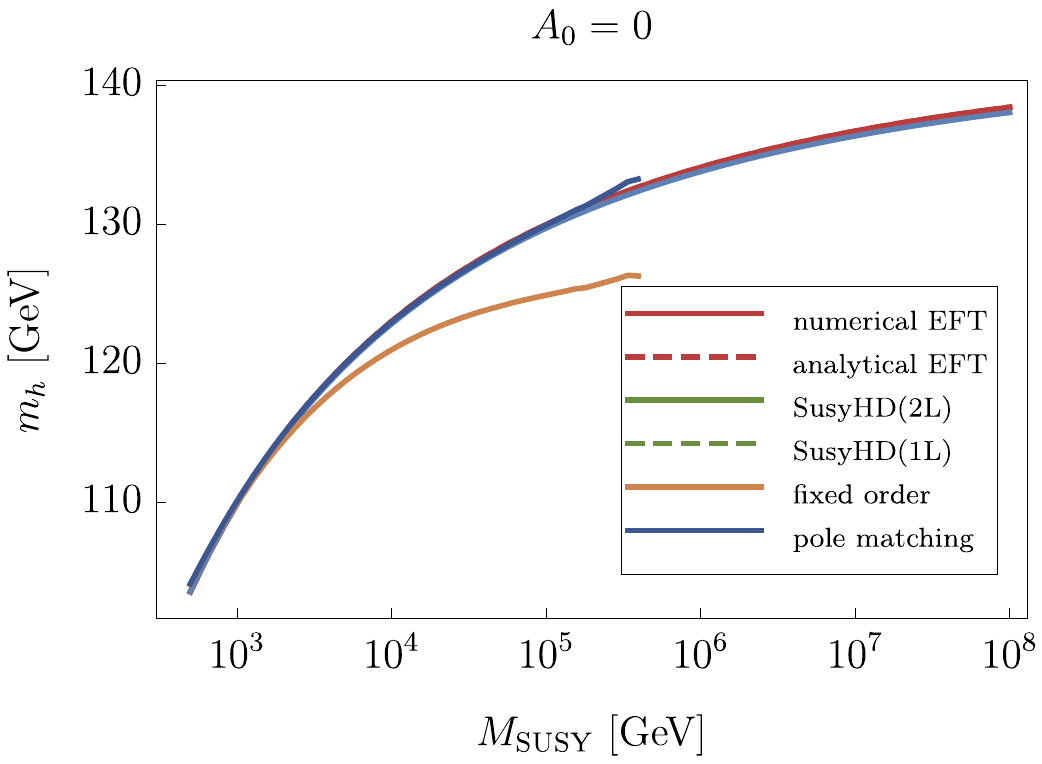} \hfill 
\includegraphics[width=0.49\linewidth]{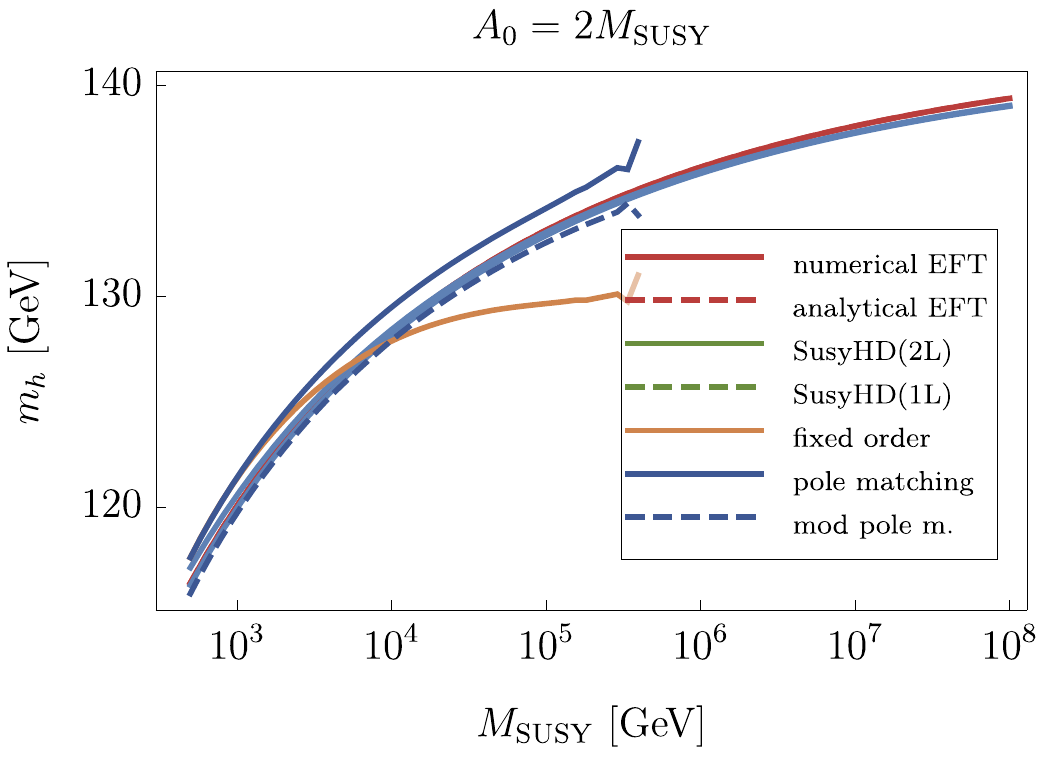} \\[5mm]
\includegraphics[width=0.49\linewidth]{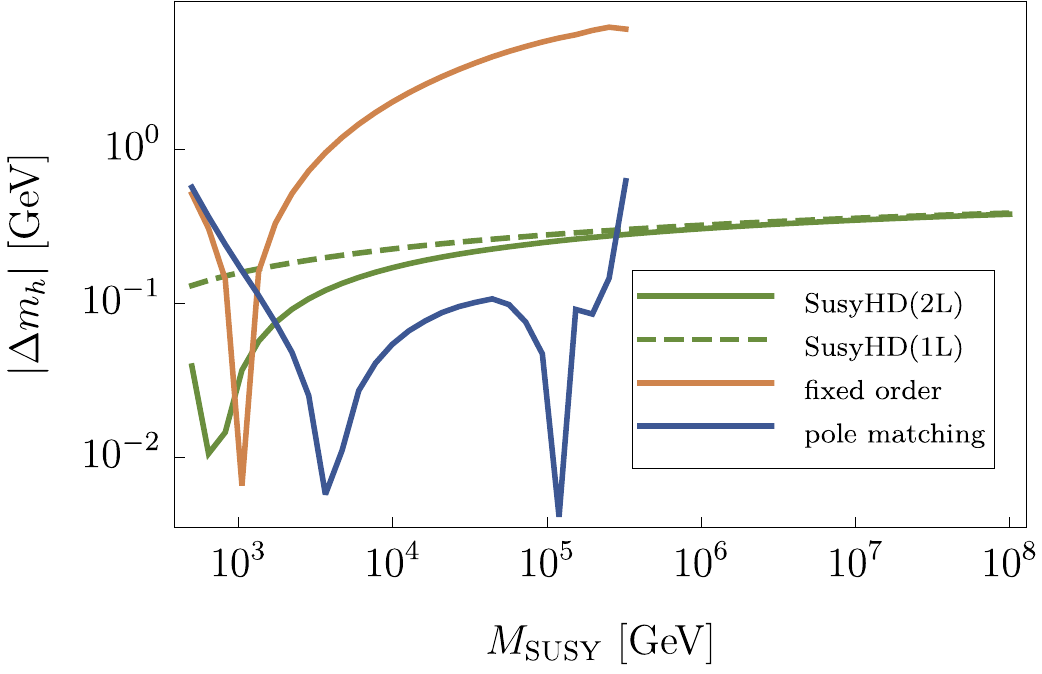} \hfill 
\includegraphics[width=0.49\linewidth]{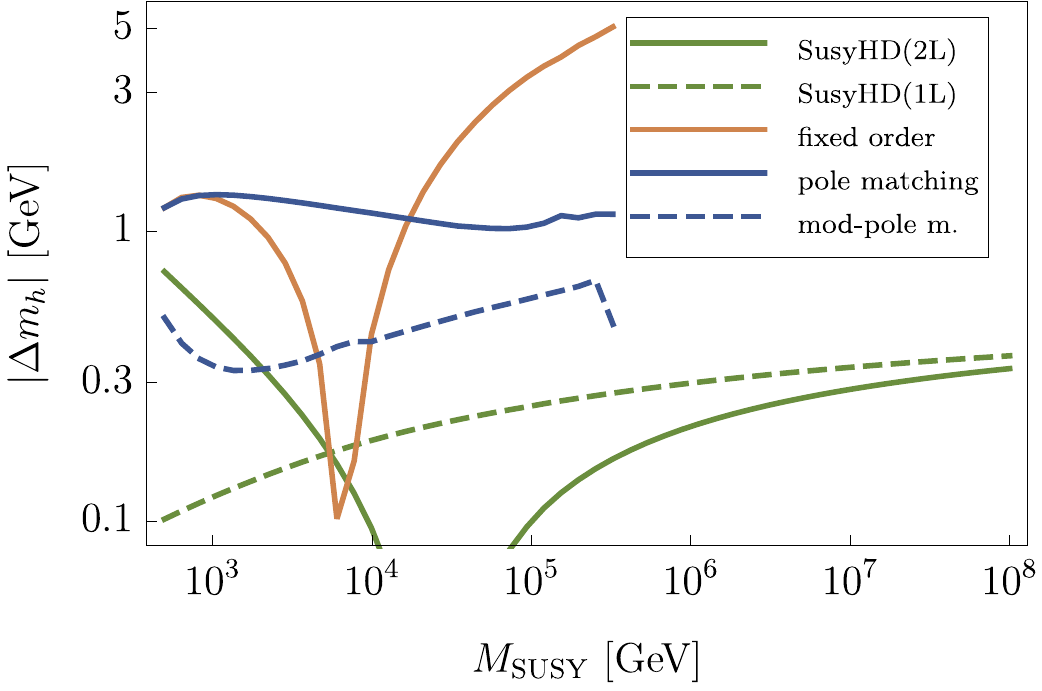}
\caption{Comparison between \SPheno and {\tt SusyHD} for high-scale SUSY as
    function of the SUSY scale $M_{\rm SUSY}=m_0=M_\chi=M_A$. Here, we used
    $\tan\beta=10$ and two different values of $A_0$. For \SPheno we give the
    results using the analytical and numerical approach to perform the matching
    at the SUSY scale. Also the results of a fixed-order calculation and using
    a pole-mass matching are included. For {\tt SusyHD} the results with and
    without the two-loop corrections in the matching of $\lambda_{SM}$ are
    depicted. The first row shows the calculated Higgs boson mass of all codes, while
the second row gives the differences between the different
calculations w.r.t. the numerical approach.}
\label{fig:HighscaleSUSY}
\end{figure}
An even more extreme setup than split-SUSY is high-scale SUSY in which all SUSY
partners are very heavy. Thus, the effective model is just the SM, i.e.
\begin{equation}
 \lagr_{\rm EFT} =  \lagr_{\rm SM} \,,
\end{equation}
and the only visible impact of SUSY is the prediction of $\lambda_{SM}$ at the
matching scale $M_{\rm SUSY} \simeq M_\chi \simeq M_A$. The matching conditions
at the SUSY scale are just the combination of the two matching conditions for
split-SUSY applied at a single matching. Thus, it is obvious that also for this
case a full agreement between our analytical results and those of
Ref.~\cite{Bagnaschi:2014rsa} exists. However, Ref.~\cite{Bagnaschi:2014rsa}
includes also the dominant two-loop corrections in the case of high-scale SUSY
which also entered the code {\tt SusyHD}. Therefore, it's worth to discuss also
the numerical differences between \SPheno and {\tt SusyHD} for the case of
high-scale SUSY. The model implemenation in \SARAH is called
\begin{center}
{\tt HighScaleSUSY/MSSM} 
\end{center}
The results are summarised in \cref{fig:HighscaleSUSY}. In addition to the
comparison to {\tt SusyHD} we also compare the results to two other
calculations: a standard fixed-order calculation as well as an EFT calculation
based on the  pole-mass matching \cite{Staub:2017jnp}. In the pole-mass
matching, the quartic coupling $\lambda_{SM}$ is calculated from the condition 
\begin{equation}
m_h^{2, \rm pole\ SM}(M_{\rm SUSY}) \equiv m_h^{2, \rm pole\ MSSM}(M_{\rm SUSY})
\end{equation}
which can be translated into 
\begin{equation}
\label{eq:lambda_pole}
\lambda_{SM}(M_{\rm SUSY}) =  \frac{1}{v^2(M_{\rm
    SUSY})}\left(m_h^{2, \rm pole\ MSSM} (M_{\rm SUSY}) - \Pi_h^{\rm SM}(M_{\rm SUSY})\right) \,,
\end{equation}
where $\Pi^{\rm SM}_h$ are the loop corrections to $m_h$ known from the SM. The
pole-mass matching has the advantage that also terms $\frac{v}{M_{\rm SUSY}}$
are included and that only two-point functions need to be calculated instead of
four-point functions, see Ref.~\cite{Athron:2016fuq} for more details. On the
other side, this approach has also some drawbacks. It is mainly restricted to
the SM as EFT, but it is not straightforward to be used in models with several
light scalars. Also a consistent matching at the two-loop level needs some
fiddling with the running parameters which enter the different parts of
\cref{eq:lambda_pole}, see Ref.~\cite{Athron:2017fvs}. While \SPheno by default
used \MS parameters to calculate $\Pi_h^{\rm SM}$ and \DR SUSY parameters in the
calculation of $m_h^{2, \rm pole\ MSSM}(M_{\rm SUSY})$, we also give the
results for using $Y_t^{\MS}$ in both calculations. This is called 'modified
pole-mass matching' in the left plot of \cref{fig:HighscaleSUSY}. The
difference between both results is a two-loop effect and could be taken as
estimate of the remaining uncertainty in the one-loop pole-mass matching. 
\\

Moreover, we find that the pole-mass matching becomes also numerically unstable
--at least in \SPheno-- once $M_{\rm SUSY} \gg v$ is used because the loop
functions used for the pole-mass calculations are not optimised for these
cases: we see in \cref{fig:HighscaleSUSY} that the pole-mass matching breaks down at $M_{\rm SUSY}
\simeq 5 \cdot 10^5$~GeV. Nevertheless, we find that the agreement between the
pole-mass matching and the direct matching procedure
 presented here is very good
for SUSY scales up to 100~TeV. One finds also that the fixed-order calculation
agrees perfectly with the pole-mass matching for $M_{\rm SUSY}$ below 1~TeV. Of
course, for larger SUSY scales, the discrepancy between the fixed-order
calculation and all EFT calculations grows very rapidly. 
\\

We come back to the comparison with {\tt SusyHD}: we see that the agreement
between \SPheno and {\tt SusyHD} is also very good and the differences are
always of the level of  1~GeV or below. The 1~GeV differences appear only for
the choice $A_0=2 M_{\rm SUSY}$ and $M_{\rm SUSY}$ around the TeV scale. In
that case, the two-loop corrections missing in \SPheno play some role. However,
for larger $M_{\rm SUSY}$ or smaller trilinear terms, these two-loop
corrections cause only a moderate shift -- or become even completely
negligible. Thus, we think that it is not a substantial drawback of our setup
that 'only' one-loop corrections are included so far.

\subsubsection{High-scale SUSY with intermediate $M_A$: MSSM $\to$ THDM}
\begin{figure}[ht]
\includegraphics[width=0.49\linewidth]{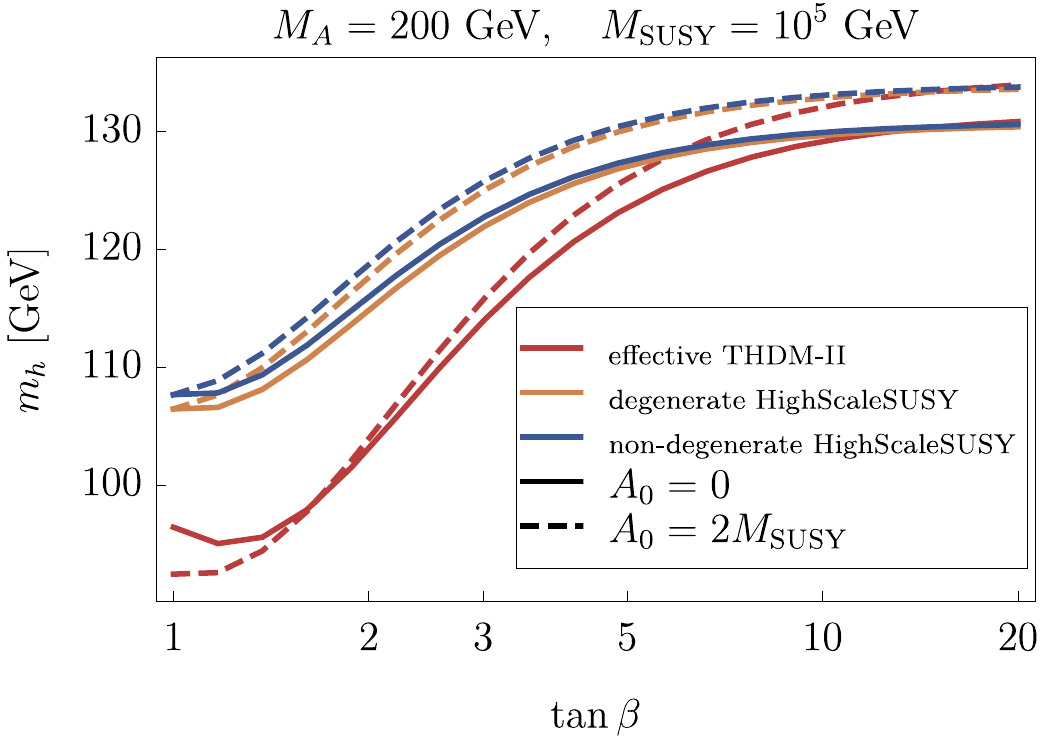} \hfill 
\includegraphics[width=0.49\linewidth]{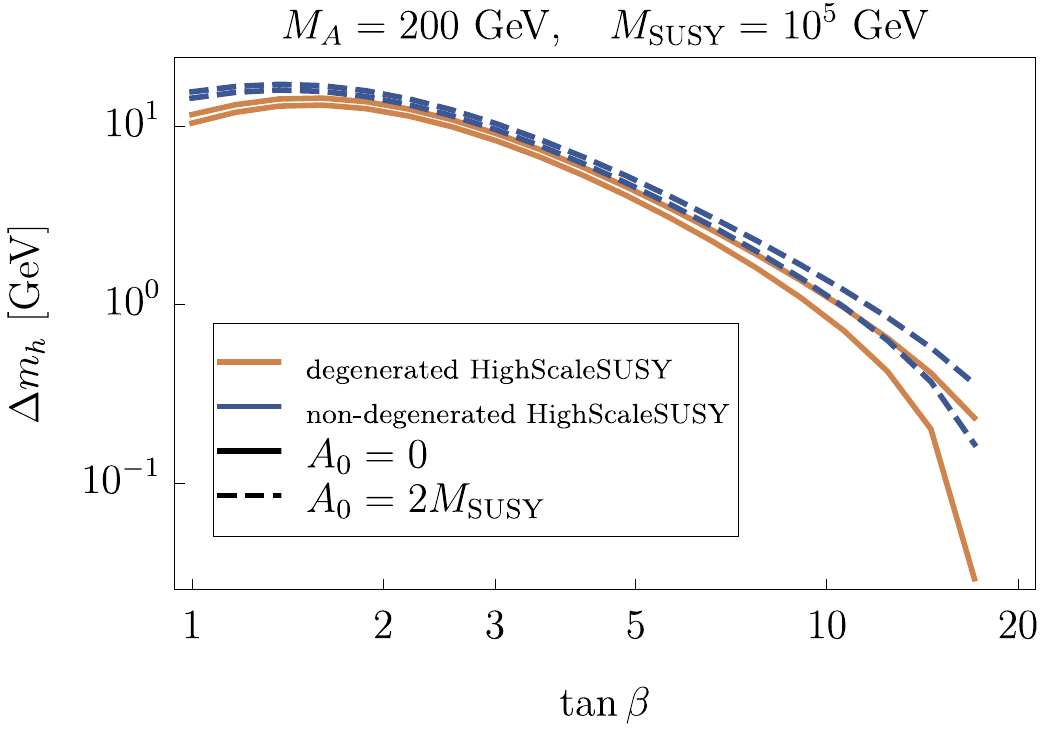} 
\caption{The Higgs boson mass prediction as function of $\tan\beta$ for a high-scale
    SUSY scenario with low $M_A$ and two different values of $A_0$. The blue
    lines show the results for a proper matching of the MSSM to the THDM-II.
    The green lines correspond to the matching of the MSSM to the SM where
    either all scalars, including the additional Higgs states, are assumed to
    be degenerate, or in which the actual value of $M_A$ has been used. The
    figure on the right show the difference between the matching to the THDM
and to the SM.}
\label{fig:lowMA_vs_HighScale}
\end{figure}
In the last example we have assumed that all BSM particles are very heavy and
degenerate. An important deviation of this ansatz is the possibility that the
second Higgs doublet remains light, i.e. only fields with negative $R$-parity
are very heavy. In this case, the low-energy theory of the MSSM is a
Two-Higgs-Doublet-Model type-II\footnote{Strictly speaking, one obtains a THDM
type-III when integrating out all SUSY fields in the MSSM because the 'wrong'
Yukawa couplings $\sim H_d^* \bar{q} u$ are loop-induced. However, this becomes
mainly important for flavour violating observables and has no visible impact on
our discussion of the Higgs boson mass prediction here.}. The Lagrangian of the EFT is
\begin{align}
-\lagr_{\rm EFT} = &  m_1^2 |H_1|^2 + m_2^2 |H_2|^2 + \lambda_1 |H_1|^4 + \lambda_2 |H_2|^4 + \lambda_3 |H_1|^2 |H_2|^2 + \lambda_4 |H_2^\dagger H_1|^2  \nonumber \\
 & + \left( m_{12} H_1^\dagger H_2 + \frac12 \lambda_5 (H_2^\dagger H_1)^2 + \lambda_6 |H_1|^2 H_1^\dagger H_2 + \lambda_7 |H_2|^2 H_1^\dagger H_2 + \text{c.c.}\right) \nonumber \\
 & + \left( Y_d H_1^\dagger \bar{d} q + Y_e H_1^\dagger \bar{e} l - Y_u H_2 \bar{u} q + \text{h.c.} \right)\, .
 \end{align}
One can make the following association between fields at the SUSY scale to
calculate the matching conditions
\begin{equation}
    \label{eq:thdmrot}
    H_1 \equiv - i \sigma_2 H_d^* \quad \text{and} \quad H_2 \equiv H_u\, .
\end{equation}
However, this choice is not unique as there is no preferred basis of Higgs
doublets in a general THDM, i.e. one could also interchange $H_1$ and $H_2$ or take any linear combination of them. 
With the common choice made in \cref{eq:thdmrot},
one can simultaneously apply a rotation into the mass basis on ($H_1,H_2$) and
($H_u$,$-i\sigma_2 H_d^*$) so that the tree-level mixing angle $\tan\beta$ of the MSSM
coincides with the effective THDM.
\begin{figure}[htb]
\includegraphics[width=0.49\linewidth]{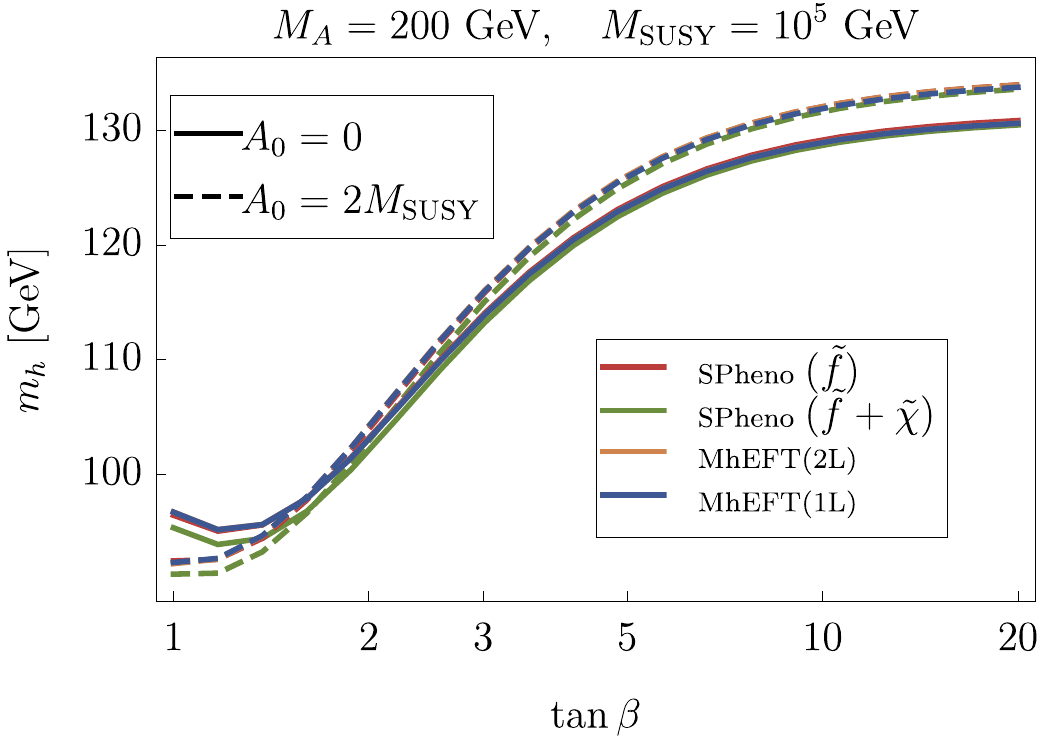} \hfill 
\includegraphics[width=0.49\linewidth]{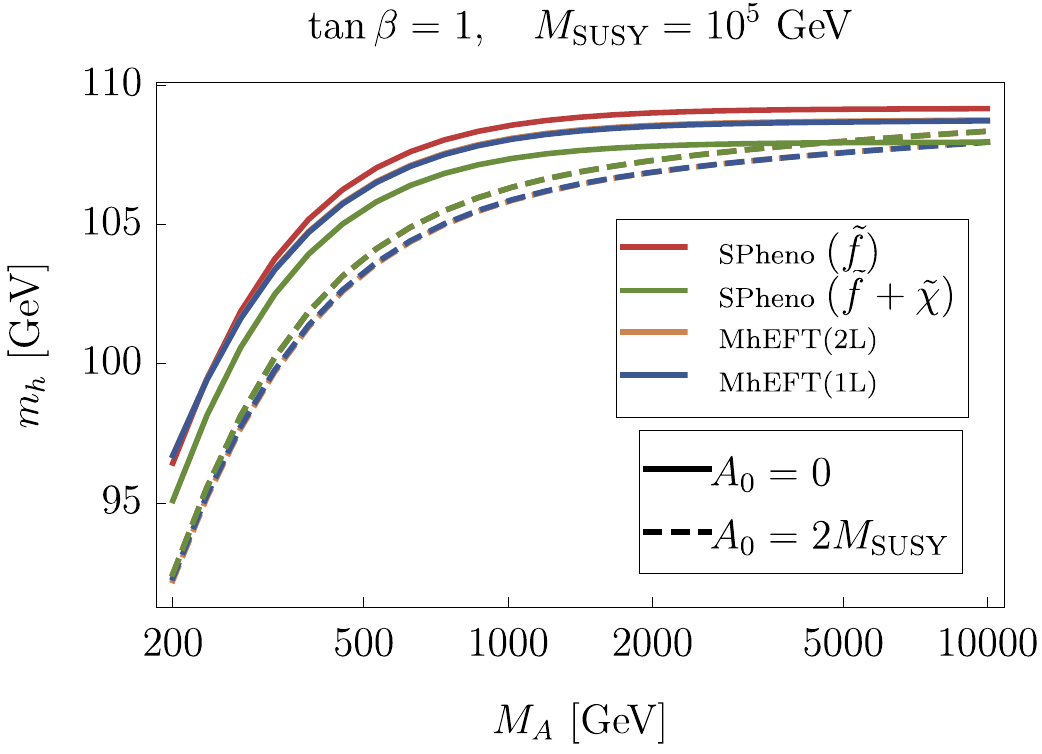} \\[5mm]
\includegraphics[width=0.49\linewidth]{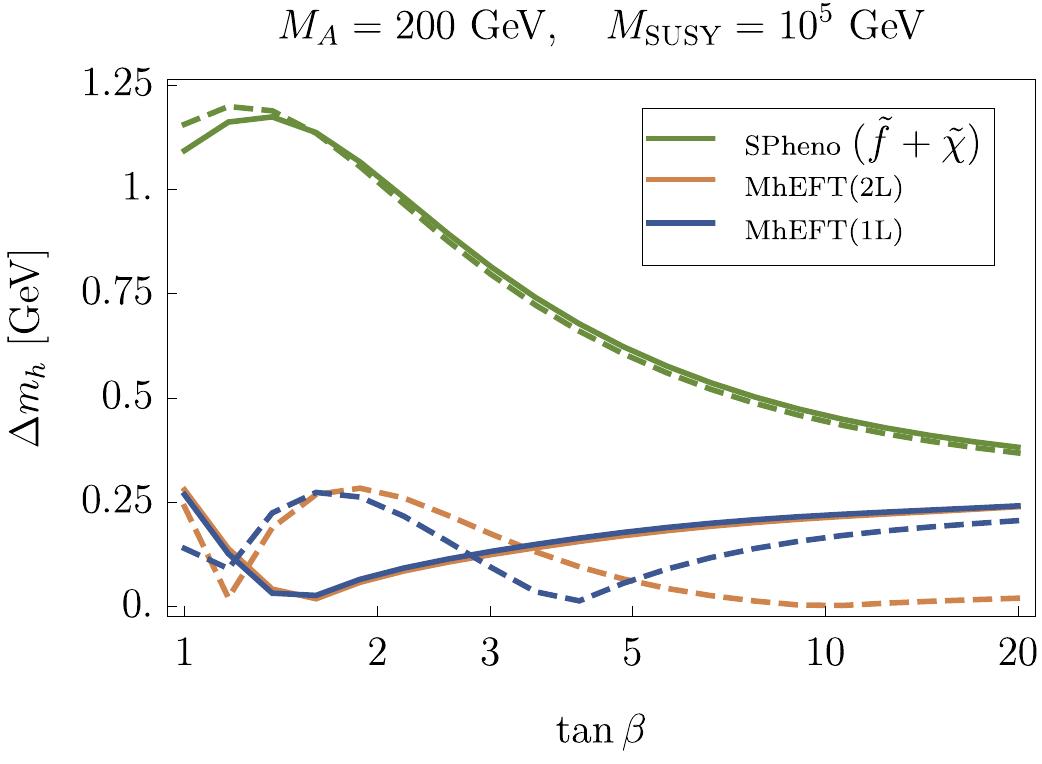} \hfill 
\includegraphics[width=0.49\linewidth]{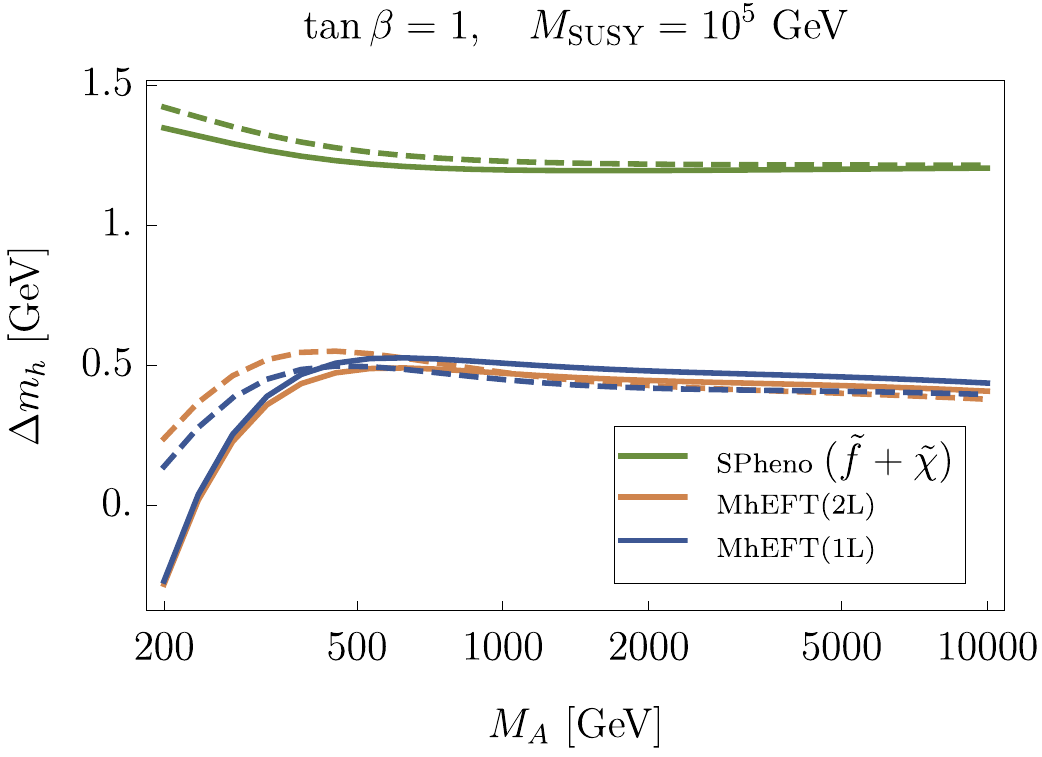} 
\caption{Comparison between {\tt MhEFT} and \SPheno for a variation of
    $\tan\beta$ (left) or $M_A$ (right). All SUSY masses are assumed to be
    degenerate and identical to $M_{\rm SUSY}$, and the results for two
different values of $A_0$ are shown.}
\label{fig:MhEFT}
\end{figure}
The dominant threshold corrections to $\lambda_1$ -- $\lambda_7$ involving
third generation Yukawa couplings are available in literature
\cite{Haber:1990aw}. We have double checked the analytical expressions derived
by \SARAH and found full agreement. 
\\

The importance of the proper matching to the THDM for the case $M_A \ll M_{\rm
SUSY}$ has been pointed out in
Ref.~\cite{Lee:2015uza}. It was found that in
particular for small $\tan\beta$ very large difference to a one-scale matching
appear. In order to demonstrate that, we compare in
\cref{fig:lowMA_vs_HighScale} the Higgs boson mass prediction using the proper
matching of the MSSM to the THDM against the simplified ansatz of decoupling
the second Higgs doublet together with all other BSM states at $M_{\rm SUSY}$.
First of all, one can check that the results for the matching to the SM change
only moderately when using the actual value of $M_A$ in the one-scale matching
compared to the fully degenerate case $M_A=m_0=M_{\rm SUSY}$. This only causes
a shift of at most $\sim$1~GeV for $\tan\beta=1$. On the other side, there are
big difference showing up when performing the matching to the THDM. For values
of $\tan\beta$ close to 1, the discrepancy can be as large as 10~GeV, while it
rapidly decreases with increasing $\tan\beta$. For $\tan\beta=10$, the
differences between both matching approaches are about 1~GeV. 
\\

Since we have demonstrated the importance of performing the matching to the THDM
properly for the case of a light second Higgs doublet, it is clear that codes
were developed to include these effects. The first tool in this direction was
{\tt MhEFT} which uses a purely EFT ansatz \cite{Lee:2015uza}. In a recent
update of {\tt FeynHiggs} a hybrid ansatz combining the fixed-order calculation
with higher-order terms was implemented \cite{Bahl:2018jom}.The overall
agreement between both codes turned out to be good once a careful translation
between the parameters in both renormalisation schemes was done. Since {\tt
MhEFT} is much closer to the ansatz of \SARAH/\SPheno we are going to compare
our results with this tool\footnote{For simplicity, we modified {\tt MhEFT} to
take $A_t$ as input instead of $X_t^{\MS}$.}. For this purpose, we have set
up the model 
\begin{center}
{\tt HighScaleSUSY/MSSMlowMA} 
\end{center}
in \SARAH. We show in \cref{fig:MhEFT} the results  of {\tt MhEFT} and \SPheno
when varying $\tan\beta$ or $M_A$ for a fixed SUSY scale of 100~TeV. The agreement 
between both codes is always good. The maximal difference
for comparable calculations is about 0.5~GeV and can be even smaller for $M_A$
below 500~GeV and arbitrary values of $\tan\beta$. The
differences are due to the three-loop RGEs which are
included in {\tt MhEFT} in the running between $m_t$ and $M_A$ while \SPheno
uses always two-loop RGEs. This explains the flattening of the difference as
the top quark Yukawa coupling runs fastest near the weak scale. 
One can also see that the impact of the additional two-loop corrections implemented in {\tt MhEFT} is very moderate. 
Thus the main source of the difference is the determination of the running top
Yukawa coupling. In contrast, the additional one-loop corrections due to
gauginos, which were presented very recently also in Ref.~\cite{Bahl:2018jom},
can be easily included in \SPheno using the numerical matching interface. 
For the considered choice of parameters
these have numerically a bigger effect than the two-loop corrections and cause
a shift of 1--1.5~GeV.

\subsection{High-scale NMSSM}
\begin{figure}[tb]
\centering
\includegraphics[width=0.49\linewidth]{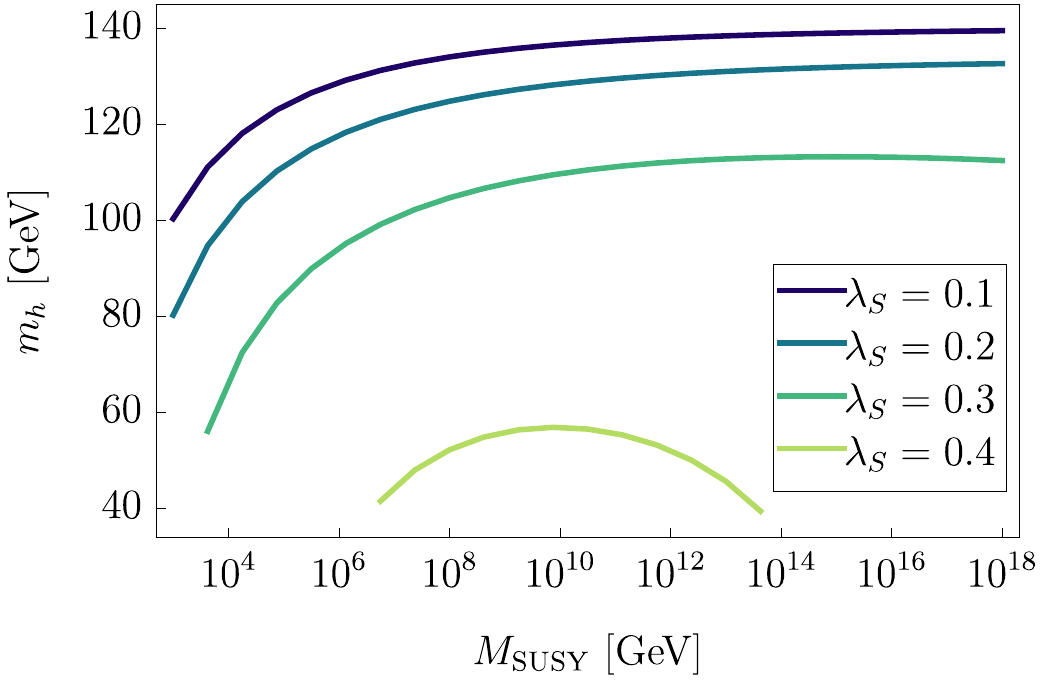}  
\caption{The Higgs boson mass in a high-scale SUSY scenario based on the NMSSM with fixed $\tan\beta=4$. All
    SUSY fields including the singlets are (nearly) degenerate with a mass of
    $M_{\rm SUSY}$. The given values of $\lambda_S$ determine the coupling
    strength of the gauge singlet and the two Higgs doublets in the NMSSM at
    the matching scale.}
\label{fig:NMSSM}
\end{figure}

Up to now we have only discussed examples of models involving very heavy BSM
particles which could already be studied with public tools like {\tt SusyHD},
{\tt MhEFT}, {\tt FlexibleSUSY} or {\tt FeynHiggs}. These are just different
low-energy limits of the MSSM. However, our framework is not restricted to this
case and in principle any SUSY or non-SUSY model could be considered as
high-scale theory. We show that crucial differences compared to the MSSM show
already up in the case of the NMSSM. The NMSSM involves an additional gauge
singlet superfield $\hat S$ which leads to the following superpotential after
imposing a $Z_3$ symmetry to forbid all dimensionful parameters
\begin{equation}
W_{\rm NMSSM} = \lambda_S \hat S \hat H_d \hat H_u +  \frac13 \kappa
\hat S^3 + W_Y \,,
\end{equation}
where $W_Y$ represents the terms involving Yukawa couplings that are identical
to the MSSM. The NMSSM-specific soft-SUSY breaking parameters are
\begin{equation}
- \lagr_{\rm SB, \ NMSSM} =  m_S^2 |S|^2 + (T_\lambda S H_d H_u +
\frac13 T_\kappa S^3 + c.c.) \,.
\end{equation}
The scalar singlet $S$ can receive a VEV even without EWSB
\begin{equation}
\langle S \rangle = \frac{1}{\sqrt{2}} v_S 
\end{equation}
which causes an effective Higgsino mass term 
\begin{equation}
\mu_{\rm eff} = \frac{1}{\sqrt{2}} \lambda_S v_S \,.
\end{equation}
We can now study what the impact of the additional gauge singlet in a
high-scale SUSY scenario is. For this purpose, we impose the following relation
among the parameters 
\begin{align}
\kappa = & \frac12 \lambda_S \\
\mu_{\rm eff} = &  m_0 \\
T_\kappa = & - \frac13 \lambda_S m_0   \\
T_\lambda = & \frac12 \lambda_S m_0 \left(\sin(2\beta) -1\right) \,.
\end{align}
This leads to a nearly degenerate spectrum of SUSY fields with masses of
$m_0$ apart from one CP-even singlet which has a mass of $\sqrt{\frac23} m_0$. Thus, the
EFT model is again the SM, i.e.
\begin{equation}
 \lagr_{\rm EFT} =  \lagr_{\rm SM}
\end{equation}
The full high-scale model has three free parameters 
\begin{eqnarray}
m_0, \quad \lambda_S, \quad \tan\beta \, .
\end{eqnarray}
The MSSM limit is obtained for $\lambda_S \to 0$. We have implemented this
model in \SARAH as
\begin{center}
{\tt HighScaleSUSY/NMSSM} 
\end{center}
The predicted mass for the SM-like Higgs boson as function of the SUSY scale $M_{\rm
SUSY}$ is shown in \cref{fig:NMSSM} for different values of $\lambda_S$. Thus,
one can see that there are significant shifts in the Higgs boson mass already for
values of $\lambda_S$ of 0.2 or 0.3. In general, one finds that the Higgs boson mass
decreases with increasing $\lambda_S$. The main reason for this are tree-level
contributions proportional to $\frac{T_\lambda^2}{m_S^2} \simeq \frac14
\lambda_S^2$ which dominate for small $\tan\beta$ over the D-term
contributions. Thus, the conclusion that the maximal possible SUSY scale in
agreement with $m_h$ is about $10^{10}$~GeV only holds for the MSSM,
while in the NMSSM one can push $M_{\rm SUSY}$ towards the Planck scale without
being in conflict with Higgs boson mass measurements. 
Of course, one could now start to consider also other low-energy limits of the
NMSSM. However, this is beyond the scope of this paper here and interesting
applications are given elsewhere
\cite{SplitNMSSM}.
\\

\section{Summary}
\label{sec:summary}
We have presented an extension of the \Mathematica package \SARAH which derives
the one-loop matching conditions for effective scalar couplings based on a UV
theory. Two different approaches exists, which are based on either an
analytical or fully numerical calculation. The full agreement between both
calculations and analytical results available in literature has been pointed
out. Furthermore, good agreement with specialised codes to study Split- or
High-scale SUSY like {\tt SusyHD} or {\tt MhEFT} was shown. Since our approach
is completely general, it can be used to study UV completions of a large variety of BSM
models with and without an extended Higgs sector.

\section*{Acknowledgements}
We thank Mark Goodsell for many fruitful discussions about the matching of scalar couplings and other related topics as well as Pietro Slavich for proof reading the manuscript. 
FS is supported by the ERC Recognition Award ERC-RA-0008 of the Helmholtz Association. MG acknowledges financal support by 
the GRK 1694 "Elementary Particle Physics at Highest Energy and highest Precision".
\begin{appendix}
\section{Generic Diagrams}
\allowdisplaybreaks
\label{app:generic}
In this appendix we provide a complete list of all possible one-loop diagrams
with 2, 3 and 4 external scalars and internal fermions or scalars. The results
were obtained in the limit of vanishing external momenta using the computer programs {\tt FeynArts} and {\tt FormCalc}
\cite{Hahn:2000kx,Hahn:2009bf}.
\\

We distinguish between \textit{topologies}, where neither the statistical nature (spin=0,1,$\nicefrac{1}{2}$) nor the
mass (light or heavy i.e. zero or non-zero) of internal fields is specified and
\textit{generic diagrams}, where the spin of all fields is specified but not
their mass hierarchies. This is still a model-independent graph but gives the
possibility to write down a model-independent \textit{generic amplitude}. 
Couplings and masses appearing in the expression of a generic amplitude are seen as generic couplings that do not have any relation 
to parameters used in other generic amplitudes.
\\

\begin{figure}
    \centering
    \begin{subfigure}[t]{0.65\textwidth}
    \centering
  	\includegraphics{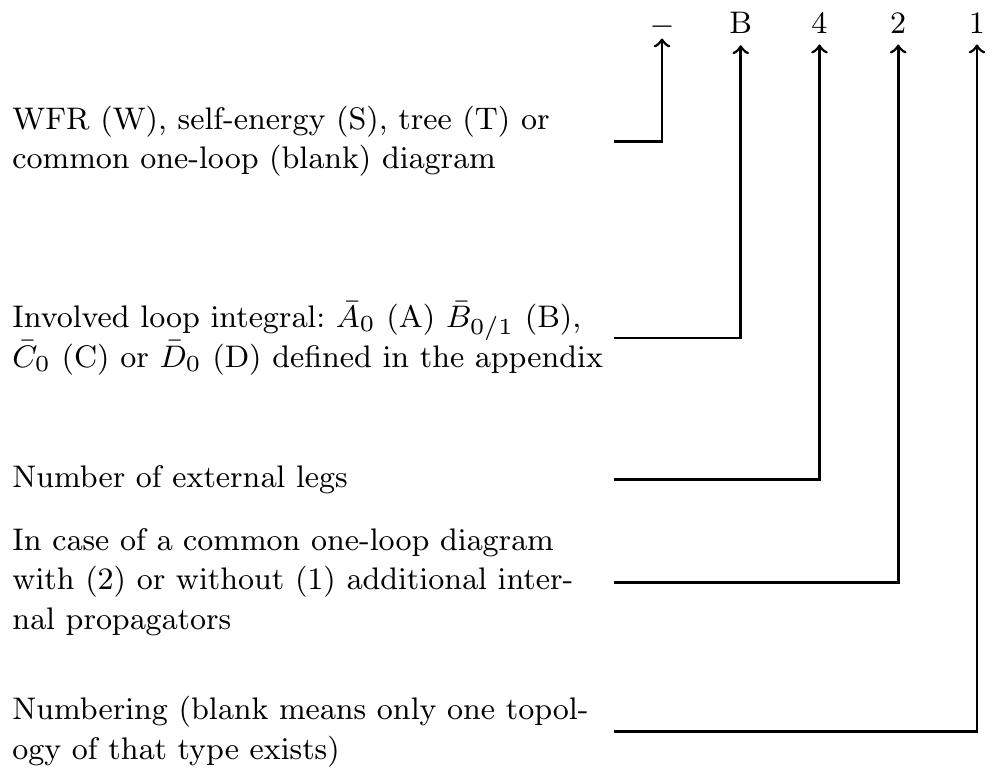}
    \caption{}
    \end{subfigure}
    \begin{subfigure}[t]{0.3\textwidth}
    \centering
    \rotatebox{90}{
  	\includegraphics{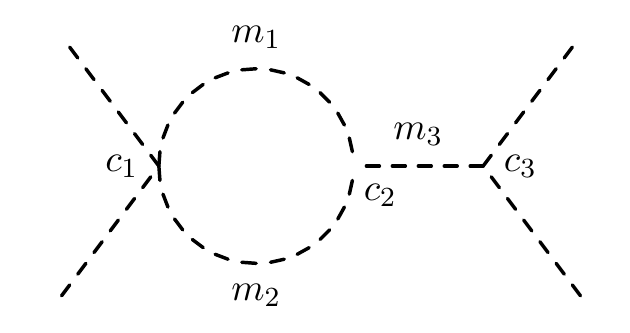}
	}
    \caption{\lstinline{B[4][2][1]}}
    \end{subfigure}
    \caption{Notation on topologies appearing in generic one-loop diagrams.
    The example expression explained in (a) corresponds for instance
    to the diagram shown in (b).}
    \label{fig:app:topo:toponot}
\end{figure}

As already mentioned in \cref{sec:implementation}, \SARAH's analytical  
matching interface does not only compute the full one-loop contribution but is
also able to compute a subset of diagrams based on e.g. a choice of topologies. For this reason,
a notation with a successive structure is introduced.
A topology is described by a string consisting of maximum four characters. It
starts with the specification of the diagram type which can be tree-level (T),
self-energy (S), WFR (W) or ordinary one-loop diagram (blank) followed by a
letter specifying the type of the involved loop integral defined in
\cref{app:loopfunctions}. In the next digit, the number of external
fields is specified. If the diagram contains an internal single propagator
additional to a loop (2) or a loop only (1) is denoted by the next digit
 (blank means that no diagrams with additional internal lines exist). 
The last digit is a counting index (blank means only one diagram of that type exists).
An example explanation of the notation is given in \cref{fig:app:topo:toponot}.
The \lstinline{Topologies} options of the matching routines can either 
select certain topologies, like \lstinline!{B[4][2][1],B[4][2][2]}! , or
successively select topology groups, like \lstinline!{B[4][2]}! or just
\lstinline!{B[4]}!. All possible topologies and groups are stored in the list \lstinline{TopoNotation}.
In the following, we list the analytical expressions for all generic amplitudes
as well as the topologies they belong to in the format used in the \SARAH
matching routines.

\subsection{Tree-Level Contributions}
\subsubsection{Quartic Couplings}
There are two tree-level topologies with four external scalars. 
The first one is a local quartic coupling which could for example be given by supersymmetric 
D-terms and/or F-terms while the second one has one
internal propagator, necessarily heavy and of bosonic nature. Thus, 
fermions can only enter one-loop and higher-order
corrections.
\begin{align}
\text{\lstinline{T[4][1]}} \equiv 
\begin{aligned}
\includegraphics{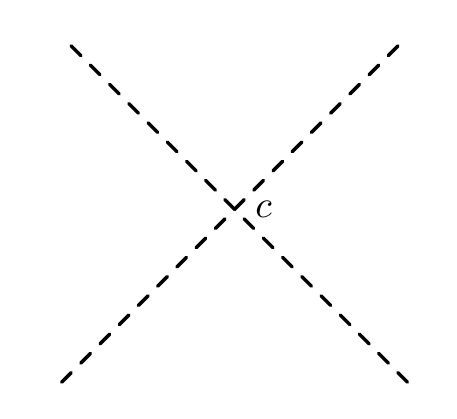} 
\end{aligned}
& \mathcal{M} = c \\
\text{\lstinline{T[4][2]}} \equiv 
\begin{aligned}
\includegraphics{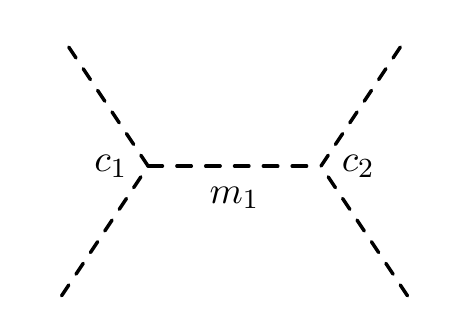}
\end{aligned}
& \mathcal{M} = -\frac{c_1 c_2}{m_1^2} 
\end{align}
\subsubsection{Cubic Couplings}
Since there is only one three-point topology, 
the matching of trilinear couplings at tree level is
trivial. 
\begin{align}
\text{\lstinline{T[3]}} \equiv 
\begin{aligned}
\includegraphics{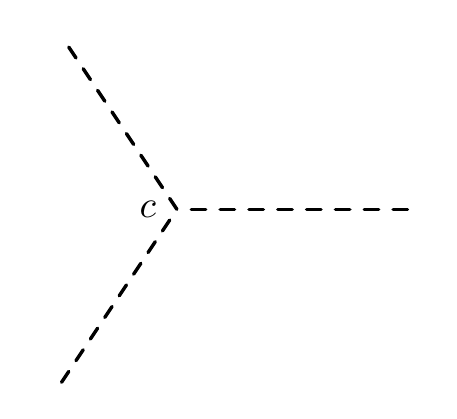}
\end{aligned}
& \mathcal{M} = c 
\end{align}
\subsubsection{Bilinear Parameters}
The two-point function is necessary for the matching of scalar sectors that involve non-Higgs scalar fields i.e. scalars that do not develop a VEV. 
In this case the scalar masses and
couplings are independent parameters and have to be matched separately.
\begin{align}
\text{\lstinline{T[2]}} \equiv 
\begin{aligned}
\includegraphics{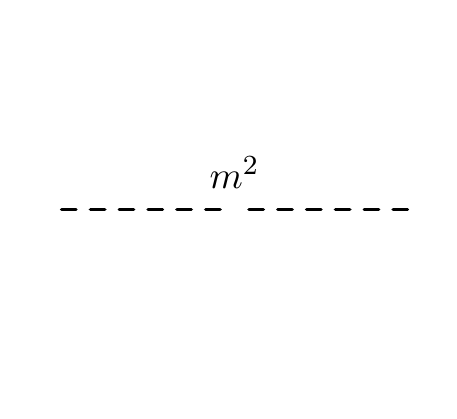}
\end{aligned}
& \mathcal{M} = m^2 
\end{align}

\subsection{One-Loop Contributions}
At the one-loop order, we distinguish between irreducible diagrams and
reducible diagrams which contain an additional internal propagator line. 
\subsubsection{Irreducible Diagrams}
\paragraph{Quartic Couplings}\mbox{}\\[2mm]
Since we consider renormalisable theories, fermions can only enter the one-loop
corrections to quartic couplings in box diagrams (topology \lstinline{D}, note that
\lstinline{D[4]} is not a valid topology as it is a reserved \Mathematica
symbol). The generic diagrams and amplitudes are 
\begin{align}
\text{\lstinline{B[4][1]}} \equiv 
\begin{aligned}
\includegraphics{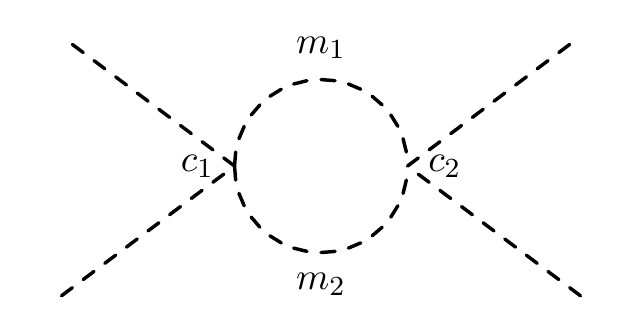}
\end{aligned}
& \mathcal{M} = \frac12 c_1 c_2 \overline{B}_0(m_1^2,m_2^2) \\
\text{\lstinline{C[4][1]}} \equiv 
\begin{aligned}
\includegraphics{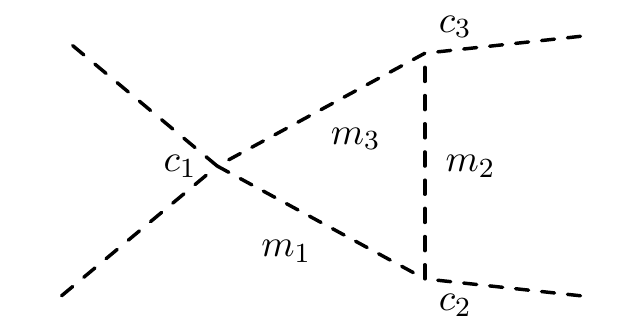}
\end{aligned}
& \mathcal{M} = -c_1 c_2 c_3 \overline{C}_0(m_1^2,m_2^2,m_3^2) \label{eq:triangle} \\
\text{\lstinline{D}} \equiv 
\begin{aligned}
\includegraphics{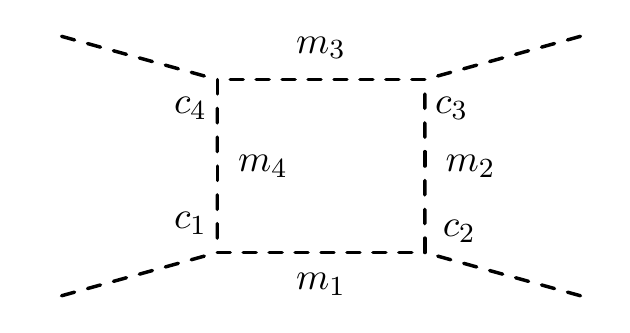}
\end{aligned}
& \mathcal{M} = c_1 c_2 c_3 c_4 \overline{D}_0(m_1^2,m_2^2,m_3^2,m_4^2)\\
\text{\lstinline{D}} \equiv 
\begin{aligned}
    \label{eq:fermionbox}
\includegraphics{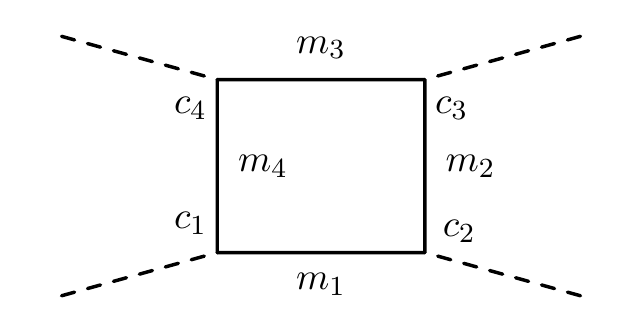}
\end{aligned}
& \mathcal{M} = \mathcal{M}_F^4 
\end{align}
with 
\begin{align}
\mathcal{M}_F^4 = & \left(c_1^R c_2^L c_3^R c_4^L + c_1^L c_2^R c_3^L c_4^R\right) \overline{B}_0(m_3^2,m_4^2)  +  \Big[(c_1^R c_2^L c_3^R c_4^L + c_1^L c_2^R c_3^L c_4^R) (m_1^2 + m_2^2)  \nonumber \\
 &+ c_4^L (c_1^L m_1 (c_2^L c_3^R m_2 + c_2^R c_3^R m_3 + c_2^R c_3^L m_4) 
  + c_1^R (c_2^L c_3^L m_3 m_4 + c_2^R m_2 (c_3^R m_3 + c_3^L m_4))) +  \nonumber \\
 &    c_4^R (c_1^R m_1 (c_2^R c_3^L m_2 + c_2^L c_3^L m_3 + c_2^L c_3^R m_4) + c_1^L (c_2^R c_3^R m_3 m_4 + c_2^L m_2 (c_3^L m_3 + c_3^R m_4)))\Big] \overline{C}_0(m_2^2,m_3^2,m_4^2) +  \nonumber \\
+&  m_1 \Big[c_1^L (c_2^L m_2 (c_3^L m_3 (c_4^R m_1 + c_4^L m_4) + c_3^R m_1 (c_4^L m_1 + c_4^R m_4)) + c_2^R m_1 (c_3^L m_1 (c_4^R m_1 + c_4^L m_4) \nonumber \\
&  + c_3^R m_3 (c_4^L m_1 + c_4^R m_4))) +  c_1^R (c_2^L m_1 (c_3^L m_3 (c_4^R m_1 + c_4^L m_4) + c_3^R m_1 (c_4^L m_1 + c_4^R m_4)) \nonumber \\
& +  c_2^R m_2 (c_3^L m_1 (c_4^R m_1 + c_4^L m_4) + c_3^R m_3 (c_4^L m_1 + c_4^R m_4)))\Big] \overline{D}_0(m_1^2,m_2^2,m_3^2,m_4^2\, ,
\end{align}
where $c_i^{R,L}$ are the left- and
right-handed fermion couplings corresponding to the vertices $c_i$ in \eqref{eq:fermionbox}. 
\paragraph{Cubic Couplings}\mbox{}\\[2mm]
The effective trilinear couplings are 
\begin{align}
\text{\lstinline{B[3]}} \equiv 
\begin{aligned}
\includegraphics{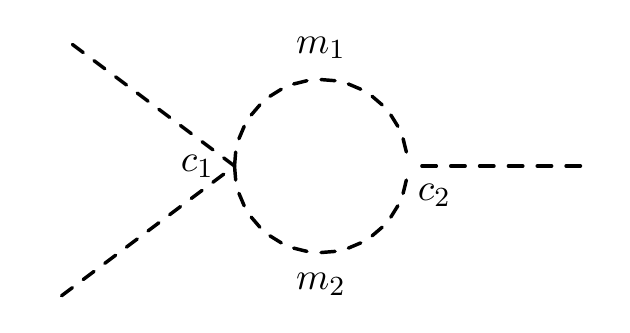}
\end{aligned}
& \mathcal{M} = \frac12 c_1 c_2 \overline{B}_0(m_1^2,m_2^2) \\
\text{\lstinline{C[3]}} \equiv 
\begin{aligned}
\includegraphics{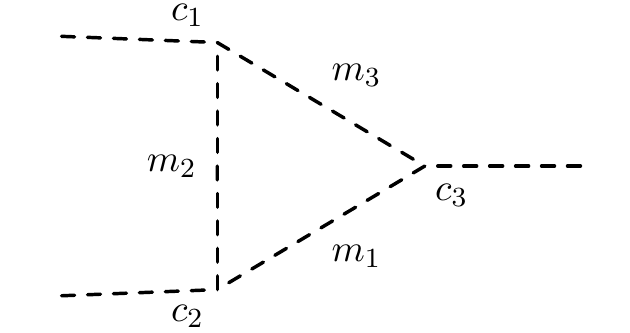}
\end{aligned}
& \mathcal{M} = -c_1 c_2 c_3 \overline{C}_0(m_1^2,m_2^2,m_3^2)\\
\text{\lstinline{C[3]}} \equiv 
\begin{aligned}
\includegraphics{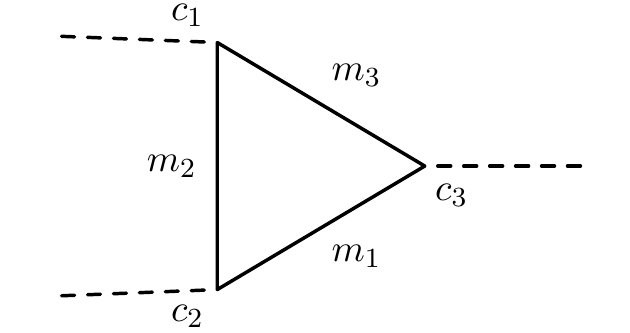}
\end{aligned}
& \mathcal{M} = \mathcal{M}_F^3 
\end{align}
with 
\begin{align}
\mathcal{M}_3^F =& (-c_1^R(c_2^Lc_3^Rm_1 + c_2^Rc_3^Lm_2 + c_2^Lc_3^Lm_3) - c_1^L(c_2^Rc_3^Lm_1 + c_2^Lc_3^Rm_2 + c_2^Rc_3^Rm_3))\overline{B}_0(m_2^2,m_3^2) \nonumber \\
& + \Big[-m_1(c_1^L(c_2^Lm_2(c_3^Rm_1 + c_3^Lm_3) + c_2^Rm_1(c_3^Lm_1 + c_3^Rm_3)) \nonumber \\
 & \hspace{2cm} +c_1^R(c_2^Lm_1(c_3^Rm_1 + c_3^Lm_3) + c_2^Rm_2(c_3^Lm_1 + c_3^Rm_3)))\Big]\overline{C}_0(m_1^2,m_2^2,m_3^2)\, .
\end{align}
\paragraph{Bilinear Parameters}\mbox{}\\[2mm]
Generic two-point amplitudes are computed with 
\begin{align}
\text{\lstinline{A[2]}} \equiv 
\begin{aligned}
\includegraphics{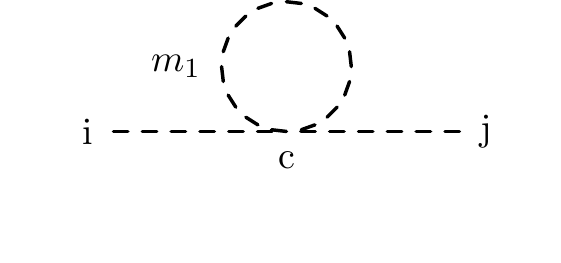}
\end{aligned}
& \mathcal{M} = c A_0(m_1^2) \\
\text{\lstinline{B[2]}} \equiv 
\begin{aligned}
\includegraphics{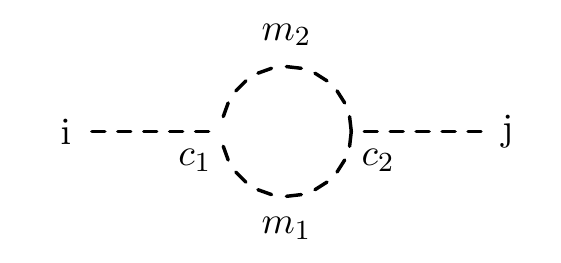}
\end{aligned}
& \mathcal{M} = c_1 c_2 \overline{B}_0(m_1^2,m_2^2) \\
\text{\lstinline{B[2]}} \equiv 
\begin{aligned}
\includegraphics{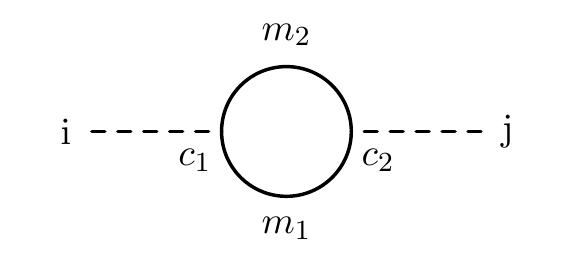}
\end{aligned}
& 
\mathcal{M} = \mathcal{M}_2^{F}
\end{align}
with 
\begin{align}
 \mathcal{M}_2^{F} = 
    \frac12 \left(c_1^R c_2^L + c_1^L c_2^R\right) \left( A_0(m_1^2) + A_0(m_2^2) +
    \left(m_1^2+ m_2^2\right)  \overline{B}_0(m_1^2,m_2^2) \right) \nonumber \\ 
+ m_1 m_2 \left(c_1^L c_2^L+c_1^R c_2^R\right) \overline{B}_0(m_1^2,m_2^2) \, .
\end{align}

\subsubsection{Reducible Diagrams}
The reducible diagrams are selected with a "2" in the second digit of the
topology. Diagrams are skipped if the additional internal propagators are not
heavy. 
\begin{align}
\text{\lstinline{B[4][2][1]}} \equiv 
\begin{aligned}
\includegraphics{FeynMF/FM_Bprop}
\end{aligned}
& \mathcal{M} = \frac{1}{m_3^2} c_1 c_2 c_3 \overline{B}_0(m_1^2,m_2^2) \\
\text{\lstinline{B[4][2][2]}} \equiv 
\begin{aligned}
\includegraphics{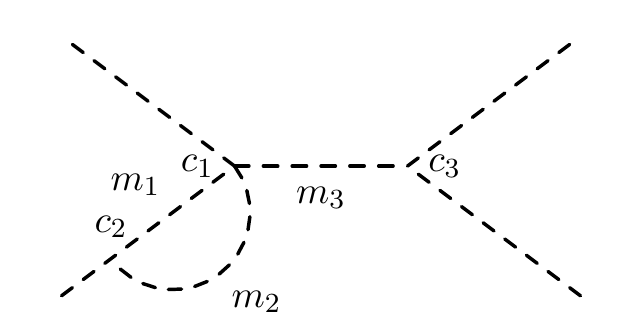}
\end{aligned}
& \mathcal{M} = \frac{1}{m_3^2} c_1 c_2 c_3 \overline{B}_0(m_1^2,m_2^2) \\
\text{\lstinline{C[4][2]}} \equiv 
\begin{aligned}
\includegraphics{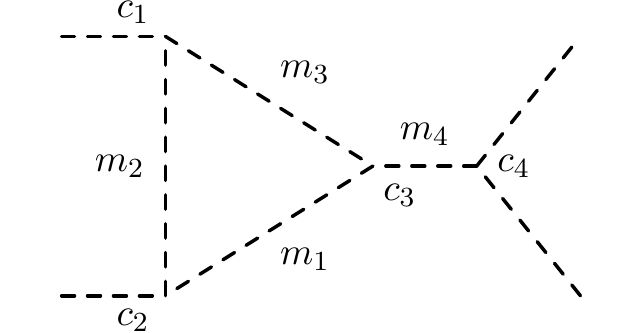}
\end{aligned}
& \mathcal{M} = - \frac{1}{m_4^2} c_1 c_2 c_3 c_4 \overline{C}_0(m_1^2,m_2^2,m_3^2) \\
\text{\lstinline{C[4][2]}} \equiv 
\begin{aligned}
\includegraphics{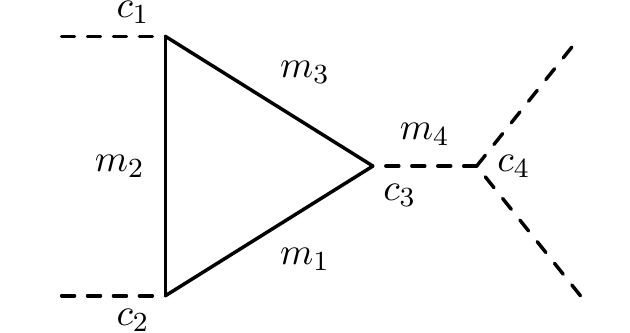}
\end{aligned}
& \mathcal{M} = \mathcal{M}_{3'}^F \\
\text{\lstinline{SB[4]}} \equiv 
\begin{aligned}
\includegraphics{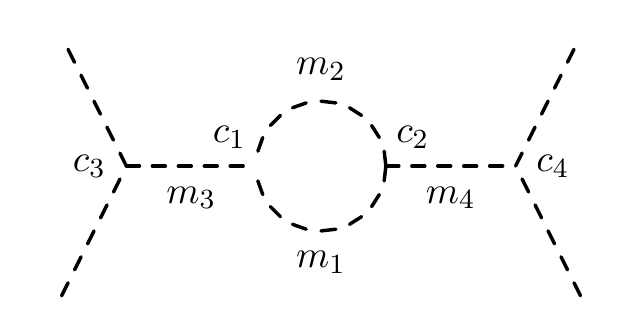}
\end{aligned}
& \mathcal{M} = \frac{1}{2 m_3^2 m_4^2} c_1 c_2 c_3 c_4 \overline{B}_0(m_1^2,m_2^2) \\
\text{\lstinline{SB[4]}} \equiv 
\begin{aligned}
\includegraphics{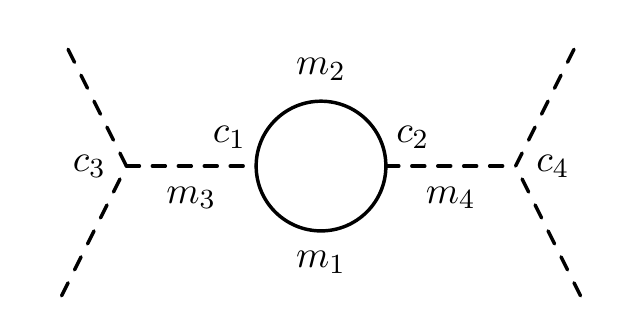}
\end{aligned}
& \mathcal{M} = \mathcal{M}_4^{SF} \\
\text{\lstinline{SA[4]}} \equiv 
\begin{aligned}
\includegraphics{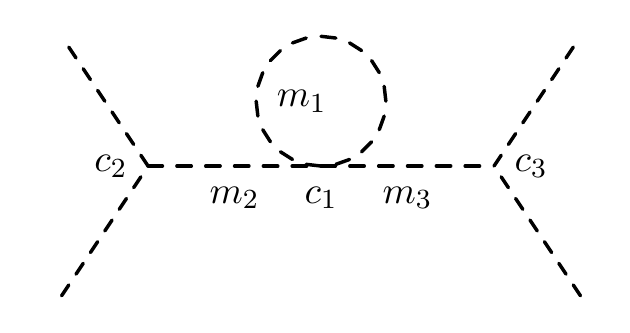}
\end{aligned}
& \mathcal{M} = -\frac{1}{m_2^2 m_3^2} c_1 c_2 c_3 A_0(m_1^2) 
\end{align}

with 
\begin{align}
\mathcal{M}_{3'}^F =& \frac{-2 c_4}{m_4^2} \Big[(-c_1^R(c_2^Lc_3^Rm_1 + c_2^Rc_3^Lm_2 + c_2^Lc_3^Lm_3) - c_1^L(c_2^Rc_3^Lm_1 + c_2^Lc_3^Rm_2 + c_2^Rc_3^Rm_3))\overline{B}_0(m_2^2,m_3^2) \nonumber \\
& + \big[-m_1 \big(c_1^L(c_2^Lm_2(c_3^Rm_1 + c_3^Lm_3) + c_2^Rm_1(c_3^Lm_1 + c_3^Rm_3)) \nonumber \\ 
& \hspace{1cm} +   c_1^R(c_2^Lm_1(c_3^Rm_1 + c_3^Lm_3) +
  c_2^Rm_2(c_3^Lm_1 +
  c_3^Rm_3))\big)\big]\overline{C}_0(m_1^2,m_2^2,m_3^2) \Big] 
\end{align}
and
\begin{align}
\mathcal{M}_4^{SF} =& 
  -\frac{2 c_3 c_4}{m_3^2 m_4^2} \Big[m_1m_2(c_1^Lc_2^L+c_1^Rc_2^R)\overline{B}_0(m_1^2,m_2^2) \nonumber \\
& + \left(\frac12(c_1^Rc_2^L+ c_1^Lc_2^R)(A_0(m_1^2)+ A_0(m_2^2) + (m_1^2 + m_2^2)  \overline{B}_0(m_1^2,m_2^2))\right) \Big]
      \, .
\end{align}

\subsubsection{Wave-Function Corrections}
Contributions on external legs are divided into diagonal and off-diagonal WFR
topologies named \lstinline{DiagonalWFRs} and \lstinline{OffdiagonalWFRs}. They
consist of a Z-factor times the tree-level amplitude times a combinatorial
factor. In the following, we give the analytical expressions for the Z
factors.

\begin{align}
\begin{aligned}
\includegraphics{FeynMF/FM_waveS1}
\end{aligned}
& Z = \left\{\begin{array}{cc} c_1 c_2 \dot{\overline{B}}_0(m_1^2,m_2^2) & i=j \\ \frac{c_1 c_2}{m_i^2-m_j^2} \overline{B}_0(m_1^2,m_2^2) & i \neq j \end{array} \right.  \\
\begin{aligned}
\includegraphics{FeynMF/FM_waveF}
\end{aligned}
& Z = \left\{\begin{array}{cc} Z_F^{\rm dia} & i=j\\ Z_F^{\rm off} & i \neq j \end{array} \right. \\
\begin{aligned}
\includegraphics{FeynMF/FM_waveS2}
\end{aligned}
& Z = \left\{\begin{array}{cc} 0 & i=j \\ \frac{c}{m_i^2-m_j^2} A_0(m_1^2) & i \neq j \end{array} \right.  
\end{align}
with
\begin{align}
    Z_F^{\rm dia} =& \frac12 (c_1^R c_2^L+ c_1^L c_2^R)\left((- m_1^2 - m_2^2) \dot{\overline{B}}_0(m_1^2,m_2^2) + \overline{B}_0(m_1^2,m_2^2) \right) \nonumber \\ 
                   &-m_1 m_2 (c_1^L c_2^L+c_1^R c_2^R) \dot{\overline{B}}_0(m_1^2,m_2^2)\\[0.8cm]
 Z_F^{\rm off} =& -\frac{1}{m_i^2-m_j^2}\Big(\frac12 (c_1^R c_2^L + c_1^L c_2^R) \left( A_0(m_1^2) + A_0(m_2^2) + (m_1^2+ m_2^2)  \overline{B}_0(m_1^2,m_2^2) \right) \nonumber \\ 
  &\hspace{2cm} + m_1 m_2 (c_1^L c_2^L+c_1^R c_2^R)
    \overline{B}_0(m_1^2,m_2^2)\Big) \, ,
\end{align}
where the dotted notation is introduced in \cref{app:loopfunctions}. 

\section{IR-safe Loop-Functions}
\label{app:loopfunctions}
In this appendix we give analytical expressions for all loop functions used
in the matching routines. In particular, we list the limits for all possible
combinations of vanishing and equal masses as they are needed to provide
numerical stability of the matching routines.
\\
The common prefactor
\begin{equation}
    \kappa_D = \frac{\left(2\pi Q\right)^{4-D}}{i\pi^2}
\end{equation}
and the integrand $I_n$ with vanishing external momenta 
\begin{equation}
I_n = \prod_{i=1}^n \left( q^2 - m_i^2 \right)^{-1}
\end{equation}
simplify the definitions of the loop integrals
\begin{equation}
\begin{aligned}
    A_0(m^2)                     & = \kappa_D \int d^Dq I_1 \,\, , \\
    B_0(m_1^2,m_2^2)             & = \kappa_D \int d^Dq I_2 \,\, , \\
    C_0(m_1^2,m_2^2,m_3^2)       & = \kappa_D \int d^Dq I_3 \,\, , \\
    D_0(m_1^2,m_2^2,m_3^2,m_4^2) & = \kappa_D \int d^Dq I_4 \,\, .
\end{aligned}
\end{equation}
The integrand $I_n$ is symmetric w.r.t to the masses and thus also the loop
functions are symmetric w.r.t. their arguments. One can reduce all integrals to
the $A_0$ integral by using partial fractioning and integration by parts. 
We define the following abbreviations for finite logarithmic terms
\begin{equation}
    t_i \equiv
    \log \frac{m_i^2}{Q^2} \,,
    \qquad t_{ij} \equiv \log\frac{m_i^2}{m_j^2},
\end{equation}
as well as for diverging terms
\begin{equation}
    \begin{aligned}
        \Delta_t &\equiv \lim_{\epsilon_{UV}^2\to 0} \log\frac{\epsilon_{UV}^2}{Q^2} ,\\
        \Delta_\epsilon &\equiv \lim_{\epsilon_{UV}^2\to 0}
        \frac{1}{-\epsilon_{UV}^2}. \\
    \end{aligned}
\end{equation}
The IR-safe loop functions implemented in \SARAH are obtained by 
\begin{equation}
    \overline{X} = \lim_{\Delta_t\to 0}\lim_{\Delta_\epsilon\to 0} X\, , \quad
    X=A_0,B_0,C_0,D_0,B_1 \,,
\end{equation}
as these divergences would cancel in the matching conditions.

\subsection{One- and Two-Point Integrals}
The tadpole integral $A_0$ is given by
\begin{align}
        A_0(m^2) &= m^2(1 - t) \,\, ,\\
        A_0(0)   &= 0\,\, ,
\end{align}
whereas the two-point integral $B_0$ is
\begin{align}
    B_0(m_1^2,m_2^2) &= \frac{A_0(m_1^2)-A_0(m_2^2)}{m_1^2-m_2^2} \\
                         &= 1 - t_2 - \frac{m_1^2}{m_1^2-m_2^2}\, t_{12} \,\, ,\\
        B_0(m^2,0) &= 1 - t  \,\, , \\
        B_0(m^2,m^2) &= -t  \,\, ,\\
        B_0(0,0) &= - \Delta_t  \,\, .
\end{align}
In addition, the tensor integral
\begin{equation}
    B_1(m_1^2,m_2^2) = \frac{1}{p^2}p^\mu B^\mu =\frac{\kappa_D}{p^2}\int  d^Dq \,\, pq \left[ \left(q^2-m_1^2\right)\left(q+p\right)^2-m_2^2\right]^{-1}
\end{equation}
in the limit of vanishing external momentum reads
\begin{align}
    B_1(m_1^2,m_2^2) &= -\frac{1}{4(m_1^2-m_2^2)}\left[ m_1^2 + m^2_2 - 2 m_1^2 B_0(m_1^2,0) \right. \nonumber\\
                     & \qquad \left. + (4 m_1^2 - 2 m_2^2) B_0(m_1^2, m_2^2)
\right] \,\, ,\\
    B_1(m^2,0)       &= -\frac{1}{4}\left(1+ 2B_0(m_1^2,0)\right)\,\, ,\\
    B_1(m^2,m^2)     &= \frac{1}{2} t \,\, ,\\
    B_1(0,0)     &= \frac{1}{2} \Delta_t \,\, .
\end{align}

\subsection{Triangle Integrals}
The three-point function can be simplified with the definitions
\begin{equation}
    q_1 \equiv \frac{m_2^2}{m_1^2} \,, \qquad q_2\equiv\frac{m_3^2}{m_1^2},
\end{equation}
which yields
\begin{align}
    C_0(m_1^2,m_2^2,m_3^2)     &= \frac{B_0(m_1^2,m_3^2)-B_0(m_2^2,m_3^2)}{m_1^2-m_2^2}\\
                               &=  -\frac{q_1\, t_{21}\,(q_2 -1) + q_2\, t_{32}\,(1-q_1)}{m_1^2\,(-1+q_1)(q_1-q_2)(-1+q_2)} \,\, ,\\
        C_0(m_1^2,m_1^2,m_2^2) &= \frac{m_2^2-m_1^2 +m_2^2\,t_{12}}{\left(m_1^2-m_2^2\right)^2}\,\, ,\\
        C_0(m^2,m^2,m^2)       &= -\frac{1}{2 m^2} \,\, ,\\
        C_0(m_1^2,m_2^2,0)     &= -\frac{t_{12}}{m_1^2-m_2^2}\,\, ,\\
        C_0(m^2,m^2,0)         &= -\frac{1}{m^2}\,\, ,\\
        C_0(m^2,0,0)           &= \frac{1}{m^2} - \frac{t}{m^2} + \frac{\Delta_t}{m^2} \,\, ,\\
        C_0(0,0,0)             &= \frac{1}{2} \Delta_\epsilon\,\, . 
\end{align}

\subsection{Box Integrals}
Analogously we compute the four-point integral in all mass combinations,
\begin{align}
        D_0(m_1^2,m_2^2,m_3^2,m_4^2) &= \frac{1}{m_1^2-m_2^2}\left(C_0(m_1^2,m_3^2,m_4^2) - C_0(m_2^2,m_3^2,m_4^2) \right)  \,\, ,\\
        D_0(m_1^2,m_1^2,m_2^2,m_3^2) &= \frac{1}{m_2^2-m_3^2}\left( \frac{-m_1^2+m_2^2+ m_2^2\, t_{12}}{\left(m_2^2-m_1^2\right)^2}   + \frac{m_1^2-m_3^2-m_3^2\, t_{13}}{\left(m_3^2-m_1^2\right)^2}  \right) \,\, ,\\
        D_0(m_1^2,m_1^2,m_2^2,m_2^2) &= \frac{1}{\left(m_1^2-m_2^2\right)^3}\left( -2m_1^2+2m_2^2+ (m_1^2+m_2^2)\,t_{12}  \right) \,\, ,\\
        D_0(m_1^2,m_1^2,m_1^2,m_2^2) &= \frac{-m_1^2 + m_2^2 + 2m_1^2\, m_2^2\, t_{12}}{2m_1^2\,\left(m_2^2-m_1^2\right)^3} \,\, , \\
        D_0(m_1^2,m_1^2,m_1^2,m_1^2) &= \frac{1}{6 m_1^4 \,\, ,}\\
\end{align}
and with at least one vanishing mass,
\begin{align}
        D_0(m_1^2,m_2^2,m_3^2,0) &= \frac{(m_2^2-m_3^2)\, t_{31} + (m_3^2-m_1^2)\, t_{32}}{(m_1^2-m_2^2)(m_1^2-m_3^2)(m_2^2-m_3^2)} \,\, , \\
        D_0(m_1^2,m_2^2,m_2^2,0) &= \frac{m_1^2-m_2^2+m_2^2\, t_{21}}{m_2^2\,\left(m_1^2-m_2^2\right)}  \,\, ,\\
        D_0(m^2,m^2,m^2,0) &= \frac{1}{2m^4} \,\, ,\\
        D_0(m_1^2,m_2^2,0,0) &= \frac{t_2}{m_1^2 m_2^2} + \frac{t_{21}}{m_1^2(m_1^2-m_2^2)} - \frac{\Delta_t}{m_1^2 m_2^2} \,\, ,\\
        D_0(m^2,m^2,0,0) &= -\frac{2}{m^4} + \frac{t}{m^4} - \frac{\Delta_t}{m^4}  \,\, ,\\
        D_0(m^2,0,0,0) &=\frac{1}{m^4} - \frac{t}{m^4} + \frac{\Delta_t}{m^4} - \frac{\Delta_\epsilon}{2 m^2}  \,\, ,\\
        D_0(0,0,0,0) &= \frac{1}{6} \Delta_\epsilon^2\,\, .
\end{align}

\subsection{Derivatives of the Loop-Functions}
\begin{align}
    \dot{B}_0(m_1^2,m_2^2) & = \kappa_D \partial_p^2 \left. \int d^Dq \left( (q+p)^2 - m_2^2 \right)^{-1}\left( q^2 - m_1^2 \right)^{-1} \right|_{p^2\to 0} \\
                              & = \frac{1}{2\left(m_1^2-m_2^2\right)^2} \left( m_1^2 + m_2^2 + \frac{2m_1^2 m_2^2 t_{21}}{m_1^2 - m_2^2}  \right) \,\, .
\end{align}
The different mass combinations read
\begin{align}
        \dot{B}_0(m^2,m^2) &= \frac{1}{6 m^2} \,\, ,\\
        \dot{B}_0(m^2,0) &= \frac{1}{2 m^2}\,\, ,\\
        \dot{B}_0(0,0) &= -\frac{1}{6} \Delta_\epsilon\,\, ,
\end{align}
and analogously for the tensor integral
\begin{align}
\dot{B}_1(m_1^2,m_2^2) =& \frac{-2 m_1^6-3 m_1^4 m_2^2 \left(2\log\frac{m_2^2}{m_1^2}+1\right)+6 m_1^2 m_2^4-m_2^6}{6 (m_1^2-m_2^2)^4} \,\, ,\\
\dot{B}_1(m_1^2,m_1^2) =& -\frac{1}{12 m_1^2}  \,\, ,\\
\dot{B}_1(0,m_2^2) =& \frac{1}{6 m_2^2}  \,\, ,\\
\dot{B}_1(m_1^2,0) =& \frac{1}{3 m_1^2}  \,\, ,\\
\dot{B}_1(0,0) =& \frac{\Delta_\epsilon}{12}  \,\, .
\end{align}

\section{{\tt SPheno.m} for the MSSM}
\label{app:sphenomssm}
A version of the {\tt SPheno.m} file to set up the MSSM with CMSSM boundary conditions is
\begin{lstlisting}[language=SLHA,title=SPheno.m]
(*----------------------------------*)
(* General information *)
(*----------------------------------*)

(* Matching conditions to the SM *)
DEFINITION[MatchingConditions]=Default[THDMII];

(* List for two- and three-body decays *)

ListDecayParticles = Automatic;
ListDecayParticles3B = Automatic;


(*----------------------------------*)
(* Information for High-Scale input *)
(*----------------------------------*)

(* CMSSM input parameters *)
MINPAR =    {{1,m0},
             {2,m12},
             {3,TanBeta},
             {4,SignumMu},
             {5,Azero}};

ParametersToSolveTadpoles = {\[Mu],B[\[Mu]]};

(* Choice for the renormalisation scale *)
RenormalizationScaleFirstGuess = m0^2 + 4 m12^2;
RenormalizationScale = Su[1]*Su[6];

(* Condition to determine the GUT scale *)
ConditionGUTscale = g1 == g2;

(* GUT conditions  *)
BoundaryHighScale={
{T[Ye], Azero*Ye},
{T[Yd], Azero*Yd},
{T[Yu], Azero*Yu},
{mq2, DIAGONAL m0^2},
{ml2, DIAGONAL m0^2},
{md2, DIAGONAL m0^2},
{mu2, DIAGONAL m0^2},
{me2, DIAGONAL m0^2},
{mHd2, m0^2},
{mHu2, m0^2},
{MassB, m12},
{MassWB,m12},
{MassG,m12}
};
\end{lstlisting}
The content of the \SPheno specific input file for \SARAH is the
following: 
\begin{enumerate}
\item Input parameter ({\tt MINPAR}, {\tt EXTPAR}): a list of parameters which
    should be read by \SPheno from the block {\tt MINPAR} or {\tt EXTPAR} in a
    LesHouches file. Note that there are no hard coded entries for {\tt MINPAR}
    or {\tt EXTPAR}. This makes it necessary to define these blocks also for
    models with already existing SLHA conventions. However, this also provides
    more freedom in varying the model and the free parameters.
    \\
\item {\tt RealParameters}: By default, all parameters defined in {\tt MINPAR}
or {\tt EXTPAR} are assumed to be complex, i.e. it is possible to use also the block
{\tt IMMINPAR} to define the imaginary part. However, some
Fortran functions like {\tt sin} cannot be used with complex numbers, therefore
is is necessary to define parameters like \(\tan\beta\) explicitly as real.
\item {\tt ParametersToSolveTadpoles}: The set of parameters that are fixed by the tadpole equations. 
\item  {\tt RenormalizationScaleFirstGuess}: For the first run of the RGEs,
before any mass has been calculated by \SPheno, one can define an approximate renormalisation scale. 
\item {\tt RenormalizationScale}: For all further runs, another renormalisation
scale can be given which can be for instance the function of the calculated masses or running parameters.
\item A condition can be defined to obtain a dynamically adjusted GUT scale.
    Here, we defined that the GUT scale is the scale at which the EW gauge
    couplings are identical.
\item Boundary Condition: It is possible to define boundary conditions at different scales:
\begin{itemize}
\item Electroweak scale: {\tt BoundaryEWSBScale}
\item Renormalisation or SUSY scale: {\tt BoundarySUSYScale}/{\tt BoundaryRenScale}
\item GUT scale: {\tt BoundaryHighScale} 
\end{itemize}
\item Decays: One can tell \SARAH that it should make use of the default conventions to write code to calculate two- and three-body decays with \SPheno. 
This includes the decays for all BSM and Higgs states, but not for SM fermions and vector bosons. 
\end{enumerate}

\section{{\tt SPheno.m} for an MSSM Version with Three Matching Scales}
\label{app:3scales}
\begin{lstlisting}[language=SLHA,title=SPheno.m]
MINPAR={{ 1, m0},
        { 2, mGaugino},
        { 3, TanBeta},
        { 5, Azero},
        {23, Muinput},
        {24, MAinput}     
        };
        
RealParameters = {TanBeta,m0,MAinput};        

RenormalizationScaleFirstGuess = 100^2;
RenormalizationScale = v^2;


UseParameterAsGUTscale = {m0};

BoundaryHighScale={
 {T[Ye], Azero*Ye},
 {T[Yd], Azero*Yd},
 {T[Yu], Azero*Yu},
 {mq2, DIAGONAL m0^2},
 {ml2, DIAGONAL m0^2},
 {md2, DIAGONAL m0^2},
 {mu2, DIAGONAL m0^2},
 {me2, DIAGONAL m0^2}
};


ParametersToSolveTadpoles = {m2};

DEFINITION[MatchingConditions]=Default[OHDM];

ListDecayParticles = Automatic;
ListDecayParticles3B = Automatic;

DefaultInputValues ={m0 -> 10^6, mGaugino -> 10^6, Azero -> 0, Muinput->10^6, MAinput->10^6, TanBeta->2};


(*---------------------------------------------------*)
(* information for matching to THDM -- THDM + EWkinos -- MSSM at three threshold scales *)
(*---------------------------------------------------*)

MatchingToModel= {"THDMnoZ2","THDM+EWkinos","MSSM"};
MatchingScale = {MAinput,mGaugino,m0};

IncludeParticlesInThresholds={
 {hh,Ah,Hpm},
 {Chi,Cha },
 {Su,Sd,Se,Sv}
};

AssumptionsMatchingScale={
 {
  {v1,epsUV*Cos[ArcTan[TanBeta]]}, 
  {v2,epsUV*Sin[ArcTan[TanBeta]]}
 },
 {
  {vd,epsUV}, 
  {vu,epsUV},
  {M12,epsUV^2}
 },
 {
  {vd,epsUV}, 
  {vu,epsUV},
  {B[\[Mu]],epsUV^2},
  {\[Mu],epsUV},
  {MassB,epsUV},
  {MassWB,epsUV}  
 }
};

BoundaryMatchingScaleUp={
  (* SM -> THDM *)
{
 {M12, -MAinput^2 TanBeta/(1+TanBeta^2)},
  {Yu, Sqrt[1+TanBeta^2]/TanBeta*Yu - Sqrt[2]*ShiftCoupNLO[bar[Fu].Fu.hh[1]][PL]/ZH[1,2]},
  {Yd, Sqrt[1+TanBeta^2]*Yd - Sqrt[2]*ShiftCoupNLO[bar[Fd].Fd.hh[1]][PL]/ZH[1,1]},
  {Ye, Sqrt[1+TanBeta^2]*Ye - Sqrt[2]*ShiftCoupNLO[bar[Fe].Fe.hh[1]][PL]/ZH[1,1]}
},
  (* THDM -> THDM +EWkinos *)
{  
 {Yu, (Yu - Sqrt[2]*ShiftCoupNLO[bar[Fu].Fu.hh[2]][PL])},
 {Yd, (Yd - Sqrt[2]*ShiftCoupNLO[bar[Fd].Fd.hh[1]][PL])},
 {Ye, (Ye - Sqrt[2]*ShiftCoupNLO[bar[Fe].Fe.hh[1]][PL])},
 {MuSplit,MuInput},
 {MG, mGaugino},
 {MW, mGaugino},
 {MB, mGaugino}
},
  (* THDM + EWkinos -> MSSM *)
{
 {Yu, (Yu - Sqrt[2]*ShiftCoupNLO[bar[Fu].Fu.hh[2]][PL])},
 {Yd, (Yd - Sqrt[2]*ShiftCoupNLO[bar[Fd].Fd.hh[1]][PL])},
 {Ye, (Ye - Sqrt[2]*ShiftCoupNLO[bar[Fe].Fe.hh[1]][PL])},
 {B[\[Mu]], -M12},
 {MassB, MB},
 {MassWB, MW},
 {MassG, MG}
 }
};

BoundaryMatchingScaleDown={
 (* THDM -> SM *)
{
 (* {M12, -MAinput^2 TanBeta/(1+TanBeta^2)}, *)
  {lamSM, -1/3 EFTcoupNLO[hh[1].hh[1].hh[1].hh[1]]}
},
 (* THDM + EWkinos-> THDM *)
{
 (* {MuSplit,MuInput},
  {MG, mGaugino},
  {MW, mGaugino},
  {MB, mGaugino},  *)
  {Lambda1, -1/6 EFTcoupNLO[hh[1].hh[1].hh[1].hh[1]]},
  {Lambda2, -1/6 EFTcoupNLO[hh[2].hh[2].hh[2].hh[2]]},
  {Lambda3, -EFTcoupNLO[hh[1].hh[1].Hm[2].conj[Hm[2]]]},
  {Lambda4,  -EFTcoupNLO[hh[1].hh[2].conj[Hm[1]].Hm[2]] - I*EFTcoupNLO[hh[1].Ah[2].conj[Hm[2]].Hm[1]]},
  {Lambda5,  -EFTcoupNLO[hh[1].hh[2].conj[Hm[1]].Hm[2]] + I*EFTcoupNLO[hh[1].Ah[2].conj[Hm[2]].Hm[1]]},
  {Lambda6, -EFTcoupNLO[hh[1].hh[2].Hm[1].conj[Hm[1]]]},
  {Lambda7, -EFTcoupNLO[hh[1].hh[2].Hm[2].conj[Hm[2]]]}
 },
 (* MSSM -> THDM + EWkinos *)
 {
 {MG, MassG},
 {MW, MassWB},
 {MB, MassB},
 {MuSplit,\[Mu]},
 {M12, -B[\[Mu]]},
 {g1u,-2*(EFTcoupNLO[Chi[fB].Chi[FHu0].hh[2]][PL])},
 {g2u,2*(EFTcoupNLO[Chi[fW0].Chi[FHu0].hh[2]][PL])}, 
 {g1d,2*(EFTcoupNLO[Chi[fB].Chi[FHd0].hh[1]][PL])},
 {g2d,-2*(EFTcoupNLO[Chi[fW0].Chi[FHd0].hh[1]][PL])},
 {Lew1, -1/6 EFTcoupNLO[hh[1].hh[1].hh[1].hh[1]]},
 {Lew2, -1/6 EFTcoupNLO[hh[2].hh[2].hh[2].hh[2]]},
 {Lew3, -EFTcoupNLO[hh[1].hh[1].Hpm[2].conj[Hpm[2]]]},
 {Lew4,  EFTcoupNLO[hh[1].hh[2].Hpm[2].conj[Hpm[1]]] + I*EFTcoupNLO[hh[1].Ah[2].Hpm[1].conj[Hpm[2]]]},
 {Lew5,  EFTcoupNLO[hh[1].hh[2].Hpm[2].conj[Hpm[1]]] - I*EFTcoupNLO[hh[1].Ah[2].Hpm[1].conj[Hpm[2]]]},
 {Lew6, -EFTcoupNLO[hh[1].hh[2].Hpm[1].conj[Hpm[1]]]},
 {Lew7, -EFTcoupNLO[hh[1].hh[2].Hpm[2].conj[Hpm[2]]]}
 }
};

ParametersToSolveTadpoleMatchingScale={
 {M112,M222}, 
 {M112,M222},
 {mHd2,mHu2}
}; 
\end{lstlisting}

\end{appendix}

\bibliographystyle{ArXiv}
\bibliography{lit}

\providecommand{\bysame}{\leavevmode\hbox to3em{\hrulefill}\thinspace}
\begin{thebibliography}{10}

\bibitem{Aad:2012tfa}
ATLAS, G.~Aad et~al., Phys. Lett. \textbf{B716} (2012), 1--29,  [1207.7214].

\bibitem{Chatrchyan:2012xdj}
CMS, S.~Chatrchyan et~al., Phys. Lett. \textbf{B716} (2012), 30--61,
  [1207.7235].

\bibitem{ArkaniHamed:2004fb}
N.~Arkani-Hamed and S.~Dimopoulos, JHEP \textbf{06} (2005), 073,
  [hep-th/0405159].

\bibitem{Giudice:2004tc}
G.~F. Giudice and A.~Romanino, Nucl. Phys. \textbf{B699} (2004), 65--89,
  [hep-ph/0406088], [Erratum: Nucl. Phys.B706,487(2005)].

\bibitem{ArkaniHamed:2004yi}
N.~Arkani-Hamed, S.~Dimopoulos, G.~F. Giudice, and A.~Romanino, Nucl. Phys.
  \textbf{B709} (2005), 3--46,  [hep-ph/0409232].

\bibitem{Antoniadis:2006eb}
I.~Antoniadis, K.~Benakli, A.~Delgado, M.~Quiros, and M.~Tuckmantel, Nucl.
  Phys. \textbf{B744} (2006), 156--179,  [hep-th/0601003].

\bibitem{Antoniadis:2005em}
I.~Antoniadis, A.~Delgado, K.~Benakli, M.~Quiros, and M.~Tuckmantel, Phys.
  Lett. \textbf{B634} (2006), 302--306,  [hep-ph/0507192].

\bibitem{Bhattacharyya:2012ct}
G.~Bhattacharyya and T.~S. Ray, JHEP \textbf{05} (2012), 022,  [1201.1131].

\bibitem{Giudice:2011cg}
G.~F. Giudice and A.~Strumia, Nucl. Phys. \textbf{B858} (2012), 63--83,
  [1108.6077].

\bibitem{Bagnaschi:2014rsa}
E.~Bagnaschi, G.~F. Giudice, P.~Slavich, and A.~Strumia, JHEP \textbf{09}
  (2014), 092,  [1407.4081].

\bibitem{Batell:2015fma}
B.~Batell, G.~F. Giudice, and M.~McCullough, JHEP \textbf{12} (2015), 162,
  [1509.00834].

\bibitem{Zarate:2016jch}
L.~Zarate, JHEP \textbf{07} (2016), 102,  [1601.05946].

\bibitem{Unwin:2012fj}
J.~Unwin, Phys. Rev. \textbf{D86} (2012), 095002,  [1210.4936].

\bibitem{Benakli:2013msa}
K.~Benakli, L.~Darmé, M.~D. Goodsell, and P.~Slavich, JHEP \textbf{05} (2014),
  113,  [1312.5220].

\bibitem{Dudas:2013gga}
E.~Dudas, M.~Goodsell, L.~Heurtier, and P.~Tziveloglou, Nucl. Phys.
  \textbf{B884} (2014), 632--671,  [1312.2011].

\bibitem{Fox:2014moa}
P.~J. Fox, G.~D. Kribs, and A.~Martin, Phys. Rev. \textbf{D90} (2014), no.~7,
  075006,  [1405.3692].

\bibitem{Benakli:2015ioa}
K.~Benakli, L.~Darmé, and M.~D. Goodsell, JHEP \textbf{11} (2015), 100,
  [1508.02534].

\bibitem{Arvanitaki:2005fa}
A.~Arvanitaki, C.~Davis, P.~W. Graham, A.~Pierce, and J.~G. Wacker, Phys. Rev.
  \textbf{D72} (2005), 075011,  [hep-ph/0504210].

\bibitem{Ibe:2013oha}
M.~Ibe, T.~T. Yanagida, and N.~Yokozaki, JHEP \textbf{08} (2013), 067,
  [1303.6995].

\bibitem{Aad:2015zhl}
ATLAS, CMS, G.~Aad et~al., Phys. Rev. Lett. \textbf{114} (2015), 191803,
  [1503.07589].

\bibitem{Bernal:2007uv}
N.~Bernal, A.~Djouadi, and P.~Slavich, JHEP \textbf{07} (2007), 016,
  [0705.1496].

\bibitem{Arvanitaki:2012ps}
A.~Arvanitaki, N.~Craig, S.~Dimopoulos, and G.~Villadoro, JHEP \textbf{02}
  (2013), 126,  [1210.0555].

\bibitem{Vega:2015fna}
J.~Pardo~Vega and G.~Villadoro, JHEP \textbf{07} (2015), 159,  [1504.05200].

\bibitem{Athron:2016fuq}
P.~Athron, J.-h. Park, T.~Steudtner, D.~Stöckinger, and A.~Voigt, JHEP
  \textbf{01} (2017), 079,  [1609.00371].

\bibitem{Staub:2017jnp}
F.~Staub and W.~Porod, Eur. Phys. J. \textbf{C77} (2017), no.~5, 338,
  [1703.03267].

\bibitem{Allanach:2018fif}
B.~C. Allanach and A.~Voigt,  (2018),  1804.09410.

\bibitem{Sasaki:1991qu}
K.~Sasaki, M.~Carena, and C.~E.~M. Wagner, Nucl. Phys. \textbf{B381} (1992),
  66--86.

\bibitem{Carena:1995bx}
M.~Carena, J.~R. Espinosa, M.~Quiros, and C.~E.~M. Wagner, Phys. Lett.
  \textbf{B355} (1995), 209--221,  [hep-ph/9504316].

\bibitem{Haber:1990aw}
H.~E. Haber and R.~Hempfling, Phys. Rev. Lett. \textbf{66} (1991), 1815--1818.

\bibitem{Haber:1996fp}
H.~E. Haber, R.~Hempfling, and A.~H. Hoang, Z. Phys. \textbf{C75} (1997),
  539--554,  [hep-ph/9609331].

\bibitem{Carena:2000dp}
M.~Carena, H.~E. Haber, S.~Heinemeyer, W.~Hollik, C.~E.~M. Wagner, and
  G.~Weiglein, Nucl. Phys. \textbf{B580} (2000), 29--57,  [hep-ph/0001002].

\bibitem{Carena:2000yi}
M.~Carena, J.~R. Ellis, A.~Pilaftsis, and C.~E.~M. Wagner, Nucl. Phys.
  \textbf{B586} (2000), 92--140,  [hep-ph/0003180].

\bibitem{Carena:2001fw}
M.~Carena, J.~R. Ellis, A.~Pilaftsis, and C.~E.~M. Wagner, Nucl. Phys.
  \textbf{B625} (2002), 345--371,  [hep-ph/0111245].

\bibitem{Espinosa:2001mm}
J.~R. Espinosa and I.~Navarro, Nucl. Phys. \textbf{B615} (2001), 82--116,
  [hep-ph/0104047].

\bibitem{Gorbahn:2009pp}
M.~Gorbahn, S.~Jager, U.~Nierste, and S.~Trine, Phys. Rev. \textbf{D84} (2011),
  034030,  [0901.2065].

\bibitem{Lee:2015uza}
G.~Lee and C.~E.~M. Wagner, Phys. Rev. \textbf{D92} (2015), no.~7, 075032,
  [1508.00576].

\bibitem{Hahn:2013ria}
T.~Hahn, S.~Heinemeyer, W.~Hollik, H.~Rzehak, and G.~Weiglein, Phys. Rev. Lett.
  \textbf{112} (2014), no.~14, 141801,  [1312.4937].

\bibitem{Bahl:2017aev}
H.~Bahl, S.~Heinemeyer, W.~Hollik, and G.~Weiglein, Eur. Phys. J. \textbf{C78}
  (2018), no.~1, 57,  [1706.00346].

\bibitem{Bahl:2016brp}
H.~Bahl and W.~Hollik, Eur. Phys. J. \textbf{C76} (2016), no.~9, 499,
  [1608.01880].

\bibitem{Bahl:2018jom}
H.~Bahl and W.~Hollik,  (2018),  1805.00867.

\bibitem{Braathen:2018htl}
J.~Braathen, M.~D. Goodsell, and P.~Slavich,  (2018),  1810.09388.

\bibitem{Staub:2008uz}
F.~Staub,  (2008),  0806.0538.

\bibitem{Staub:2009bi}
F.~Staub, Comput.Phys.Commun. \textbf{181} (2010), 1077--1086,  [0909.2863].

\bibitem{Staub:2010jh}
F.~Staub, Comput.Phys.Commun. \textbf{182} (2011), 808--833,  [1002.0840].

\bibitem{Staub:2012pb}
F.~Staub,  (2012),  1207.0906.

\bibitem{Staub:2013tta}
F.~Staub, Comput. Phys. Commun. \textbf{185} (2014), 1773--1790,  [1309.7223].

\bibitem{Porod:2003um}
W.~Porod, Comput.Phys.Commun. \textbf{153} (2003), 275--315,  [hep-ph/0301101].

\bibitem{Porod:2011nf}
W.~Porod and F.~Staub,  (2011),  1104.1573.

\bibitem{Bagnaschi:2017xid}
E.~Bagnaschi, J.~Pardo~Vega, and P.~Slavich, Eur. Phys. J. \textbf{C77} (2017),
  no.~5, 334,  [1703.08166].

\bibitem{Martin:1993yx}
S.~P. Martin and M.~T. Vaughn, Phys. Lett. \textbf{B318} (1993), 331--337,
  [hep-ph/9308222].

\bibitem{Machacek:1983tz}
M.~E. Machacek and M.~T. Vaughn, Nucl. Phys. \textbf{B222} (1983), 83--103.

\bibitem{Machacek:1983fi}
M.~E. Machacek and M.~T. Vaughn, Nucl. Phys. \textbf{B236} (1984), 221--232.

\bibitem{Machacek:1984zw}
M.~E. Machacek and M.~T. Vaughn, Nucl. Phys. \textbf{B249} (1985), 70--92.

\bibitem{Martin:1993zk}
S.~P. Martin and M.~T. Vaughn, Phys. Rev. \textbf{D50} (1994), 2282,
  [hep-ph/9311340], [Erratum: Phys. Rev.D78,039903(2008)].

\bibitem{Luo:2002ti}
M.-x. Luo, H.-w. Wang, and Y.~Xiao, Phys. Rev. \textbf{D67} (2003), 065019,
  [hep-ph/0211440].

\bibitem{Fonseca:2011vn}
R.~M. Fonseca, M.~Malinsky, W.~Porod, and F.~Staub, Nucl. Phys. \textbf{B854}
  (2012), 28--53,  [1107.2670].

\bibitem{Goodsell:2012fm}
M.~D. Goodsell, JHEP \textbf{01} (2013), 066,  [1206.6697].

\bibitem{Fonseca:2013bua}
R.~M. Fonseca, M.~Malinský, and F.~Staub, Phys. Lett. \textbf{B726} (2013),
  882--886,  [1308.1674].

\bibitem{Sperling:2013eva}
M.~Sperling, D.~Stöckinger, and A.~Voigt, JHEP \textbf{07} (2013), 132,
  [1305.1548].

\bibitem{Sperling:2013xqa}
M.~Sperling, D.~Stöckinger, and A.~Voigt, JHEP \textbf{01} (2014), 068,
  [1310.7629].

\bibitem{Schienbein:2018fsw}
I.~Schienbein, F.~Staub, T.~Steudtner, and K.~Svirina,  (2018),  1809.06797.

\bibitem{Goodsell:2014bna}
M.~D. Goodsell, K.~Nickel, and F.~Staub, Eur. Phys. J. \textbf{C75} (2015),
  no.~1, 32,  [1411.0675].

\bibitem{Goodsell:2015ira}
M.~Goodsell, K.~Nickel, and F.~Staub, Eur. Phys. J. \textbf{C75} (2015), no.~6,
  290,  [1503.03098].

\bibitem{Braathen:2017izn}
J.~Braathen, M.~D. Goodsell, and F.~Staub, Eur. Phys. J. \textbf{C77} (2017),
  no.~11, 757,  [1706.05372].

\bibitem{Goodsell:2017pdq}
M.~D. Goodsell, S.~Liebler, and F.~Staub, Eur. Phys. J. \textbf{C77} (2017),
  no.~11, 758,  [1703.09237].

\bibitem{Porod:2014xia}
W.~Porod, F.~Staub, and A.~Vicente, Eur. Phys. J. \textbf{C74} (2014), no.~8,
  2992,  [1405.1434].

\bibitem{Goodsell:2015yca}
M.~D. Goodsell, K.~Nickel, and F.~Staub, Phys. Lett. \textbf{B758} (2016),
  18--25,  [1511.01904].

\bibitem{Allanach:2008qq}
B.~Allanach, C.~Balazs, G.~Belanger, M.~Bernhardt, F.~Boudjema, et~al.,
  Comput.Phys.Commun. \textbf{180} (2009), 8--25,  [0801.0045].

\bibitem{Athron:2014yba}
P.~Athron, J.-h. Park, D.~Stöckinger, and A.~Voigt, Comput. Phys. Commun.
  \textbf{190} (2015), 139--172,  [1406.2319].

\bibitem{Athron:2017fvs}
P.~Athron, M.~Bach, D.~Harries, T.~Kwasnitza, J.-h. Park, D.~Stöckinger,
  A.~Voigt, and J.~Ziebell, Comput. Phys. Commun. \textbf{230} (2018),
  145--217,  [1710.03760].

\bibitem{SplitNMSSM}
M.~Gabelmann, M.~M{\"u}hlleitner, and F.~Staub,  in preparation.

\bibitem{Hahn:2000kx}
T.~Hahn, Comput. Phys. Commun. \textbf{140} (2001), 418--431,
  [hep-ph/0012260].

\bibitem{Hahn:2009bf}
T.~Hahn, PoS \textbf{ACAT08} (2008), 121,  [0901.1528].

\end{thebibliography}

\end{document}